\documentclass[10pt,dvipsnames]{article}
\usepackage[preprint]{tmlr}
\usepackage{blindtext}
\usepackage{lscape}

\usepackage{tabularray}
\usepackage{paralist}
\usepackage{todonotes}
\usepackage{booktabs}
\usepackage{adjustbox}
\usepackage{multicol}
\usepackage{multirow}
\usepackage{pifont}
\usepackage{pgfplots}
\pgfplotsset{compat=1.17} 
\usepackage[edges]{forest}
\usepackage{caption}
\usepackage{subcaption}
\usepackage{stmaryrd}
\usepackage{makecell}
\usepackage{tikz}
\usepackage{wrapfig}
\usetikzlibrary{trees}
\usetikzlibrary{positioning}

\definecolor{citecolor}{HTML}{0071bc}
\usepackage[colorlinks=true,allcolors=citecolor]{hyperref}
\hypersetup{linktocpage}

\newtheorem{definition}{Definition}[section]

\definecolor{bloodred}{RGB}{153,0,0}

\newcommand\encirclered[1]{%
	\tikz[baseline=(X.base)]
	\node (X) [draw, fill=bloodred, text=white, font=\bfseries, shape=circle, inner sep=0] {\strut #1};}

\makeatletter
\DeclareRobustCommand\citeps
{\begingroup\footnotesize\NAT@swatrue\let\NAT@ctype\z@\NAT@partrue
    \@ifstar{\NAT@fulltrue\NAT@citetp}{\NAT@fullfalse\NAT@citetp}}
\makeatother

\usepackage{lastpage}

\usepackage{amsmath,amsfonts,bm}

\newcommand*\rot{\rotatebox{90}}
\newcommand*\OK{\ding{51}}

\def\Secref#1{Section~\ref{#1}}

\def\eqref#1{equation~\ref{#1}}

\def\1{\bm{1}}

\def\rmF{{\mathbf{F}}}

\def\vzero{{\bm{0}}}

\def\vmu{{\bm{\mu}}}
\def\vtheta{{\bm{\theta}}}

\def\mE{{\bm{E}}}

\def\mR{{\bm{R}}}

\DeclareMathAlphabet{\mathsfit}{\encodingdefault}{\sfdefault}{m}{sl}
\SetMathAlphabet{\mathsfit}{bold}{\encodingdefault}{\sfdefault}{bx}{n}

\def\gA{{\mathcal{A}}}

\def\gC{{\mathcal{C}}}

\def\gE{{\mathcal{E}}}

\def\gG{{\mathcal{G}}}

\def\gL{{\mathcal{L}}}

\def\gN{{\mathcal{N}}}
\def\gO{{\mathcal{O}}}

\def\gQ{{\mathcal{Q}}}
\def\gR{{\mathcal{R}}}
\def\gS{{\mathcal{S}}}
\def\gT{{\mathcal{T}}}

\def\gV{{\mathcal{V}}}

\def\fQ{{\mathfrak{Q}}}

\newcommand{\Em}[1]{\mathbf{#1}}

\newcommand{\dqtr}{\llbracket q\rrbracket_\texttt{train}}
\newcommand{\dqv}{\llbracket q\rrbracket_\texttt{val}}
\newcommand{\dqte}{\llbracket q\rrbracket_\texttt{test}}
\newcommand{\gtrain}{\gG_{\textit{train}}}
\newcommand{\gval}{\gG_{\textit{val}}}
\newcommand{\gtest}{\gG_{\textit{test}}}
\newcommand{\ginf}{\gG_{\textit{inf}}}
\newcommand{\etrain}{\gE_{\textit{train}}}

\newcommand{\einf}{\gE_{\textit{inf}}}
\newcommand{\rtrain}{\gR_{\textit{train}}}

\newcommand{\rinf}{\gR_{\textit{inf}}}
\newcommand{\strain}{\gS_{\textit{train}}}

\newcommand{\sinf}{\gS_{\textit{inf}}}

\newcommand{\CC}{\mathbb{C}} 
\newcommand{\RR}{\mathbb{R}} 
\newcommand{\clqa}{CLQA\xspace}
\newcommand{\longclqa}{Complex logical query answering\xspace}

\newcommand{\ngdb}{NGDB\xspace}

\newcommand{\abox}{\textsc{ABox}}
\newcommand{\tbox}{\textsc{TBox}}
\newcommand{\con}{\texttt{Con}}
\newcommand{\term}{\texttt{Term}}
\newcommand{\variable}{\texttt{Var}}
\newcommand{\CQn}{CQ\textsubscript{neg}}
\newcommand{\CQ}{CQ}
\newcommand{\scoredomain}{\ensuremath{\mathbb{R}}}
\newcommand{\UCQn}{UCQ\textsubscript{neg}}
\newcommand{\UCQ}{UCQ}

\DeclareMathOperator{\join}{join}

\newcommand{\ie}{\textit{i.e.}}
\newcommand{\eg}{\textit{e.g.}}

\begin{document}

\title{Neural Graph Reasoning: Complex Logical Query Answering Meets Graph Databases}

\author{%
\renewcommand*{\thefootnote}{\arabic{footnote}}%
\name Hongyu Ren\textsuperscript{*}\,\footnotemark[1] \email hyren@cs.stanford.edu
       \AND
       \name Mikhail Galkin\textsuperscript{*}\,\footnotemark[2]\email mikhail.galkin@intel.com
        \AND
       \name Michael Cochez\footnotemark[3] \email m.cochez@vu.nl
       \AND
       \name Zhaocheng Zhu\footnotemark[4] \,\footnotemark[5] \email zhaocheng.zhu@umontreal.ca
       \AND
       \name Jure Leskovec\footnotemark[1] \email jure@cs.stanford.edu
       \AND \\
    \textsuperscript{*} \textmd{\textbf{Equal contribution}} \hspace{.2cm} \\
    \footnotemark[1]       \addr Stanford University
    \footnotemark[2]       \addr Intel AI Lab
    \footnotemark[3]       \addr Vrije Universiteit Amsterdam and Elsevier discovery lab, Amsterdam, the Netherlands
    \footnotemark[4]       \addr Mila - Qu\'ebec AI Institute
    \footnotemark[5]       \addr Universit\'e de Montr\'eal
}


\maketitle

\begin{abstract}

\longclqa (\clqa) is a recently emerged task of graph machine learning that goes beyond simple one-hop link prediction and solves a far more complex task of multi-hop logical reasoning over massive, potentially incomplete graphs in a latent space. 
The task received a significant traction in the community; numerous works expanded the field along theoretical and practical axes to tackle different types of complex queries and graph modalities with efficient systems. 
In this paper, we provide a holistic survey of \clqa with a detailed taxonomy studying the field from multiple angles, including graph types (modality, reasoning domain, background semantics), modeling aspects (encoder, processor, decoder), supported queries (operators, patterns, projected variables), datasets, evaluation metrics, and applications.  

Refining the \clqa task, we introduce the concept of \emph{Neural Graph Databases} ({\ngdb}s). 
Extending the idea of graph databases (graph DBs), \ngdb consists of a \emph{Neural Graph Storage} and a \emph{Neural Graph Engine}. 
Inside Neural Graph Storage, we design a graph store, a feature store, and further embed information in a latent embedding store using an encoder. 
Given a query, Neural Query Engine learns how to perform query planning and execution in order to efficiently retrieve the correct results by interacting with the Neural Graph Storage.
Compared with traditional graph DBs, {\ngdb}s allow for a flexible and unified modeling of features in diverse modalities using the embedding store.
Moreover, when the graph is incomplete, they can provide robust retrieval of answers which a normal graph DB cannot recover. 
Finally, we point out promising directions, unsolved problems and applications of \ngdb for future research.

\end{abstract}


\tableofcontents

\section{Introduction}

Graph databases (graph DBs) are key architectures to capture, organize and navigate structured relational information over real-world entities. 
Unlike traditional relational DBs storing information in tables with a rigid schema, graph DBs store information in the form of heterogeneous graphs, where nodes represent entities and edges represent relationships between entities.
In graph DBs, a relation (i.e. heterogeneous connection between entities) is the first-class citizen. With the graph structure and a more flexible schema, graph DBs allow for a more efficient and expressive way to handle higher-order relationships between distant entities, especially navigating through multi-hop hierarchies.
While traditional DBs require expensive join operations to retrieve information, graph DBs can directly traverse the graph and navigate through links more efficiently with the adjacency matrix.
Due to its capabilities, graph databases serve as the backbone of many critical industrial applications including question answering in virtual assistants \citep{alexa,siri}, recommender systems in marketplaces \citep{amazon,uber}, social networking in mobile applications \citep{facebook}, and fraud detection in financial industries~\citep{ibmfraud,pourhabibi2020fraud}.

\begin{figure}[t]
	\centering
	\includegraphics[width=\linewidth]{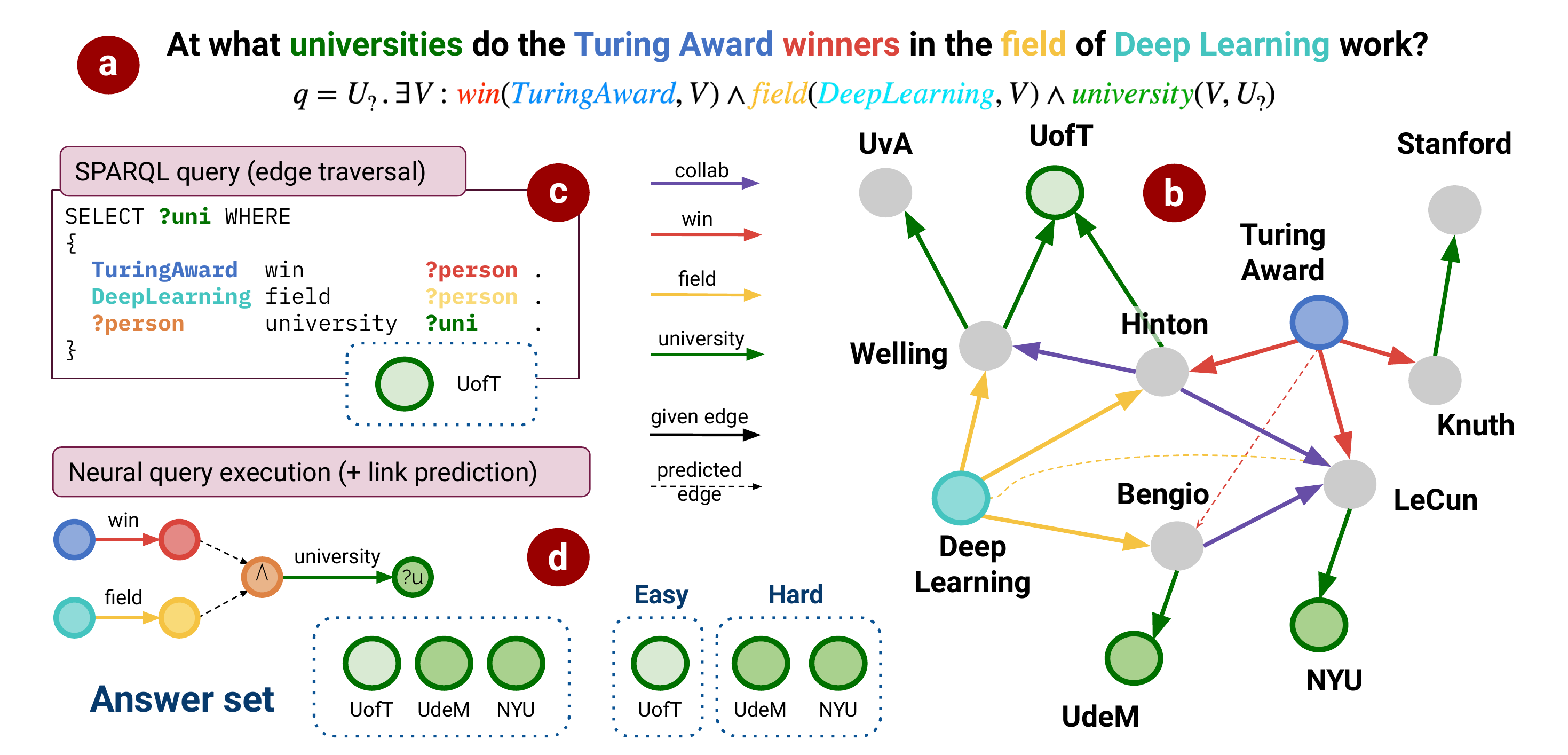}
	\caption{A complex logical query \protect\encirclered{a} and its execution over an incomplete graph \protect\encirclered{b}. Symbolic engines like  SPARQL \protect\encirclered{c} perform edge traversal and retrieve an incomplete set of \emph{easy} answers directly reachable in the graph, i.e., \{\texttt{UofT}\}. Neural query execution \protect\encirclered{d} recovers missing ground truth edges (dashed) and returns an additional set of \emph{hard} answers \{\texttt{UdeM}, \texttt{NYU}\} unattainable by symbolic methods.}
	\label{fig:clqa_1}
 \vspace{-1em}
\end{figure}

Given a downstream task, one of the most important tasks of graph DBs is to perform complex query answering. The goal is to retrieve the answers of a given input query of interest from the graph database. Given the query, graph DBs first translate and optimize the query into a more efficient graph traversal pattern with a query planner, and then execute the pattern on the graph database to retrieve the answers from the graph storage using the query executor.
The storage compresses the graphs into symbolic indexes suitable for fast table lookups. 
Querying is thus fast and efficient under the assumption of completeness, i.e., stored graphs have no missing edges.

However, most real-world graphs are notoriously incomplete, e.g., in Freebase, 93.8\% of people have no place of birth and 78.5\% have no nationality~ \citep{mintz2009distant}, about 68\% of people do not have any profession~\citep{DBLP:conf/www/WestGMSGL14}, while in Wikidata, about 50\% of artists have no date of birth~\citep{zhang2022enriching}, and only 0.4\% of known buildings have information about height~\citep{DBLP:conf/www/Ho0MSW22}.
Na\"ively traversing the graph in light of incompleteness leads to a significant miss of relevant results and the issue further exacerbates with an increasing number of hops.
This inherently hinders the application of graph databases.
Link prediction is a challenging task, prior works predict links by learning a latent representation of each link \citep{transe,distmult,complex,rotate} or mining rules~\citep{Galrraga2013AMIEAR,Xiong2017DeepPathAR,Lin2018MultiHopKG, qu2021rnnlogic}. However, it is always a trade-off between possibly incomplete results and decidability -- with a denser graph, some SPARQL entailment regimes~\citep{sparql2013} do not guarantee that query execution terminates in finite time.

On the other hand, recent advances in graph machine learning enabled expressive reasoning over large graphs in a latent space without facing decidability bottlenecks. 
The seminal work of \citet{gqe} on Graph Query Embedding (GQE) laid foundations of answering complex, database-like logical queries over incomplete KGs where inferring missing links during query execution is achieved via parameterization of entities, relations, and logical operators with learnable vector representations and neural networks.
For the incomplete knowledge graph in (\autoref{fig:clqa_1}), given a complex query \emph{``At what universities do the Turing Award winners in the field of Deep Learning work?''}, traditional symbolic graph DBs (SPARQL- or Cypher-like) would return only one answer (\texttt{UofT}), reachable by edge traversal. 
In contrast, neural query embedding parameterizes the graph and the query with learnable vectors in the embedding space.
Neural query execution is akin to \emph{traversing the graph and executing logical operators in the embedding space} that infers missing links and enriches the answer set with two more relevant answers \texttt{UdeM} and \texttt{NYU} unattainable by symbolic DBs. 

Since then, the area has seen a surge of interest with numerous improvements of supported logical operators, query types, graph modalities, and modeling approaches.
In our view, those improvements have been rather scattered, without an overall aim.
There still lacks a unifying framework to organize the existing works and guide future research.
To this end, we propose to present one of the first holistic studies about the field. We devise a taxonomy classifying existing works along three main axes, \ie, 
\begin{inparaenum}[(i)]
\item \textbf{Graphs} (\autoref{sec:graphs} --
logical formalisms behind the underlying graph and its schema)
\item \textbf{Modeling} (\autoref{sec:learning} -- what are the neural approaches to answer queries)
\item \textbf{Queries} (\autoref{sec:queries} -- what queries can be answered).
\end{inparaenum}
We then discuss \textbf{Datasets and Metrics} (\autoref{sec:datasets} -- how we measure performance of \ngdb engines).
Each of these dimensions is further divided into fine-grained aspects.
Finally, we list \clqa applications (\autoref{sec:apps}) and summarize open challenges for future research (\autoref{sec:future}).

We further propose a novel framework \emph{Neural Graph Databases} (NGDB, \autoref{sec:ngdb}), where complex approximate query answering (\autoref{sec:prelim}) is the core task and approached by the neural query embedding methods.
NGDB consists of a neural graph storage and a neural query engine, which correspond to a neural version of both counterparts in a regular graph DB respectively.
Besides a graph store and feature store that store the graph structure as well as the multimodal node-/edge-level features, neural graph storage assumes an encoder that further compresses the raw information in a semantic-preserving latent embedding space, \ie, similar entities/relations are mapped to similar embeddings. During retrieval, such a design choice instantly allows for both exact identity match as well as approximate nearest neighbor search.
As another key component, the neural query engine is responsible for optimizing, planning and executing a given input query in the embedding space, and finally returning the query answers by interacting with the neural graph storage.
We provide a detailed conceptual scheme for an NGDB with various levels of design principles and assumptions over the two components.
We hope this sheds light on the current state of the art and provides a roadmap for future research on NGDBs.

\textbf{Related Work.} 
While there exist insightful surveys on general graph machine learning~\citep{chami2022survey}, simple link prediction in KGs~\citep{ali2021light,chen2023generalizing}, and logic-based link prediction~\citep{zhang2022_logic_emb_survey,delong2023Neurosymbolic}, the complex query answering area remained uncovered so far.
With our work, we close this gap and provide a holistic view on the state of affairs in this emerging field.
\section{Preliminaries}\label{sec:prelim}

\begin{wrapfigure}[14]{R}{0.50\textwidth}
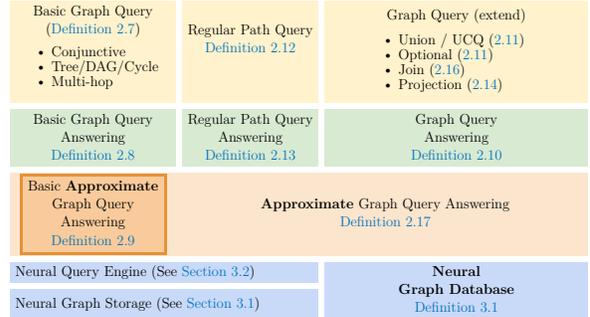

      \begin{adjustbox}{width=0.50\textwidth}

\definecolor{row1_color}{HTML}{fff2cc}
\definecolor{row2_color}{HTML}{d9ead3}
\definecolor{row3_color}{HTML}{fce5cd}
\definecolor{row3b_color}{HTML}{f9cb9c}
\definecolor{row3_split_color}{HTML}{e69138}
\definecolor{row4_color}{HTML}{c9daf8}

{
\Large

\begin{tblr}{
    	colspec={Q[c,m]|Q[c,m]|Q[c,m]},
		row{1} = {row1_color},
		row{2} = {row2_color},
		row{3} = {row3_color},
		row{4} = {row4_color},
		row{5} = {row4_color},
		hlines = {white,5px},
		vlines = {white, 5px},
		vline{2} = {3-3}{row3_color, 5px},
	}
{
Basic Graph Query\\ 
(\autoref{def:basic_graph_query})\\
	\parbox{0.33\textwidth}{
\centering
\begin{itemize}
	\itemsep-0.5em
	\item Conjunctive 
	\item Tree/DAG/Cycle
	\item Multi-hop
\end{itemize}
}
}
 & 
{
Regular Path Query\\
\autoref{def:regular_path_query} 
}
&
{
Graph Query (extend)\\
\parbox{0.4\textwidth}{
\centering
\begin{itemize}
		\itemsep-0.5em
\item Union / UCQ 	  (\ref{def:union}) %
\item Optional 	  (\ref{def:union})
\item Join 		  (\ref{def:join})
\item Projection 	  (\ref{def:projection})
\end{itemize}}
}
\\
{
Basic Graph Query \\
Answering \\
\autoref{def:basic_graph_query_answering}
}
&
{
Regular Path Query \\
Answering \\
\autoref{def:regular_path_query_answering}
}
&
{
Graph Query \\
Answering \\
\autoref{def:graph_query_answering}
}
\\
{
		\begin{tblr}{
    	      colspec={Q[c,m]},
				hlines = {row3_split_color,2px},
				vlines = {row3_split_color, 2px},
				row{1}={row3b_color},	
		}
		{
			Basic \textbf{Approximate} \\ 
			Graph Query \\ 
			Answering \\
			\autoref{def:basic_approx_gqa}
		}
		\end{tblr}
}
&
{
 \SetCell[c=2]{c} 
\textbf{Approximate} Graph Query Answering\\	
\autoref{def:approx_gqa}
}
\\
{
\SetCell[c=2]{l}
Neural Query Engine	(See \autoref{sec:neural_qe})
}
&
 
&
{
\SetCell[r=2]{c}
\textbf{Neural \\
Graph Database \\}
\autoref{def:ngdb}
}
\\
{
\SetCell[c=2]{l}

Neural Graph Storage	(See \autoref{sec:storage})	
}
&
&

\end{tblr}

}
     
      \end{adjustbox}
\definecolor{row1_color}{HTML}{fff2cc}
\definecolor{row2_color}{HTML}{d9ead3}
\definecolor{row3_color}{HTML}{fce5cd}
\definecolor{row3b_color}{HTML}{f9cb9c}
\definecolor{row3_split_color}{HTML}{e69138}
\definecolor{row4_color}{HTML}{c9daf8}  
\vspace{-\baselineskip}
     \caption{
A hierarchy of definitions from \colorbox{row1_color}{Graph Query} to \colorbox{row2_color}{Graph Query Answering}, a more general
     \colorbox{row3b_color}{Approximate Graph Query Answering}
     leading to \colorbox{row4_color}{Neural Graph Database, \ngdb}. }
     \label{fig:def_hierarchy}
 \end{wrapfigure}

Here, we discuss the foundational terms and preliminaries often seen in the database and graph learning literature 
before introducing neural graph databases. This includes definitions of different types of graphs, graph queries and their mapping to structured languages like SPARQL, formalizing approximate graph query answering, logical operators, and fuzzy logic. The full hierarchy of definitions is presented in \autoref{fig:def_hierarchy}.

\subsection{Types of Graphs}  

\begin{figure}[!ht]
	\centering
	\includegraphics[width=\linewidth]{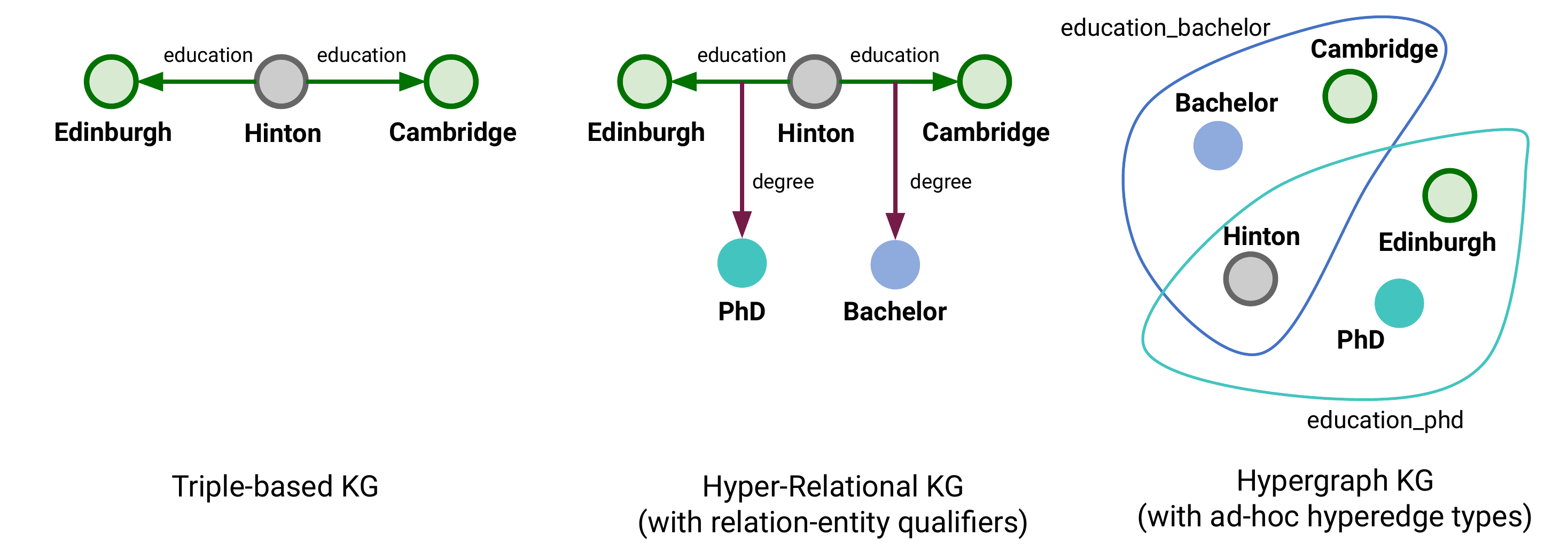}
	\caption{An example of KG modalities. Triple-only graphs have directed binary edges between nodes. Hyper-relational graphs allow key-value relation-entity attributes over edges. Hypergraphs consist of hyperedges composed of multiple nodes.
 }
	\label{fig:modality}
\end{figure}

Here we introduce different types of graphs relevant for this research area. 
These serve as the core data structure of various query reasoning works and also our neural graph storage module (\Secref{sec:storage}). Our definitions are adaptations from those in \cite[ch. 2]{hogan2021knowledge}. 

We begin by defining a set with all elements used in our graph
\begin{definition}[\con]
$\con$ is an infinite set of constants.\footnote{
In case the constants do not include representations for the real numbers, this set would be countably infinite.
}
\end{definition}

\begin{definition}[(Standard / Triple based) Knowledge Graph]
\label{def:kg}
We define a knowledge graph (KG) $\gG=(\gE, \gR, \gS)$, where $\gE \subset \con$ represents a set of nodes (entities), $\gR \subset \con$ is a set of relations (edge types), and $\gS \subset (\gE \times \gR \times \gE)$ is a set of edges (triples)~\footnote{There might be an overlap between $\gE$ and $\gR$, i.e., some constants might be used as an edge and as a node.}.
Each edge $s\in\gS$ on a KG $\gG$ denotes a statement (or fact) $(e_s, r, e_o)$ in a triple format or $r(e_s,e_o)$ as a formula in first-order logic (FOL) form, where $e_s, e_o \in \gE$ denote the subject and object entities, and $r\in\gR$ denotes the relationship between the subject and the object.
Often, the graph is restricted such that the set of nodes and edges must be finite, we will make this assumption unless indicated otherwise.
\end{definition}
For example, one statement on the KG in \autoref{fig:clqa_1}\encirclered{b} is \texttt{(Hinton, university, UofT)} or \texttt{university(Hinton, UofT)}. Note all relations are binary on KGs, \ie, each relation involves exactly two entities. 
This is also the choice made for the resource description framework (RDF)~\citep{brickley2014rdf} that defines a triple as a basic representation of a fact, yet RDF allows a broader definition of triples, where the object can be literals including numbers and timestamps (detailed in the following paragraphs).

Hyper-relational KGs generalize triple KGs by allowing edges to have relation-entity qualifiers. We define them as follows:
\begin{definition}[Hyper-relational Knowledge Graph - derived from \cite{starqe}]
With $\gE$ and $\gR$ defined as in \autoref{def:kg}, let $\fQ = 2^{(\gR \times \gE)}$, we define a hyper-relational knowledge graph $\gG=(\gE,\gR,\gS)$. In this case, the edge set $\gS \subset (\gE \times \gR \times \gE \times \fQ)$ consists of qualified statements, where each statement $s=(e_s, r, e_o, qp)$ includes $qp$ that represent contextual information for the main relation $r$.
This set $qp = \{q_1, \ldots\} = \{(qr_1, qe_1), (qr_2, qe_2) \ldots\} \subset \mathcal{R} \times \mathcal{E}$ is the set of \emph{qualifier pairs}, where $\{qr_1, qr_2, \ldots\} $ are the \emph{qualifier relations} and $\{qe_1, qe_2, \ldots\} $ the \emph{qualifier entities}.
\label{def:hyper_relational_kg}
\end{definition}
We note that if $\gE$ and $\gR$ are finite sets, then also $\fQ$ is finite and, there are only a finite number of ($r, qp$) combinations possible.
As a consequence, we find that with these conditions, and by defining a canonical ordering over $\fQ$, we can represent the hyper-relational graph using first order logic by coining a new predicate $r_{qp}$ for each combination.
The statement from the definition can then be written as $r_{qp}(e_s,e_o)$.

For example, one statement on a hyper-relational KG in (\autoref{fig:modality}) is \texttt{(Hinton, education, Cambridge, \{(degree, Bachelor)\})}. The qualifier pair \texttt{\{(degree, Bachelor)\}} provides additional context to the main triple and helps to distinguish it from other \texttt{education} facts. 
If the conditions mentioned above are met, we can write this statement as $\mathtt{education_{\{(degree:Bachelor)\}}(Hinton, Cambridge)}$. 

Hyper-relational KGs can capture a more diverse set of facts by using the additional qualifiers for facts on the graph.
This concept is closely connected to the RDF-star (previously known as RDF*) format introduced in \citet{hartig2022rdfstar}, as an extension of RDF. 
RDF-star explicitly separates metadata (context) from data (statements).
RDF-star supports a superset of hyper-relational graphs; it extends the values allowed in the object and qualifier value position to include literals, which we will discuss below. 

\begin{definition}[Hypergraph Knowledge Graph]
With $\gE$ and $\gR$ defined as in \autoref{def:kg}, a hypergraph KG is a graph $\gG=(\gE, \gR, \gS)$ where statements $\gS \subset (\gR \times 2^\gE )$ are hyperedges.
Such a hyperedge $s = (r, e) = (r, \{e_1, \dots, e_k\})$ has one relation type $r$ and links the $k$ entities together, where the order of these entities is not significant. $k$, the size of $e$ is called the \emph{arity} of the hyperedge.
\label{def:hypergraph_kg}
\end{definition}

An example hyperedge in \autoref{fig:modality} is $(\texttt{education\_degree}, \{\texttt{Hinton}, \texttt{Cambridge}, \texttt{Bachelor}\})$. 
This is a 3-ary statement (being comprised of three entities).
In contrast to hyper-relational KGs, hyperedges consist of only one relation type and cannot naturally incorporate more fine-grained relational compositions. 
Instead, the type of the hyperedge is rather a merged set of statement relations and every composition leads to a new relation type on hypergraph KGs. It is thus not as compositional as the hyper-relational KG. 
That is, when describing, for instance, a \texttt{major} relation of the \texttt{education(Hinton, Cambridge)} fact, a hyper-relational KG can simply use it as a qualifier and retain both relations whereas a hypergraph model has to come up with a new hyperedge type \texttt{education\_major}. 
With a growing number of relations, such a strategy of creating new hyperedge relation types might lead to a combinatorial explosion. 
However, the hypergraph model is more suitable for representing relationships of varying arity, with equal contributions of all entities such as \texttt{partnership(companyA, companyB, companyC)}.

Each of the graph types introduced above can be extended to also support literals.
This happens by allowing nodes or qualifier values to contain a literal value. 
Some graphs, like property graphs allow nodes to contain attributes, but this is equivalent with creating extra nodes with these attribute values and adding relations to those.
In theory, also the edge type could be a literal, but we exclude these from this work. 
\begin{definition}[KG with Literals]
With $\gE$ and $\gR$ as for one of the graph types above, a corresponding KG with literals $\gG_{\gL}=(\gE, \gR, \gS_{\gL}, \gL)$ has an additional set of literals $\gL \subset \con$ representing numerical, categorical, textual, images, sound waves, or other continuous values ($\gL$ is disjoint from $\gE$ and $\gR$). 
If we extend standard KGs to RDF graphs, literals can only be used in the objects position, that is $\gS_{\gL} \subset (\gE \times \gR \times (\gE \cup \gL))$. 
In RDF-star graphs, literals can be objects or qualifier values, that is, $\fQ = 2^{(\gR \times (\gE \cup \gL))}$ and $\gS_{\gL} \subset (\gE \times \gR \times (\gE \cup \gL) \times \fQ)$.\footnote{
Both RDF and RDF-star also allow blank nodes used to indicate entities without a specified identity in the subject and object position
\cite{brickley2014rdf}
.
Besides, they also have support for named graphs (and in some cases for quadruples). 
We do not support these explicitly in our formalism, but all of these can be modeled using hyper-relational graphs.
}
For both of these, we could also define graph types with literals in other positions of the triples, as necessary, or introduce more complex substructures in the elements of the triple (see e.g.,~\cite{cochez2012semantic}).
In hypergraph KGs, literals can be introduced in the elements of a hyperedge, $\gS_{\gL} \subset (\gR \times 2^{\gE \cup \gL} )$
\label{def:kg_literals}
\end{definition}
If the set of possible statements of the KG with literals has the same finiteness properties as the one without literals, then the properties regarding expressing it using first order logic do not change.

An example triple with a literal object (\autoref{fig:reasoning_domain}) is \texttt{(Cambridge, established, 1209)}. 
Literals contain numerical, categorical, discrete timestamps, or text data that cannot be easily discretized in an exhaustive set of possible values, like entities.
Similarly to the Web Ontology Language (OWL,~\cite{motik2009owl}) that distinguishes \emph{object properties} that connect two entities from \emph{datatype properties} that connect an entity and a literal, it is common to use the term \emph{relation} for an edge between two entities, and \emph{attribute} for an edge between an entity and a literal.
For example, in \texttt{education(Hinton, Cambridge)}, \texttt{education} is a relation, whereas \texttt{established(Cambridge, 1209)} is an attribute.

The knowledge graph definitions above are not exhaustive. 
It is possible to create graphs with other properties.
Examples include graphs with undirected edges, without edge labels, with time characteristics, with probabilistic or fuzzy relations, etc.
Besides, it is also possible to have graphs or edges with combined characteristics. 
One could, for instance, define a hyperedge with qualifying information. 
Because of this plethora of options, we decided to limit ourselves to the limited set of options above.
In the next section we introduce how to query these graphs. 

\subsection{Basic Graph Query Answering}
We base our definition of basic graph queries on the one from \cite[section 2.2.1 basic graph patterns]{hogan2021knowledge}, but adapt it to our graph formalization.

\begin{definition}[Term and Var]
Given a knowledge graph $\gG=(\gE, \gR, \gS)$ (or equivalent $\gG=(\gE, \gR, \gS, \gL)$), 
we define the set of variables $\variable  = \{v_1, v_2, \dots \}$ which take values from $\con$, but is strictly disjoint from it.
We call the union of the variables and the constants the terms: $\term = \con \cup \variable$
\end{definition}

With these concepts in place a Basic Graph Query is defined as follows:

\begin{definition}[Basic Graph Query]
\label{def:basic_graph_query}
A basic graph query is a 4-tuple $\gQ=(\gE', \gR', \gS', \overline{\gS'})$
 (equivalently a 5-tuple $\gQ=(\gE', \gR', \gS', \overline{\gS'}, \gL')$, with literals), 
with $\gE' \subset \term$ a set of node terms, $\gR' \subset \term$ a set of relation terms, and $\gS', \overline{\gS'} \subset \gE' \times \gR' \times \gE'$, two sets of edges (or equivalent to how graph edges are defined with literals).
The query looks like two (small) graphs; one formed by the edges in $\gS'$, and another by the edges in $\overline{\gS'}$.
The former set includes edges that must be matched in the graph to obtain answers to the query, while the latter contains edges that \textbf{must not} be matched (\ie, atomic negation). 
\end{definition}

With $\variable_Q = \gE' \cap \variable$ (all variables in the query), the answer to the query is defined as follows. 

\begin{definition}[Basic Graph Query Answering]
\label{def:basic_graph_query_answering}
Given a knowledge graph $\gG$ and a query $\gQ$ as defined above, an answer to the query is any mapping 
$\mu : \variable_Q \to \con$ 
such that replacing each variable $v_i$ in the edges in $\gS'$ with $\mu(v_i)$ results in an edge in $\gS$, \emph{and} replacing each variable $v_j$ in the edges in $\overline{\gS'}$ with $\mu(v_j)$ results in an edge that is \textbf{not} in $\gS$.
The set of all answers to the query is the set of all such mappings. 
\end{definition}

For triple-based KGs, each edge $(a, R, b)$ of the edge set $\gS$ (respectively $\gS'$) can be seen as relation projections $R(a,b)$ (resp. $\neg R(a,b)$), \ie, binary functions. Now, because the conjunction (\ie, all) of these relation projections (resp. the negation) have to be true, we also call these \emph{conjunctive} queries with atomic negation~(\CQn). 
The SPARQL query language defines basic graph patterns (BGPs). 
These closely resemble our basic graph query for RDF graphs, but with $\overline{\gS'} = \emptyset$, \ie, they do not support atomic negation but only conjunctive queries (\CQ).
We could analogously create {\CQ} and {\CQn} classes for the other KG types. 

In \autoref{fig:clqa_1}\encirclered{c} we provide an example of a Basic Graph Query expressed in the 
form of a SPARQL query.
This corresponds to our formalism: 
\begin{align*}
\gQ=(&\{\texttt{TuringAward}, \textit{?person}, \texttt{DeepLearning}, \textit{?uni}\}, \tag{\texttt{\# Node Terms } $\gE'$}  \\
&\{\textit{win}, \textit{field}, \textit{university}\}, \tag{\texttt{\# Relation Terms } $\gR'$} \\
&\{(\texttt{TuringAward},\textit{win}, \textit{?person}), (\texttt{DeepLearning}, \textit{field}, \textit{?person}), (\textit{?person},\textit{university}, \textit{?uni})\}, \tag{$\gS'$}\\
&\emptyset \tag{$\overline{\gS'}$})
\end{align*}
A possible answer to this query is the following partial mapping:
$
\mu_1 =\{(\textit{?person}, \texttt{Hinton}), (\textit{?uni}, \texttt{UofT})\}
$
This is also the only answer, so the set of all answers is $\{\mu_1\}$

If we want to exclude people who have worked together with \texttt{Welling}, then we modify the query as follows:
\begin{align*}
	\gQ=(&\{\texttt{TuringAward}, \textit{?person}, \texttt{DeepLearning}, \textit{?uni}\},\\
	&\{\textit{win}, \textit{field}, \textit{university}\}, \\
	&\{(\texttt{TuringAward},\textit{win}, \textit{?person}), (\texttt{DeepLearning}, \textit{field}, \textit{?person}), (\textit{?person},\textit{university}, \textit{?uni})\},\\
	&\pmb{\{(\textit{?person}, \textit{collab}, \texttt{Welling})\}})
\end{align*}
Note the key difference here is that we add \{(\textit{?person}, \textit{collab}, \texttt{Welling})\} to $\overline{\gS'}$. The answer set to this query is empty.

\subsection{Basic Approximate Graph Query Answering}

Until now, we have assumed that our KG is complete, \ie, it is possible to exactly answer the queries.
However, we are interested in the setting where we do not have the complete graph.
The situation can be described as follows:
Given a knowledge graph $\gG$ (subset of a complete, but not observable graph $\hat{\gG}$) and a basic graph query $\gQ$.
Basic Approximate Graph Query Answering is the task of answering $\gQ$, without having access to $\hat{\gG}$.
Depending on the setting, an approach to this could have access to $\gG$, or to a set of example (query, answer) pairs, which can be used to produce the answers (see also \autoref{sec:graph_training}).
In the example from \autoref{fig:clqa_1}, we also have several edges drawn with dashes. These are edges which are true, but only there in the non-observable part of the graph $\hat{\gG}$.

The goal is to find the answer set as if $\hat{\gG}$ was known.
In this case, the complete answer set becomes
$
\{
\mu_1, 
\{(\textit{?person}, \texttt{Bengio}), (\textit{?uni}, \texttt{UdeM})\}, 
\{(\textit{?person}, \texttt{LeCun}), (\textit{?uni}, \texttt{NYU})\} 
\}
$, which includes the answer $\mu_1$ of the non-approximate version. 

The query which includes the negation would have the following answers if our graph was complete:
$
\{
\{(\textit{?person}, \texttt{Bengio}), (\textit{?uni}, \texttt{UdeM})\}, 
\{(\textit{?person}, \texttt{LeCun}), (\textit{?uni}, \texttt{NYU})\} 
\}
$, which does not include $\mu_1$.

Approximate Query Answering Systems provide a score $\in \scoredomain$ for \emph{every possible} mapping\footnote{	
This score can indicate the rank of the mapping, a likelihood or a truth value, or be a binary indicating value.
}.
Hence, the answer provided by these systems is a function from mappings (\ie, a $\mu$) to their corresponding score in \scoredomain.

\begin{definition}[Basic Approximate Graph Query Answering]
\label{def:basic_approx_gqa}
Given a knowledge graph $\gG$ (subgraph of a complete, but not observable knowledge graph $\hat{\gG}$), a basic graph query $\gQ$, and the scoring domain \scoredomain,
a \emph{basic approximate graph query answer} to the query $\gQ$ is a function $f$ which maps every possible  mapping ($\mu : \variable_Q \to \con$) to \scoredomain.
\end{definition}

\begin{wraptable}[10]{R}{5cm}
    \vspace{-\baselineskip}
	\centering
	\caption{Ordered scored mappings for 
    the example query.
    The two wrong mappings are in red.}
	\label{tab:basic_approx_gqa_example}
	\begin{tabular}{ll|l}
		\toprule
		?person  & ?uni & score \\
		\midrule
		Hinton & UofT & 40 \\
		Bengio & UdeM & 35 \\
		{\color{red}Welling} & {\color{red} UofT} & 34 \\
		{\color{red} UdeM} & {\color{red} Hinton} & 33  \\
		LeCun & NYU & 32 \\
		\dots & \dots & \dots \\
		\bottomrule
	\end{tabular}
\end{wraptable}

The objective is to make it such that the correct answers according to the graph $\hat{\gG}$ get a better score (are ranked higher, have a higher probability, have a higher truth value) than those which are not correct answers. 
However, it is not guaranteed that answers which are in the observable graph $\gG$ will get a high score.

For our example, a possible mapping is visualized in \autoref{tab:basic_approx_gqa_example}. 
Each row in the table corresponds to one mapping $\mu$. 
Ideally, all correct mappings should be ranked on top, and all others below.
However, in this example we see several correct answers ranked high, but also two which are wrong.

\subsection{Graph Query Answering}

In the previous sections, we introduced the basic graph query and how it can be answered either exactly, or in an approximate graph querying setting.
However, there is variation in what types of queries methods can answer; some only support a subset  of our basic graph queries, while others support more complicated queries. 
In this section we will again focus on \emph{exact (non-approximate) query answering} and look at some possible restrictions and then at extensions. We will also highlight links to FOL fragments.

\begin{definition}[Graph Query Answering]
\label{def:graph_query_answering}
Given a knowledge graph $\gG$, a query formalism, and a graph query $\gQ$ according to that formalism. 
An answer to the query 	is any mapping $\mu: \variable_Q \to \con$ 
such that the requirements of the query formalism are met.
The set of all answers is the set of all such mappings. 	
\end{definition}
For our basic graph queries introduced in \autoref{def:basic_graph_query}, the query formalism sets requirements as to what edges must and must not exist in the graph (\autoref{def:basic_graph_query_answering}).
In that context we already mentioned conjunctive queries, which exist either with (\CQn) or without (\CQ) negation.
If the conditions for writing our graphs using first order logic (FOL) hold, we can equivalently write our basic graph queries in first order logic. Each variable in the query becomes existentially quantified, and the formula becomes a conjunction of 
\begin{inparaenum}[1)]
	\item the terms formed from the triples in $S'$, and
	\item the negation of the terms formed from the triples in $\overline{S'}$%
\end{inparaenum}.
If there are variables on the edges of the query graph, then we can rewrite the second order formula in first order logic, by interpreting them as a disjunction over all possible predicates from the finite number of options.
Our example query from above becomes the following FOL sentence.
\begin{multline*}
	q = \exists \textit{?person}, \textit{?uni} :  \textit{win}(\texttt{TuringAward}, \textit{?person}) \wedge \textit{field}(\texttt{DeepLearning}, \textit{?person}) \wedge \textit{university}(\textit{?person},\textit{?uni})
\end{multline*}
The answer to the query is the variable assignment function. For several of the more restrictive fragments, it is useful to formulate the queries using this FOL syntax.

\begin{figure}[t]
	\centering
	\includegraphics[width=0.7\linewidth]{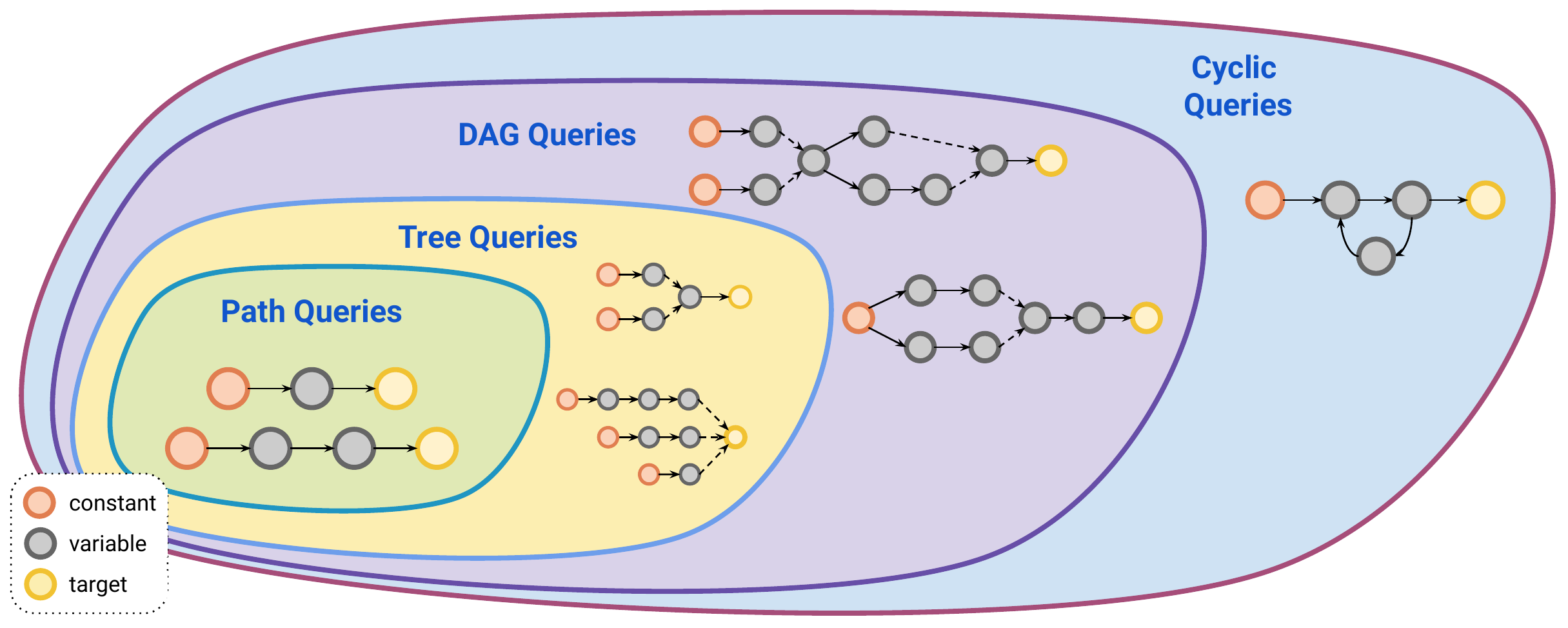}
	\caption{A space of query patterns and their relative expressiveness. 
		Multi-hop reasoning systems tackle the simplest \emph{Path} queries. Existing complex query answering systems support \emph{Tree} queries whereas \emph{DAG} and \emph{Cyclic} queries remain unanswerable by current neural models.}
	\label{fig:query_patterns}
\end{figure}

\paragraph{Restrictions}

\label{sec:multihop_vs_complex}
The first restriction one can introduce are \textbf{multi-hop queries}, known in the NLP and KG literature for a long time~\citep{guu-2015-traversing, das-2017-chains, Asai2020Learning} mostly in the context of question answering.
Formally, multi-hop queries (or \emph{path} queries) are \CQ\ which form a chain, where the tail of one projection is a head of the following projection, \ie,

\begin{equation*}
	q_{\text{path}} = V_k, \exists V_1, \dots, V_{k-1}: r_1(v, V_1) \wedge r_2(V_1, V_2) \wedge \dots \wedge r_k(V_{k-1}, V_k)
\end{equation*}

where $v \in \gE$, $\forall i \in [1, k]: r_i \in \gR, V_i \in \variable$ and all $V_i$ are existentially quantified variables.
In other words, path queries do not contain branch intersections and can be solved iteratively by fetching the neighbors of the nodes.
One could also define multi-hop queries which allow negation.

Other ways of restricting \CQ\ and on \CQn, resulting in more expressive queries than the multi-hop ones exist.
One can define families of logical queries shaped as a \emph{Tree}, a \emph{DAG}, and allowing \emph{cyclic} parts.
Illustrated in \autoref{fig:query_patterns}, path (multi-hop) queries form the least expressive set of logical queries. 
Tree queries add more reasoning branches connected by intersection operators, DAG queries drop the query tree requirement and allow queries to be directed acyclic graphs, and, finally, cyclic queries further drop the acyclic requirement.
Note that these queries do not allow variables in the predicate position. 
Besides, all entities (in this context referred to as anchors) must occur before all variables in the topological ordering.

We elaborate more on these query types in \autoref{sec:query_pat} and note that the majority of surveyed neural \clqa methods in \autoref{sec:learning} are still limited to Tree-like queries, falling behind the expressiveness of many graph database query languages. 
Bridging this gap is an important avenue for future work.

\paragraph{Extensions}

\begin{figure}
	\centering
	\includegraphics[width=0.7\textwidth]{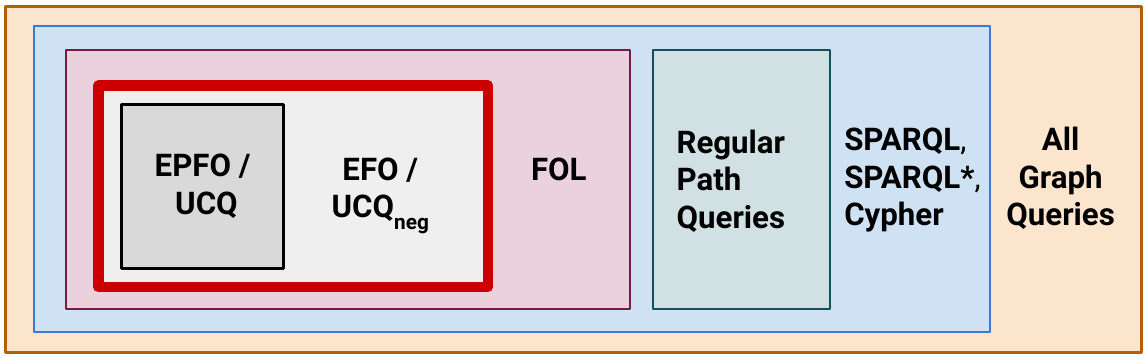}
	\caption{Current query answering models cover Existential Positive First Order (EPFO) and Existential First Order (EFO) logic fragments (marked in a red rectangle). EPFO and EFO are equivalent to unions of conjunctions (UCQ), and those with atomic negation (\UCQn), respectively. These, in turn, are a subset of first order logic (FOL). FOL queries, in turn, are only a subset of queries answerable by graph database languages. For example regular path queries cannot be expressed in FOL. Languages like SPARQL, SPARQL*, or Cypher, encompass all the query types and more.}
	\label{fig:fol_sparql_venn}
\end{figure}

The first extension we introduce is the \textbf{union}.

\begin{definition}[Union of Sets of Mappings]
	\label{def:union}
Given two sets of mappings (like $\mu$ from \autoref{def:basic_graph_query_answering}), we can create a new set of mappings by taking their union.
This union operator is commutative and associative, we can hence also talk about the union of three or more mappings. 
It is permissible that the domains of the mappings in the input sets are not the same.
\end{definition}
We can define a new type of query by applying the union operation on the outcomes of two or more underlying queries.
If these underlying queries are basic graph queries, we will call this new type of queries \textbf{Unions of Conjunctive Queries with negation (\UCQn)} and if the basic queries did not include negation, as \textbf{Union of Conjunctive Queries (\UCQ)}.
These classes are also familiar from FOL, and indeed correspond to a disjunction of conjunctions. 

As an example, the following query is in \UCQn\xspace because it is a disjunction of conjunctive terms which consist of atoms $a_i$ that are relation projections or their negations:
\begin{equation*}
	q = v_?.\exists v_1, \dots, v_n : \Big(\underbrace{r_1(c, v_1)}_{a_1} \wedge \underbrace{r_2(v_1, v_2)}_{a_2}\Big) \vee \Big(\underbrace{\lnot r_3(v_2, v_3)}_{a_3}\Big) \vee \dots \vee \underbrace{r_k(v_n, v_?)}_{a_m} 
\end{equation*}

Moreover, there are FOL fragments that are equivalent to these fragments.
Specifically, all queries in \textbf{EPFO}, which are Existential Positive First Order sentences, have an equivalent query in \UCQ, and all queries in \textbf{EFO}, Existential First Order sentences, have an equivalent query in \UCQn.
The reason is that EPFO and EFO sentences can be written in the Disjunctive Normal Form (DNF) as a union of conjunctive terms. 

Some query languages, like SPARQL, allow an optional part to a query. In our formalism, we can define the \textbf{optional} part using the union operator.
Assuming there are $n$ optional parts in the query, create $2^n$ different queries, in which other combinations of optional parts are removed.
The answer to the query is then the union of the answers to all those queries.
If the query language already allowed unions, then optionals do not make it more expressive.

Beyond these extensions, one could extend further to \textbf{all queries one could express with FOL}, which requires either universal quantification, or negation of conjunctions. These are, however, still not all possible graph queries.
An example of interesting queries, which are not in FOL, are \textbf{conjunctive regular path queries}.
These are akin to the path queries we discussed above, but without a specified length.

\begin{definition}[Regular Path Query]
	\label{def:regular_path_query}
A regular path query is a 3-tuple $(s, R, t)$, where $s \in \term$ the starting term of the path,  $R \in \term$ the relation term of the path, and $t\in \term$ the ending term of the path.
The query represents a path starting in $s$, traversing an arbitrary number of edges of type $R$  ending in $t$.
\end{definition}

Because this kind of query is a conjunction of an arbitrary length, it cannot be represented in FOL.
If one wants to express paths with a fixed length, this would be a multi-hop path like the one described above.
If one wants to express a maximum length, then this could be done using a union of all allowed lengths.
For the two latter cases, the query can still be expressed in $EPFO$.

\begin{definition}[Regular Path Query Answering]
	\label{def:regular_path_query_answering}
Given a knowledge graph $\gG$ and a regular path query $\gQ=(s, R, t)$.  
An answer to the query is any mapping $\mu : \variable_Q \to \con$ , such that if we replace all variables $v$ in $\gQ$ with $\mu(v)$, obtaining $(\hat{s}, \hat{R}, \hat{t})$, there exists a path in the graph that starts at node $\hat{s}$, then traverses one or more edges of type $\hat{R}$, and ends in $\hat{t}$.
The set of all answers to the query is the set of all such mappings.
\end{definition}

There exist several variations on regular path queries and they can also be combined with the above query types to form new ones.
In \autoref{fig:fol_sparql_venn} we illustrate how the fragments relate to other query classes.
Most methods fall strictly within the EFO fragment, \ie, they only support a subset with restrictions as we discussed above.
We will discuss further limitations of these methods in \autoref{sec:queries}. 

Two final aspects we want to highlight here are \emph{projections} and \emph{joins}.

\begin{definition}[Projection of a Query Answer]
	\label{def:projection}
	Given a query answer $\mu$, and a set of variables $\gV \in \variable$.
	The projection of the query answer on the variables $\gV$ is
	$\{  (var, val | (var, val) \in \mu, var \in \gV  \}$.
\end{definition}
In other words, it is the answer but restricted to a specific set of variables. 
The query forms introduced above can all be augmented with a projection to only obtain values for specific variables. 
This corresponds to the \texttt{SELECT ?var} clause in SPARQL. Alternatively, it is possible to project all variables in $\gV$ which is equivalent to the \texttt{SELECT *} clause.
A query without any projected variable is a Boolean subgraph matching problem equivalent to the \texttt{ASK} query in SPARQL.

A join is used to combine the results of two separate queries into one.
Specifically,

\begin{definition}[Join of Query Answers]
Given two query answers $\mu_A$ and $\mu_B$, and $\variable_A$, and $\variable_B$ the variables occurring in $\mu_A$ and $\mu_B$, respectively.
The join of these two answers only exists in case they have the same value for all variables they have in common, \ie, $\forall var \in \variable_A \cap \variable_B: \mu_A(var) = \mu_b(var) $.
In that case, $\join(\mu_A, \mu_B) = \mu_A \cup \mu_B$. 
\end{definition}

Given two sets of answers, their join is defined as follows.
\begin{definition}[Join of Query Answer Sets]
	\label{def:join}
Given two sets of query answers A and B, the join of these two is a new set 
$\join(A, B) = \{ \join(a,b) | a \in A, b \in B, \text{and} \: \join(a,b) \: \text{exists} \}$.
\end{definition}
This operation enables us to combine multiple underlying queries, potentially of multiple types into a single one.
For example, given the set of answers from our example basic graph query above:
	\[
	A=\{
	\{(\textit{?person}, \texttt{Hinton}), (\textit{?uni}, \texttt{UofT})\}, 
	\{(\textit{?person}, \texttt{Bengio}), (\textit{?uni}, \texttt{UdeM})\}, 
	\{(\textit{?person}, \texttt{LeCun}), (\textit{?uni}, \texttt{NYU})\} 
	\}
	\]
	and another set of answers 
	\[
	B=\{
	\{(\textit{?person}, \texttt{Hinton}), (\textit{?born}, \texttt{1947})\}, 
	\{(\textit{?person}, \texttt{Bengio}), (\textit{?born}, \texttt{1964})\}, 
	\{(\textit{?person}, \texttt{Welling}), (\textit{?born}, \texttt{1968})\} 
	\}
	\]
	The join of these becomes:
	\[
	\join(A, B) = \{
	\{(\textit{?person}, \texttt{Hinton}), (\textit{?uni}, \texttt{UofT}), (\textit{?born}, \texttt{1947})\}, 
	\{(\textit{?person}, \texttt{Bengio}), (\textit{?uni}, \texttt{UdeM}), (\textit{?born}, \texttt{1964})\}
	\}
	\]

We will discuss joins further in \autoref{sec:query_ops}, where we will use these basic building blocks to define a broader set of query operators, aiming to cover all operations that exist in SPARQL. 
This includes 
Kleene plus/star (+/*) for building property paths, FILTER, OPTIONAL, and different aggregation functions.

\subsection{Approximate Graph Query Answering}

Now that we have defined what Basic Approximate Graph Query Answering is, we can define what we mean by the more general Approximate Graph Query Answering.
The definition of Approximate Graph Query Answering is the same as the basic case, but rather than only providing answers to basic queries, it is about answering a broader set of query types.
\begin{definition}[Approximate Graph Query Answering]\label{def:approx_gqa}
Given a knowledge graph $\gG$, subgraph of a complete, but not observable knowledge graph $\hat{\gG}$, a query formalism, \textbf{any} graph query $\gQ$ according to that formalism, and the scoring domain \scoredomain.

An approximate graph query answer to the query $\gQ$ is a function $f$ which maps every possible mapping ($\mu : \variable_Q \to \con$) to \scoredomain.
\end{definition}
Note there that the variables are not always mapped to nodes which occur in the graph.
It is well possible that the query contains an aggregation function which results in a literal value.

In the literature, we can find the concepts of \emph{easy} and \emph{hard} answers. This refers to whether the answers can be found by only having access to $\gG$ or not.

\begin{definition}[Easy and Hard answers]\label{def:easy_hard_ans}
    Given a knowledge graph $\gG$, subgraph of a larger unobservable graph $\hat{\gG}$, and a query $\gQ$.
    Easy and hard answers are defined in terms of exact query answering (\autoref{def:graph_query_answering}).
    The set of \textbf{easy} answers is the intersection of the answers obtained from $\gG$, and those from $\hat{\gG}$.
    The set of \textbf{hard} answers is the set difference between the answers from $\hat{\gG}$ and those from $\gG$.
\end{definition}

Note the asymmetry in the definitions.
Easy answers are those that can be found in both $\gG$ and $\hat{\gG}$.
Hard answers are those that can be found only in $\hat{\gG}$ but not in $\gG$.
For Basic Graph Queries (\autoref{def:basic_graph_query}), all easy answers can also be found from $\hat{\gG}$.
However, for some more complex query types (\eg, these which allow negation) there could also be answers which  in $\gG$ which are not in $\hat{G}$.
We call these answers \textbf{false positives} in the context of answering over $\hat{\gG}$.

\subsection{Triangular Norms and Conorms}
\label{sec:tnorms}

Answering logical queries implies execution of logical operators. 
Approximate query answering, in turn, implies continuous vector inputs and output truth values that are not necessarily binary.
Besides, the methods often require that the logical operators are smooth and differentiable.
Triangular norms (T-norms) and triangular conorms (T-conorms) define functions that generalize logical conjunction and disjunction, respectively, to the continuous space of truth values and implement fuzzy logical operations.

T-norm defines a continuous function $\top: [0,1]\times [0,1] \to [0,1]$ with the following properties $\top(x,y)=\top(y,x)$ (commutativity), $\top(x, \top(y,z))=\top(\top(x,y),z)$ (associativity), and $y\leq z \to \top(x,y)\leq \top(x,z)$ (monotonicity). Also, we have 1 to be the identity element for $\top$, \ie, $\top(x,1)=1$. The goal of t-norms is to generalize logical conjunction with a continuous function. T-conorm can be seen as a duality of a t-norm that similarly defines a function $\bot$ with the same domain and range $\bot: [0,1]\times [0,1]\to [0,1]$. T-conorms use the continuous function $\bot$ to generalize disjunction to fuzzy logic. The function $\bot$ satisfies the same commutativity, associativity, and monotonicity properties as $\top$ with the only difference being 0 is the identity element, \ie, $\bot(x,0)=x$.

There exist many triangular norms, conorms, and  fuzzy negations~\citep{tnorm,vankrieken_fuzzy} that stem from the corresponding logical formalisms, \eg, 
(1) \emph{G\"odel logic} defines t-norm: $\top_{\text{min}}(x,y)=\text{min}(x,y)$, t-conorm: $\bot_\text{max}(x,y)=\max(x, y)$; (2) \emph{Product logic} with t-norm: $\top_\text{prod}(x,y)=x\cdot y$, t-conorm: $\bot_\text{prod}(x,y)=x+y-x\cdot y$; (3) in the \emph{\L{}ukasiewicz logic} t-norm: $\top_{\text{\L{}uk}}(x, y)= \max(x+y-1, 0)$, t-conorm: $\bot_{\text{\L{}uk}}(x,y)=\min(x+y, 1)$. 
Using fuzzy negation $N(x)= 1-x$, it is easy to verify that 
$\bot(x,y)=N(\top(N(x), N(y)))$
(De Morgan's laws) naturally obtaining a pair of $(\top, \bot)$. 

\subsection{Graph Representation Learning}
\label{sec:grl} 

Graph Representation Learning (GRL) is a subfield of machine learning aiming at learning low-dimensional vector representations of graphs or their elements such as single nodes~\citep{hamilton2020graph}.
For example, $\Em{h}_v \in \RR^d$ denotes a $d$-dimensional vector associated with a node $v$. 
Conceptually, we want nodes that share certain structural and semantic features in the graph to have similar vector representations (where similarity is often measured by a distance function or its modifications).

\paragraph{Shallow Embeddings}
The first GRL approaches focused on learning shallow node embeddings, that is, learning a unique vector per node directly used in the optimization task. 
For homogeneous (single-relation) graphs, DeepWalk~\citep{deepwalk} and node2vec~\citep{node2vec} trained node embeddings on the task of predicting walks in the graph whereas in multi-relational graphs TransE~\citep{transe} trained node and edge type embeddings in the autoencoder fashion by reconstructing the adjacency matrix.

\paragraph{Graph Neural Networks}

The idea of \emph{graph neural networks} (GNNs)~\citep{scarselli2008graph} implies learning an additional neural network encoder with shared parameters on top of given (or learnable) node features by performing neighborhood aggregation. 
This framework can be generalized to \emph{message passing}~\citep{Gilmer2017} where at each layer $t$ a node $v$ receives messages from its neighbors (possibly adding edge and graph features), aggregates the messages in the permutation-invariant way, and updates the representation:

\begin{align*}
    \Em{h}_v^{(t)} = \textsc{Update}\Big(\Em{h}_v^{(t-1)}, \textsc{Aggregate}_{u \in \gN(v)}\big( \textsc{Message}(\Em{h}_v^{(t-1)}, \Em{h}_u^{(t-1)}, \Em{e}_{uv})) \big)  \Big)
\end{align*}

Here, $\Em{h}_u$ is a feature of the neighboring node $u$, $\Em{e}_{uv}$ is the edge feature, \textsc{Message} function builds a message from node $u$ to node $v$ and can be parameterized with a neural network. 
As the set of neighbors $\gN(v)$ is unordered, \textsc{Aggregate} is often a permutation-invariant function like \emph{sum} or \emph{mean}. 
The \textsc{Update} function takes the previous node state and aggregated messages of the neighbors to produce the final state of the node $v$ at layer $t$ and can be parameterized with a neural network as well.

Classical GNN architectures like GCN~\citep{Kipf2017}, GAT~\citep{Velickovic2018}, and GIN~\citep{Xu2019} were designed to work with homogeneous, single-relation graphs. Later, several works have developed GNN architectures that work on heterogeneous graphs with multiple relations~\citep{rgcn,compgcn,zhu2021neural}.
GNNs and message passing paved the way for \emph{Geometric Deep Learning}~\citep{bronstein2021geometric} that leverages symmetries and invariances in the input data as inductive bias for building deep learning models.

\section{Neural Graph Databases}
\label{sec:ngdb}

Traditional databases (including graph databases) are designed around two crucial modules: the storage layer for data and the query engine to process queries over the stored data.
From that perspective, neural database \emph{per se} is not a novel term and many machine learning systems already operate in this paradigm when data is encoded into model parameters and querying is equivalent to a forward pass that can output a new representation or prediction for a downstream task.
One of the first examples of neural databases is \emph{vector databases}. In vector databases, the storage module consists of domain-agnostic vector representations of the data, which can be muliti-modal \eg, text paragraphs or images.
Vector databases belong to the family of storage-oriented systems commonly built around approximate nearest neighbor libraries (ANN) like Faiss~\citep{faiss} or ScaNN~\citep{scann} to answer distance-based queries (like maximum inner product search, MIPS). 
Being encoder-independent (that is, any encoder yielding vector representations can be a source), vector databases lack graph reasoning and complex query answering capabilities. 
Still, ANN systems are a convenient choice for implementing certain layers of Neural Graph Databases as we describe below.

With the recent rise of large-scale pretrained models (\ie, \emph{foundation models}~\citep{Bommasani2021FoundationModels}), we have witnessed their huge success in natural language processing and computer vision tasks. We argue that such foundation models are also a prominent example of neural databases. 
In foundation models, the ``storage module'' might be presented directly with model parameters or outsourced to an external index often used in retrieval-augmented models~\citep{rag, realm, alon2022neuro} since encoding all world knowledge even into billions of model parameters is hard. The ``query module'' performs in-context learning either via filling in the blanks in encoder models (BERT or T5 style) or via prompts in decoder-only models (GPT-style) that can span multiple modalities, \eg, learnable tokens for vision applications~\citep{uvim} or even calling external tools~\citep{mialon2023augmented}. 
Applied to the text modality, \citet{thorne2021acl, thorne2021vldb} devise \emph{Natural Language Databases (NLDB)} where atomic elements are textual facts encoded to a vector via a pre-trained language model (LM). 
Queries to NLDB are sent as natural language utterances that get encoded to vectors and query processing employs the \emph{retriever-reader} approach.
First, a dense neural \emph{retriever} returns a support set of candidate facts, and a fine-grained neural \emph{reader} performs a \emph{join} operation over the candidates (as a sequence-to-sequence task). 

The amount of graph-structured data is huge and spans numerous domains like general knowledge with Freebase~\citep{bollacker2008freebase}, DBpedia~\citep{lehmann2015dbpedia}, Wikidata~\citep{wikidata}, YAGO~\citep{yago}, commonsense knowledge with ConceptNet~\citep{conceptnet}, ATOMIC~\citep{atomic}, and biomedical knowledge such as 
Bio2RDF~\citep{bio2rdf} and PrimeKG~\citep{primekg}. 
With the growing sizes, the incompleteness of those graphs grows simultaneously.
At this scale, symbolic methods struggle to provide a meaningful approach to deal with incompleteness. 
Therefore, we argue that neural reasoning and graph representation learning methods are capable of addressing incompleteness at scale while maintaining high expressiveness and supporting complex queries beyond simple link prediction.
We propose to study those methods under the framework of \emph{Neural Graph Databases (\ngdb)}.
The concept of \ngdb extends the ideas of neural databases to the graph domain. \ngdb combines the advantages of traditional graph databases (graphs as a first-class citizen, efficient storage, and uniform querying interface) with modern graph machine learning (geometric and physics-inspired vector representations, ability to work with incomplete and noisy inputs by default, large-scale pre-training and fine-tuning on downstream applications).
In contrast to the work of \citet{lpg2vec} that proposed LPG2Vec, a featurization strategy for labeled property graphs to be then used in standard graph database pipelines, we design the \ngdb concept to be \emph{neural-first}. 

Using the definition of Approximate Graph Query Answering (\autoref{def:approx_gqa}), we define Neural Graph Databases as follows.

\begin{figure}[t]
	\centering
	\includegraphics[width=0.7\linewidth]{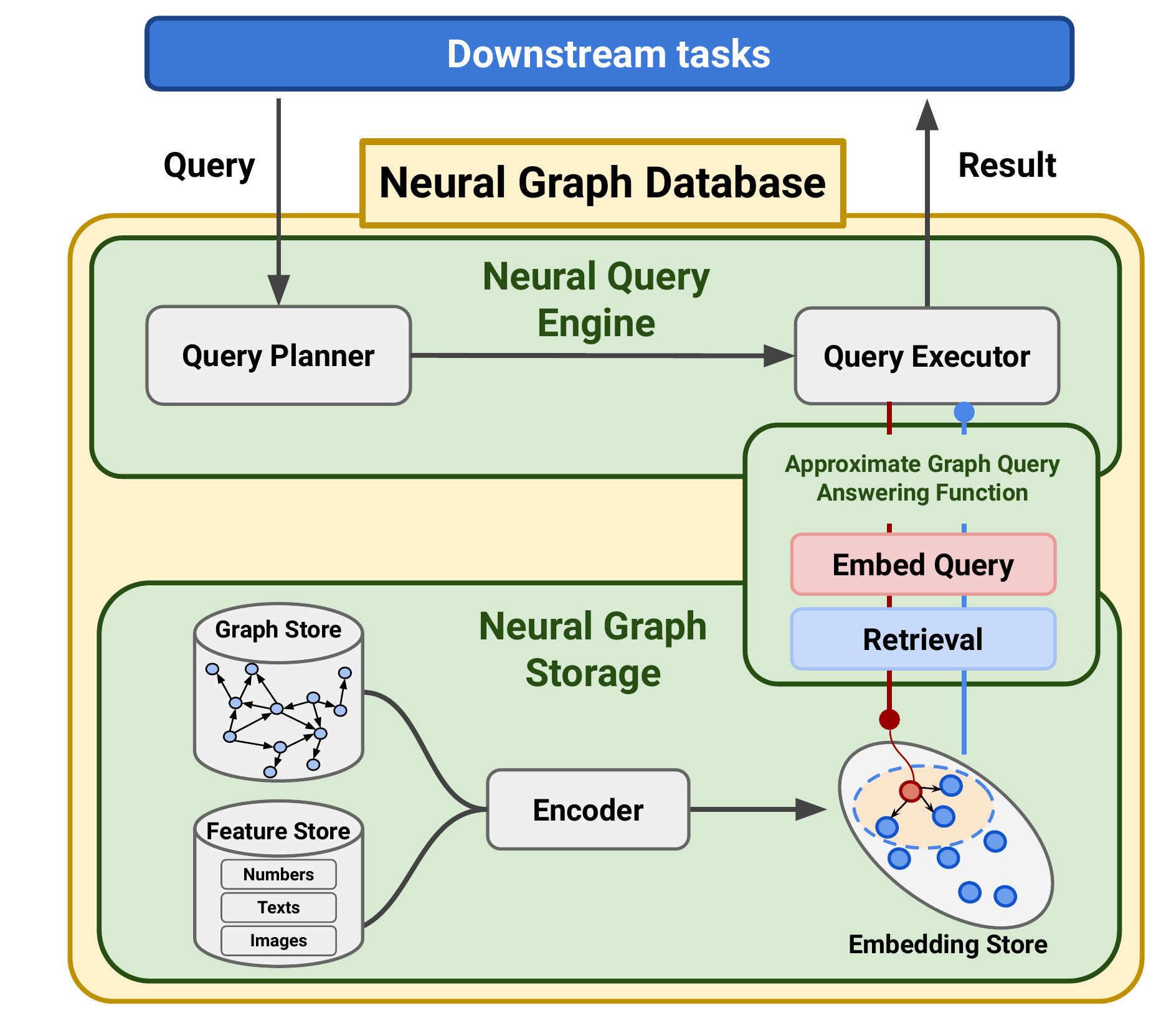}
	\caption{A conceptual scheme of Neural Graph Databases. An input query is processed by the \emph{Neural Query Engine} where the \emph{Planner} derives a computation graph of the query and the \emph{Executor} executes the query in the latent space. \emph{Neural Graph Storage} employs \emph{Graph Store} and \emph{Feature Store} to obtain latent representations in the \emph{Embedding Store}. The \emph{Executor} communicates with the embedding store via the \emph{Approximate Graph Query Answering Function $f_\theta$ (\autoref{def:ngdb})} to retrieve and return results.}
	\label{fig:ngd}
\end{figure}

\begin{definition}[Neural Graph Database, \ngdb]\label{def:ngdb}
	A Neural Graph Database (see \autoref{fig:ngd}) is a tuple $(S, E, f_{\theta})$.
$S$ is a Neural Graph Storage (see \autoref{sec:storage}), E is a Neural Query Engine (see \autoref{sec:neural_qe}), and $f_{\theta}$ is a
 parameterized Approximate Graph Query Answering function, where $\theta$ represents a set of parameters. 
\end{definition}

In particular, our \ngdb design agenda includes:
\begin{itemize}
    \item The \textbf{data incompleteness assumption}, \ie, the underlying data might have missing information on node-, link-, and graph-levels which we would like to infer and leverage in query answering;
    \item \textbf{Inductiveness and updatability}, \ie, similar to traditional databases that allow updates and instant querying, representation learning algorithms for building graph latents have to be inductive and generalize to unseen data (new entities and relation at inference time) in the zero-shot (or few-shot) manner to prevent costly re-training (for instance, of shallow node embeddings);
    \item \textbf{Expressiveness}, \ie, the ability of latent representations to encode logical and semantic relations in the data akin to FOL (or its fragments) and leverage them in query answering. Practically, the set of supported logical operators for neural reasoning should be close to or equivalent to standard graph database languages like SPARQL or Cypher;
    \item \textbf{Multimodality} beyond KGs, \ie, any graph-structured data that can be stored as a node or record in classical databases (consisting, for example, of images, texts, molecular graphs, or timestamped sequences) and can be imbued with a vector representation is a valid source for the Neural Graph Storage and Neural Query Engine.
\end{itemize}

The key methods to address the \ngdb design agenda are:
\begin{itemize}
    \item \textbf{Vector representation as the atomic element}, \ie, while traditional graph DBs hash the adjacency matrix (or edge list) in many indexes, the incompleteness assumption implies that both given edges \textbf{and} graph latents (vector representations) become the \emph{sources of truth} in the \emph{Neural Graph Storage};
    \item \textbf{Neural query execution in the latent space}, \ie, basic operations such as edge traversal cannot be performed solely symbolically due to the incompleteness assumption. Instead, the \emph{Neural Query Engine} operates on both adjacency and graph latents to incorporate possibly missing data into query answering;
\end{itemize}

A conceptual scheme of \ngdb is presented in \autoref{fig:ngd}. 
On a higher level, \ngdb contains two main components, \emph{Neural Graph Storage} and \emph{Neural Query Engine} that we describe in the following \autoref{sec:storage} and \autoref{sec:neural_qe}, respectively.
The processing pipeline starts with the query sent by some application or downstream task already in a structured format (obtained, for example, via semantic parsing~\citep{drozdov2022compositional} if an initial query is in natural language).
The query first arrives to the \emph{Neural Query Engine}, and, in particular, to the \emph{Query Planner} module. 
The task of the Query Planner is to derive an efficient computation graph of atomic operations (\eg, projections and logical operations)  with respect to the query complexity, prediction tasks, and underlying data storage such as possible graph partitioning. We elaborate on similarities and differences of the planning mechanism to standard query planners in classic databases in \autoref{sec:neural_qe}.
The derived plan is then sent to the \emph{Query Executor} that encodes the query in a latent space, executes the atomic operations over the underlying graph and its latent representations, and aggregates the results of atomic operations into a final answer set.
The execution is done via the \emph{Retrieval} module that communicates with the \emph{Neural Graph Storage}. 
The storage layer consists of (1) \emph{Graph Store} for keeping the multi-relational adjacency matrix in space- and time-efficient manner (\eg, in various sparse formats like COO and CSR; (2) \emph{Feature Store} for keeping node- and edge-level multimodal features associated with the underlying graph; 
(3) \emph{Embedding Store} that leverages an \emph{Encoder} module to produce graph representations in a latent space based on the underlying adjacency and associated features. 
The Retrieval module queries the encoded graph representations to build a distribution of potential answers to atomic operations. 
In the following subsections, we describe the Neural Graph Storage and Neural Query Engine in more detail.

\subsection{Neural Graph Storage}
\label{sec:storage}

In traditional graph databases, storage design often depends on the graph modeling paradigm. 
The two most popular paradigms are Resource Description Framework (RDF) graphs~\citep{brickley2014rdf} and Labeled Property Graphs (LPG)~\citep{lpg2vec}.
While the detailed comparison between those frameworks is out of scope of this work, the principal difference consists in that RDF is a triple-based model allowing for some formal \emph{semantics} suitable for symbolic reasoning whereas LPG allows literal attributes over edges but has no formal semantics.
We posit that the new RDF-star paradigm~\citep{hartig2022rdfstar} would be a convergence point of graph modeling combining the best of both worlds, \ie, attributes over edges (enabling other nodes and relations to be in key-value attributes) together with expressive formal semantics. 
We note that hyper-relational KGs (\autoref{def:hyper_relational_kg}) and Wikidata Statement Model~\citep{wikidata} are conceptually close to the RDF-star paradigm.
On a physical level, graph databases store edges employing various indexes and data structures optimized for read (\eg, B+ trees~\citep{rdf3x} or HDT~\citep{hdt}) or write applications (such as LSM trees~\citep{sagi2022design}).

\begin{figure}[t]
	\centering
	\includegraphics[width=0.9\linewidth]{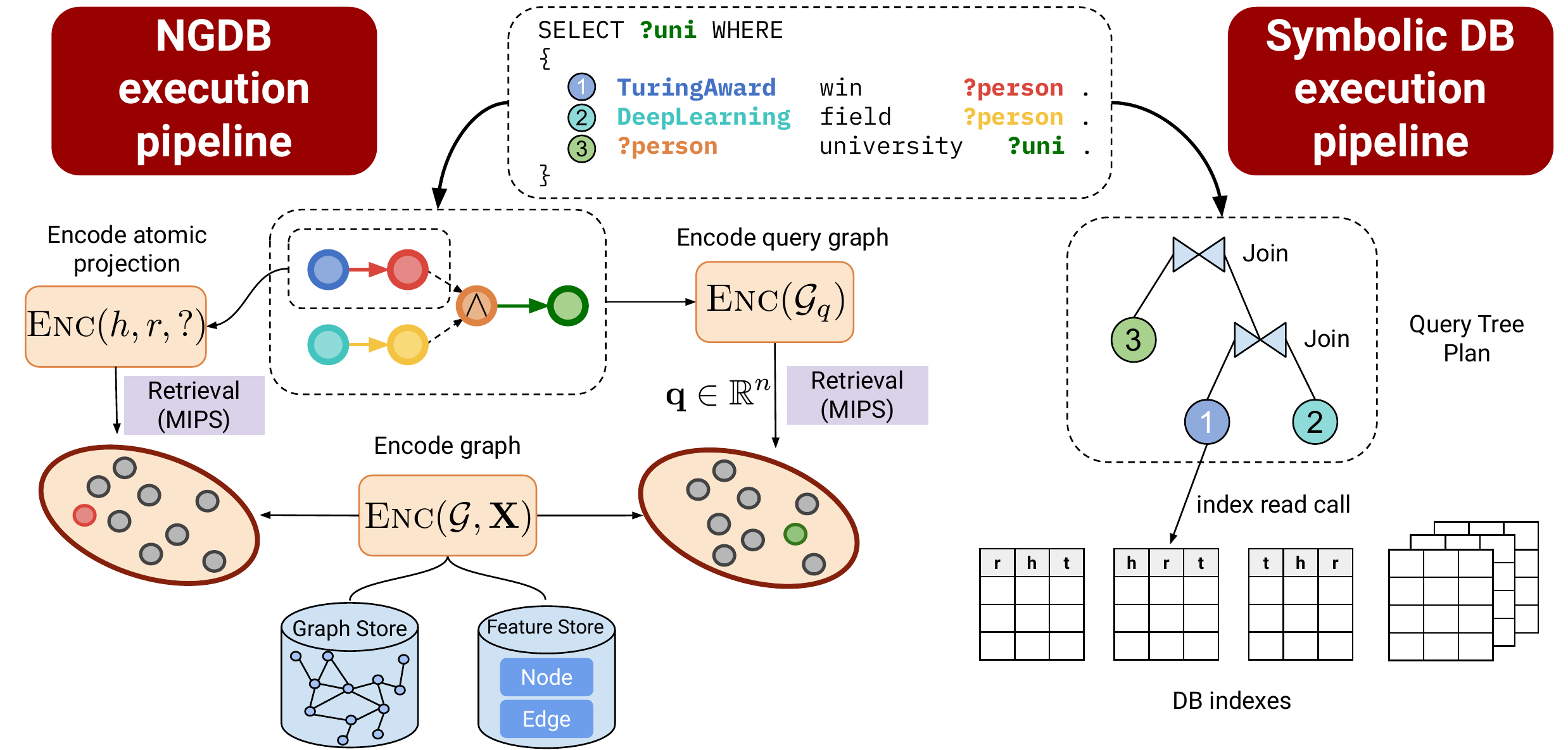}
	\caption{The storage and execution pipeline of \ngdb (left) and traditional databases (right). Traditional DBs store the graph in a collection of lookup indexes and each query pattern from the query tree plan is answered by some of those indexes. In {\ngdb}s, due to graph incompleteness, the graph and its features are first encoded in a latent space. Queries (or their atomic patterns) are encoded in a latent space either and probed using the Retrieval module (\eg, can be implemented with MIPS).}
	\label{fig:ngdb_storage}
\end{figure}

However, unlike the above methods, we propose the concept of a \emph{neural} graph storage (\autoref{fig:ngdb_storage}) where both the input graph and its vector representations are sources of truth. 
Physically, it consists of the (1) \emph{Graph Store} that stores a (multi-relational) adjacency matrix that in the basic form can be implemented with sparse COO, CSR, or more efficient compressed formats, (2) \emph{Feature Store} that stores node- and edge-level feature vectors of various formats and modalities, \eg, numerical, categorical, text data, or already given vectors.
(3) \emph{Embedding Store} that leverages an \emph{Encoder} to produce graph representations in a latent space based on the underlying adjacency and associated features. 
The embedding store is one of the biggest differences between neural graph storage and its counterpart in traditional graph DBs.
The embedding process can be viewed as a compression step but the semantic and structure similarity of entities/relations is kept. The distance between entities/relations in the embedding space should be positively correlated with the semantic/structure similarity. There are several options for the architecture of the encoder.

First, we may implement the encoder as a matrix lookup, of which each row corresponds to the embedding of an entity/relation. The benefit of such modeling is that it provides the most flexibility and the most number of free parameters -- any entry of the embedding matrix is trainable. However, it comes at the cost that it is challenging to edit the storage. We cannot easily add new entities/relations to the storage because any novel entity/relation may need a new embedding that requires learning from scratch. If we remove an entity and later add it back, such embedding matrix modeling also cannot recover the original learned embedding.

Another option for the encoder is a graph neural network (GNN) that can be analyzed through the lens of message passing and neighborhood aggregation (\autoref{sec:grl}). The idea is that we may learn the entity/relation embeddings by aggregating its neighbor information. The GNN parameters are shared for all the entities and relations on the graph. 
Such modeling provides clear benefit with much better generalization capability then shallow embedding matrices.

One important process in the neural graph storage is the \textbf{retrieval} step. 
Traditional graph databases directly perform identity-based exact match retrieval from the indexes.
In contrast, in a neural graph storage, since we store each entity and relation in the latent space, besides performing retrieval by the entity/relation id, users can also input an vector, and the retrieval process may return ``relevant'' entities/relations by measuring the distance in the embedding space. The retrieval process can be seen as a nearest neighbor search of the input vector in the embedding space. There are three direct benefits of a neural graph storage compared with traditional storages. The first advantage is that since the retrieval is operated in the embedding space with a predefined distance function, each retrieved item naturally comes with a score which may represent the confidence/relevance of the input vector in a retrieval step. 
Besides, NGDB allows for different definitions of the latent space and the distance function (which will be detailed in \autoref{sec:models}), such that NGDB is  flexible and users may customize the latent space and distance function based on different desired properties and user need.
Lastly, with the whole literature on efficient nearest neighbor search, we have the opportunity to implement the retrieval step on extremely large graphs with billions of nodes and edges with high efficiency and scalability. Existing frameworks including Faiss~\citep{faiss}, ScaNN~\citep{scann} provide scalable implementation of nearest neighbor search. 
A major limitation, however, is that existing frameworks are only applicable to L1, L2 and cosine distance functions. It is still an open research problem how to design efficient scalable nearest neighbor search algorithms for more complex distance functions such as KL divergence so that we can retrieve with much better efficiency for different \clqa methods.

\subsection{Neural Query Engine}
\label{sec:neural_qe}

In traditional databases, a typical query engine~\citep{rdf3x, endris2018querying} performs three major operations. (1) Query parsing to verify syntax correctness (often enriched with a deeper semantic analysis of query terms); (2) Query planning and optimization to derive an efficient query plan (usually, a tree of relational operators) that minimizes computational costs; (3) Query execution that scans the storage and processes intermediate results according to the query plan.

In {\ngdb}s, the sources of truth are both graph adjacency (possibly incomplete) and latent graph representations (\eg, vector representations of entities and relations). 
Similarly, queries (or their atomic operations) are encoded into a latent space. 
To answer encoded queries over latent spaces, we devise \emph{Neural Query Engines} that include two modules: (1) \emph{Query Planner} to derive an efficient query plan of atomic operations (\eg, projections and logical operators) maximizing completeness (all answers over existing edges must be returned) and inference (of missing edges predicted on the fly) taking into account query complexity, prediction tasks, and partitioning of sources; (2) \emph{Query Executor} that encodes the derived plan in a latent space, executes the atomic operations (or the whole query) over the graph and its latent representations, and aggregates intermediate results into a final answer set with possible postprocessing.

Broadly in Deep Learning, prompting large language models (LLMs) can be seen as a form of a neural query engine where queries are  unstructured texts. 
Recent prompting techniques like Chain of Thought~\citep{chain_of_thought} or Program of Thought~\citep{program_of_thought} achieved remarkable progress in natural language reasoning~\citep{prompt_survey} by providing few prompts with ``common sense'' examples of solving a given problem step-by-step. 
Such step-by-step instructions resemble a  \emph{query plan} designed manually by a prompt engineer and optimized for \emph{query execution} against a particular LLM where query execution  is framed as the language model objective (predicting next few tokens). 
Prompting often does not require any additional LLM fine-tuning or training and works in the inference regime (often called \emph{in-context learning}). 
Pre-trained LLMs can thus be seen as \emph{neural databases} that can be queried with a sequence of prompts organized in a domain-specific query language~\citep{lmql}.
Recently, a similar prompting technique~\citep{drozdov2022compositional} demonstrated strong systematic generalization skills by solving a long-standing semantic parsing task of converting natural language questions to SPARQL queries. 
Several major drawbacks of such querying techniques are: (1) a limited context window (usually, less than 8192 tokens) that does not allow prompting the whole database content; (2) issues with factual correctness and hallucinating of the generated reponses.

\begin{figure}[t]
	\centering
	\includegraphics[width=0.9\linewidth]{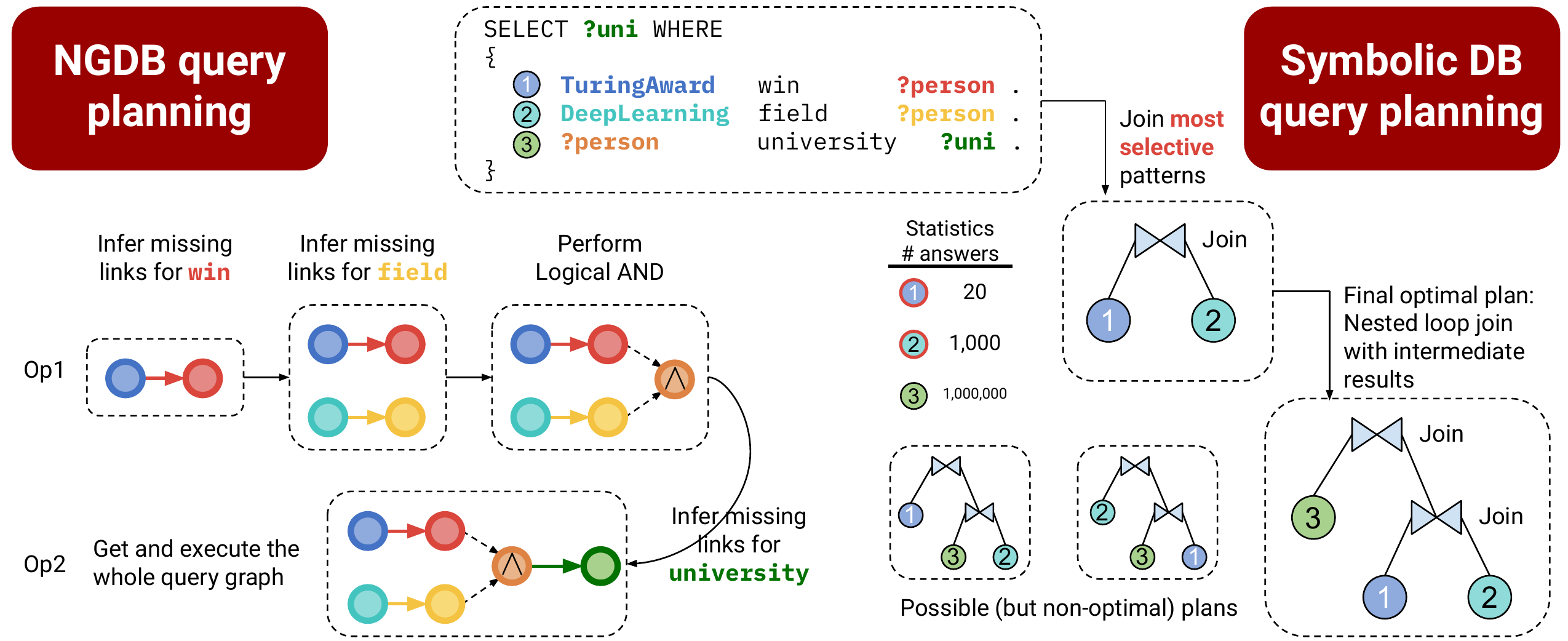}
	\caption{Query planning in {\ngdb}s (left) and traditional graph DBs (right). The \ngdb planning (assuming incomplete graphs) can be performed autoregressively step-by-step (1) or generated entirely in one step (2). The traditional DB planning is cost-based and resorts to metadata (assuming complete graphs and extracted from them) such as the number of intermediate answers to build a tree of join operators.
 }
	\label{fig:ngdb_query_planning}
\end{figure}

\paragraph{Query Planner} 
In {\ngdb}s, an input query is expected to arrive in the structured form, \eg, as generic as FOL -- we leave the discussion on possible query languages out of scope of this work emphasizing the breadth of general and domain-specific languages. 
For illustrating example queries, however, we use the SPARQL notation as one of the standard query languages for graph databases. 
Broadly, principal differences between symbolic and neural query planning are illustrated in \autoref{fig:ngdb_query_planning}.
While traditional graph DBs are symbolic and deductive, {\ngdb}s are neural and inductive (\ie, can generalize to unseen data). 
As we assume the graph data is incomplete, neural reasoning is expected to process more intermediate results that makes query planning even more important in deriving the most efficient plan.

The task of the Query Planner is to optimize the execution process by deriving the most efficient query plan, \ie, the execution order of atomic and logical operations, given the query complexity, expected prediction task, and configuration of the storage.
We envision the storage configuration to play an important role in very large graphs that cannot be stored entirely in the main memory. 
A typical configuration would 
resort to partitioning of graphs and latent representations across many devices in a distributed fashion. 
Therefore, the planner has to identify which parts of the input query to send to a relevant partition. 
Training and inference of ML models in distributed environments (with possible privacy restrictions) is at the core of \emph{federated learning}~\citep{li2020federated} and \emph{differential privacy}~\citep{diff_privacy} and we posit they would be important components of very large {\ngdb}s.

Common metrics of evaluating the quality of plans are \emph{execution time} (in units of time) and \emph{throughput} (number of queries per unit of time). 
In traditional graph DBs, different query plans return the same answers but require different execution time ~\citep{endris2018querying} due to the order of join operators (\eg, nested loop joins or hash joins), the number of calls to indexes, and intermediate results to process.
In contrast, {\ngdb}s execute queries in the latent space together with link prediction and, depending on the approach, time complexity of atomic operations (relation projection and logical operators) might depend on the hidden dimension $\gO(d)$ or the number of entities $\gO(\gE)$ (we elaborate on complexity in \autoref{sec:complexity}).

To date, query planning is still a challenge for neural query answering methods as all existing approaches (\autoref{sec:learning}) execute an already given sequence of operators and all existing datasets (\autoref{sec:datasets}) provide such a hardcoded sequence without optimizations. 
Furthermore, some existing approaches, \eg, CQD~\citep{cqd}, are susceptible to the issue when changing a query plan might result in a different answer set.

We hypothesize that principled approaches for deriving efficient query plans taking into account missing edges and incomplete graphs might be framed as \emph{Neural Program Synthesis}~\citep{neural_program_synth} often used to derive query plans for LLMs to solve complex numerical reasoning tasks~\citep{neural_module_nets, nerd}.
In the complex query answering domain, the first attempt to apply neural program synthesis for generating query plans was made by the Latent Execution-Guided Reasoning (LEGO) framework~\citep{lego}. 
LEGO iteratively expands the initial query tree root by sampling from the space of relation projection and logical operators. 
The sampling is based on learnable heuristics and pruning mechanisms.

\paragraph{Query Executor} 
Once the query plan is finalized, the \emph{Query Executor} module encodes the query (or its parts) into a latent space, communicates with the Graph Storage and its Retrieval module, and aggregates intermediate results into the final answer set. Following the \autoref{def:easy_hard_ans}, there exists easy answers and hard answers. Query Executors are expected to return the compelte set of easy answers and recover missing edges to return the missing hard answers.

There exist two common mechanisms for neural query execution described in \autoref{sec:learning}: 
(1) \emph{atomic}, resembling traditional DBs, when a query plan is executed sequentially by encoding atomic patterns (such as relation projections), retrieving their answers, and executing logical operators as intermediate steps; (2) \emph{global}, when the entire query graph is encoded and executed in a latent space in one step. 
For example (\autoref{fig:ngdb_storage}), given a query plan $q = U_? . \exists V: \textit{win}(\mathtt{TuringAward}, V) \land \textit{field}(\mathtt{DeepLearning}, V) \land \textit{university}(V, U_?)$, the atomic mechanism executes separate relation projections, \eg, $\textit{win}(\mathtt{TuringAward}, V)$, sequentially while the global mechanism encodes the whole query graph and probes it against the latent space of the graph.

Direct implications of the chosen mechanism include computational complexity (\autoref{sec:complexity}) and supported logical operators (\autoref{sec:queries}), \ie, fully latent mechanisms are mostly limited to conjunctive and intersection queries and have worse generalization qualities. 
To date, most methods follow the atomic mechanism.

The main challenge for neural query execution is matching query expressiveness to that of symbolic languages like SPARQL or Cypher.
The challenge includes three aspects:
(1) handling more expressive query operators such as \texttt{FILTER} or aggregations like \texttt{COUNT} or \texttt{SUM}; (2) supporting  query patterns more complex than trees; (3) supporting several projected variables and operations on them.
We elaborate in \autoref{sec:queries}.

\paragraph{A Taxonomy of Query Reasoning Methods}
\label{sec:taxonomy}

\begin{figure}[!ht]
    \centering
    \begin{adjustbox}{height=0.9\textheight}
      \input{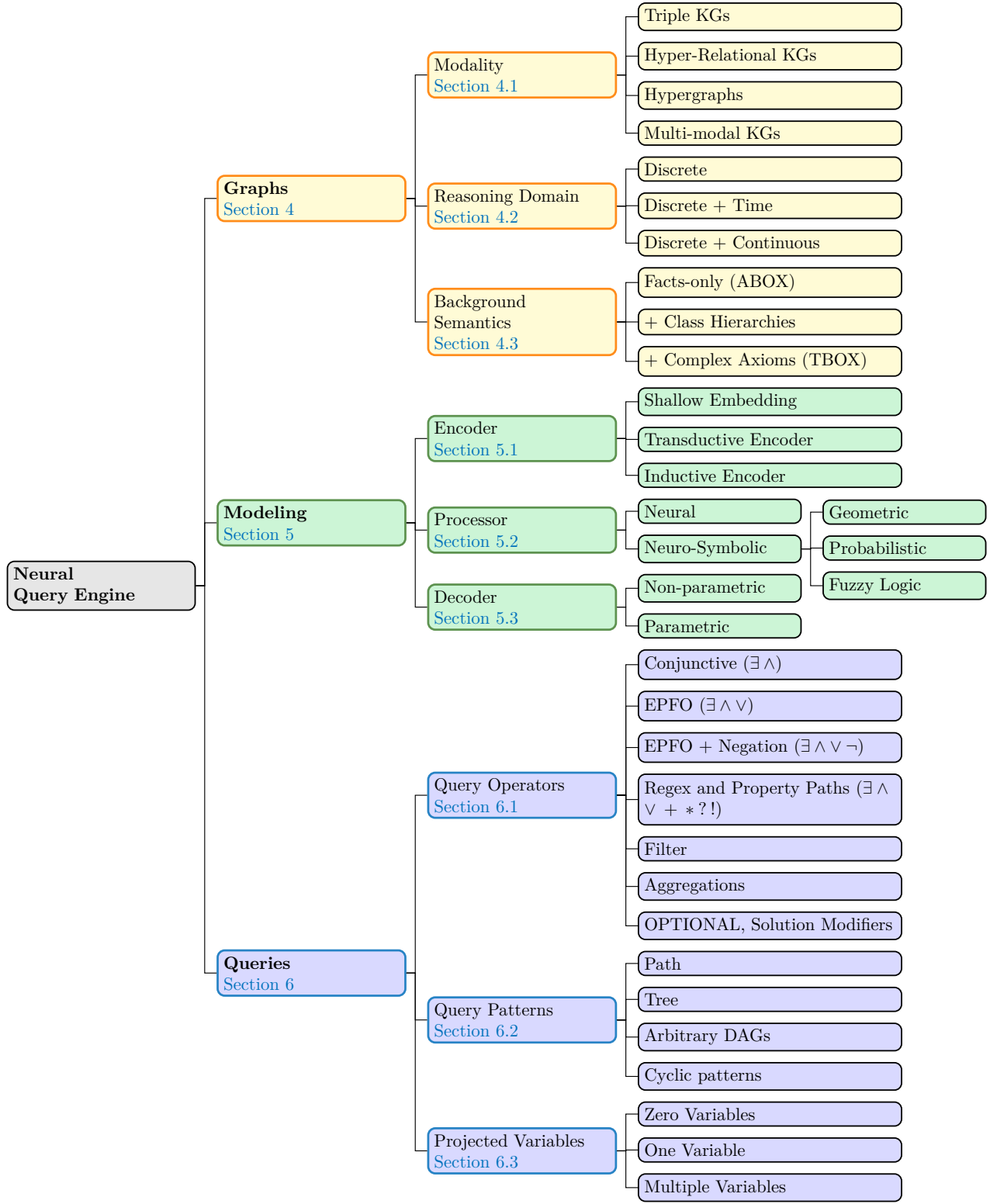}
    \end{adjustbox}
    \caption{The Neural Query Engine Taxonomy consists of three main branches -- Graphs, Modeling, and Queries. We describe each branch in more detail including prominent examples in the relevant sections.}
    \label{fig:taxonomy}
\end{figure}

In the following sections, we devise a taxonomy of query answering methods as a component of the \emph{Neural Query Engine (NQE)}. 
We categorize existing and future approaches along three main directions: (i) \textbf{Graphs} -- what is the underlying structure against which we answer queries; (ii) \textbf{Modeling} -- how we answer queries and which inductive biases are employed; (iii) \textbf{Queries} -- what we answer, what are the query structures and what are the expected answers. 
The taxonomy is presented in \autoref{fig:taxonomy}. 
In the following sections, we describe each direction in more detail and illustrate them with examples covering the whole existing literature on complex query answering (more than 40 papers).

\clearpage

\section{Graphs}
\label{sec:graphs}

The \emph{Graphs} category covers the underlying graph structure ($\gG$ in Definition~\ref{def:basic_approx_gqa}) against which complex queries are sent and the answers are produced. 
Understanding the graph, its contents and modeling paradigms is crucial for designing query answering models. 
To this end, we propose to analyze the underlying graph from three aspects: \emph{Modality}, \emph{Reasoning Domain}, and \emph{Background Semantics}.

\subsection{Modality}
\label{sec:modality}

We highlight four modalities common for KGs and graph databases: standard \textbf{triple-based KGs} adhering to the RDF data model, \textbf{hyper-relational KGs} following the RDF-star or Labeled Property Graph (LPG) formats, and \textbf{hypergraph KGs} (we elaborate on choosing an appropriate modeling paradigm in \autoref{sec:storage}).
The difference among the three modalities is illustrated in \autoref{fig:modality}.
Additionally, we outline \textbf{multi-modal KGs} that contain not just a graph of nodes and edges, but also text, images, audio, video, and other data formats linked to the underlying graph explicitly or implicitly.

We categorize the literature along the Modality aspect in \autoref{tab:modality}.
To date, most query answering approaches operate solely on \textbf{triple-based} graphs.
Among approaches supporting \textbf{hyper-relational} graphs, we are only aware of StarQE~\citep{starqe} that incorporates entity-relation \emph{qualifiers} over labeled edges and its extension NQE~\citep{nqe}. 
We posit that the hyper-relational model might serve as a theoretical foundation of temporal query answering approaches since temporal attributes are in fact continuous key-value edge attributes.
To date, we are not aware of complex query answering models supporting \textbf{hypergraphs} or \textbf{multi-modal} graphs. 
We foresee them as possible area of future research in the area.

\begin{table}[!ht]
\centering
\caption{Complex Query Answering approaches categorized under \emph{Modality}.}\label{tab:modality}
\resizebox{\columnwidth}{!}{%
\begin{tabular}{cccc}\toprule
Triple-only & Hyper-Relational & Hypergraph & Multi-modal \\
\midrule
 \multicolumn{1}{p{11cm}}{
  GQE \citeps{gqe}, GQE w hash \citeps{gqe_bin}, CGA \citeps{cga}, TractOR \citeps{tractor2020}, Query2Box \citeps{q2b}, BetaE \citeps{betae}, EmQL \citeps{emql}, MPQE \citeps{mpqe}, Shv \citeps{sheaves}, Q2B Onto \citeps{q2b_onto}, RotatE-Box \citeps{regex}, BiQE \citeps{biqe}, HyPE \citeps{hype}, NewLook \citeps{newlook}, CQD \citeps{cqd}, PERM \citeps{perm}, ConE \citeps{cone}, LogicE \citeps{logic_e}, MLPMix \citeps{mlpmix}, FuzzQE \citeps{fuzz_qe}, GNN-QE \citeps{gnn_qe}, GNNQ \citeps{gnnq}, SMORE \citeps{smore}, KGTrans \citeps{kgtrans2022}, LinE \citeps{line2022}, Query2Particles \citeps{query2particles2022}, TAR \citeps{abin_abductive2022}, TeMP \citeps{temp2022}, FLEX \citeps{lin2022flex}, TFLEX \citeps{lin2022tflex}, NodePiece-QE \citeps{galkin2022}, ENeSy \citeps{enesy}, GammaE \citeps{gammae}, NMP-QEM \citeps{nmp_qem}, QTO \citeps{qto}, SignalE \citeps{signale}, LMPNN \citeps{lmpnn}, Var2Vec \citeps{var2vec}, $\text{CQD}^{\gA}$ \citeps{cqda}, Query2Geom \citeps{query2geom}, SQE \citeps{sqe}
 }
 & 
 \multicolumn{1}{p{5cm}}{
 StarQE \citeps{starqe}, NQE \citeps{nqe}
 }
 & None & None \\
\bottomrule
\end{tabular}
}
\end{table}

\subsection{Reasoning Domain}
\label{sec:domain}

\begin{figure}[!ht]
	\centering
	\includegraphics[width=\linewidth]{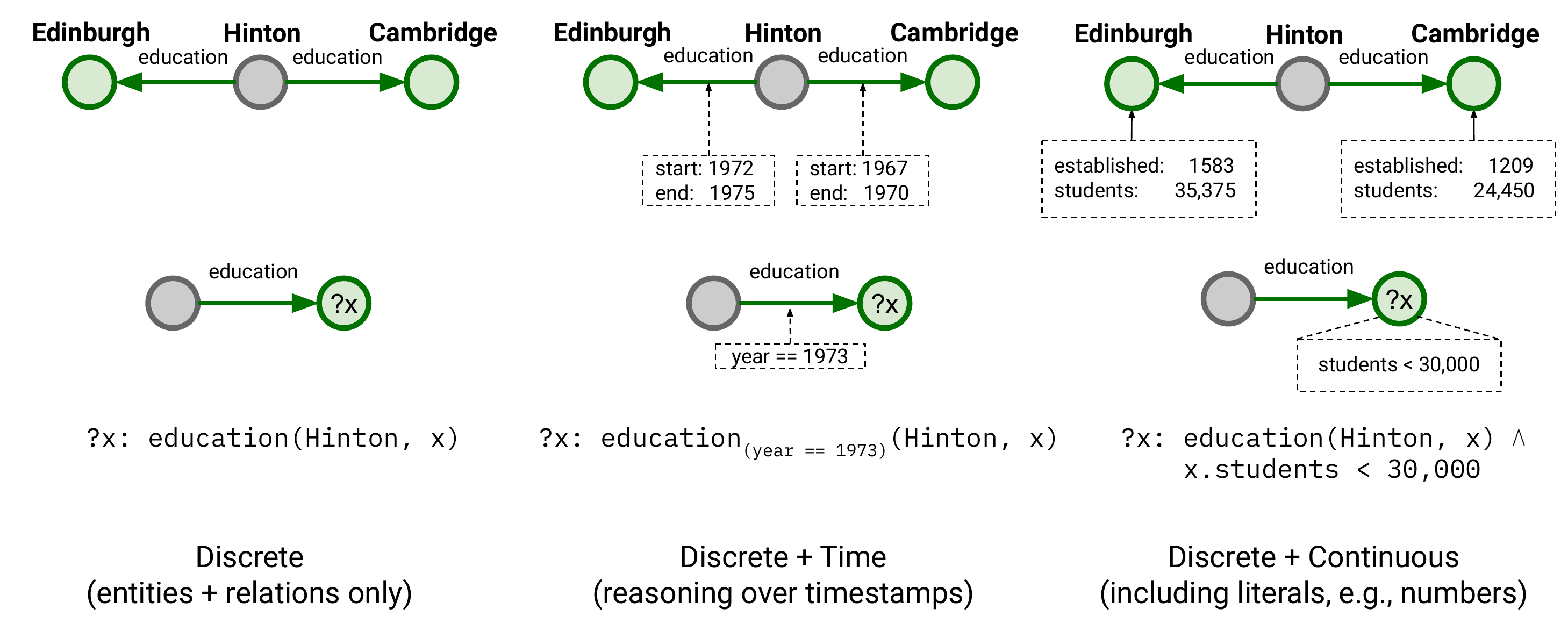}
	\caption{Reasoning Domains. The \emph{Discrete} domain only allows entities and relations as constants, variables, and answers. The \emph{Discrete + Time} domain extends the space to discrete timestamps and time-specific operators. The \emph{Discrete + Continuous} domain allows continuous inputs (literals) and outputs.}
	\label{fig:reasoning_domain}
\end{figure}

Following Definition~\ref{def:basic_graph_query_answering}, a query $\gQ$ includes constants $\con$, variables $\variable$, and returns \emph{answers} as mappings $\mu: \variable_Q \rightarrow \con$.
By \emph{Reasoning Domain} we understand a space of possible constants, variables, and answers, that is, what query answering models can reason about.
We highlight three common domains (\emph{Discrete}, \emph{Discrete + Time}, \emph{Discrete + Continuous}), illustrate them in \autoref{fig:reasoning_domain}, and categorize existing works in \autoref{tab:reasoning_domain}. 
Each subsequent domain is a superset of the previous domains, \eg, \emph{Discrete + Continuous} includes the capabilities of \emph{Discrete} and \emph{Discrete + Time} and expands the space to continuous inputs and outputs.

In the \emph{Discrete} domain, constants, variables, and answers can be entities $\gE$ and (or) relation types $\gR$ of the KG, $\con \subseteq \gE \cup \gR$, $\variable \subseteq \gE \cup \gR$, $\mu \subseteq \gE \cup \gR$.
That is, input queries may only contain entities (or relations) and their answers are only entities (or relations). 
For example, a query $?x: \mathtt{education}(\mathtt{Hinton}, x)$ in \autoref{fig:reasoning_domain} can only return two entities $\mathtt{\{Edinburgh, Cambridge\}}$ as answers.
Conceptually, the framework allows relation types $r \in \gR$ to be variables as well describing, for example, a SPARQL graph pattern \texttt{\{Hinton ?r Cambridge\}} -- or $?r: r(\mathtt{Hinton}, \mathtt{Cambridge})$ in the functional form -- that returns all relation types between two nodes \texttt{Hinton} and \texttt{Cambridge}.
However, to the best of our knowledge, the majority of existing query answering literature and datasets limit the space of constants, variables, and answers to entities-only, $\con \subseteq \gE$, $\variable \subseteq \gE$, $\mu \subseteq \gE$ and all relation types are given in advance (we discuss queries structure in more detail in \autoref{sec:queries}).
To date, most of the literature in the field 
belongs to the \emph{Discrete} reasoning domain (\autoref{tab:reasoning_domain}).

Some nodes and edges might have \emph{timestamps} from a set of discrete timestamps $t \in \mathcal{TS}$ indicating a validity period of a certain statement. 
In a more general case, certain subgraphs might be timestamped. 
We define the \emph{Discrete + Time} domain when queries include temporal data. 
In this domain, the set of constants is extended with timestamps $\con \subseteq \gE \cup \gR \cup \mathcal{TS}$ and relation projections might be instantiated with a certain timestamp $R_t(a,b)$.
For instance (\autoref{fig:reasoning_domain}), given a timestamped graph and a query $?x: \mathtt{education}_{\mathtt{year}==1973}(\mathtt{Hinton}, x)$, the answer set includes only \texttt{Edinburgh} as the timestamp $1973$ falls into the validity period of only one edge.

It is possible to extend the domain of variables and query answers with timestamps as well, $\variable, \mu \subseteq \gE \cup \gR \cup \mathcal{TS}$, such that queries might employ timestamps as intermediate existentially quantified variables and possible answers can be entities or timestamps.
This approach is followed by the Temporal Feature-Logic Embedding Framework (TFLEX) by \citet{lin2022tflex} that defines additional operators \textbf{before}, \textbf{after}, \textbf{between} over edges with discrete timestamps. 

Finally, the most expressive domain is \emph{Discrete + Continuous} that enables reasoning over continuous inputs (such as numbers, texts, continuous timestamps) often available as node and edge attributes or \emph{literals}. 
Formally, for numerical data, the space of constants, variables, and answers is extended with real numbers $\RR$, \ie, $\con, \variable, \mu \subseteq \gE \cup \gR \cup \mathcal{TS} \cup \RR$.
An example query in \autoref{fig:reasoning_domain} $?x: \mathtt{education}(\mathtt{Hinton}, x) \land x.\mathtt{students} < 30000$ includes a conjunctive term $x.\mathtt{students} < 30000$ that requires numerical reasoning over the \texttt{students} attribute of a variable $x$ to produce the answer  \texttt{Cambridge}. 
In a similar fashion, extending the answer set to continuous outputs can be framed as a regression task.
To date, we are not aware of any complex query answering approaches capable of working in the continuous domain. %
Still, the ability to reason over continuous data is crucial for query answering given that most real-world KGs heavily rely on literals.

\begin{table}[!ht]
\centering
\caption{Complex Query Answering approaches categorized under \emph{Reasoning Domain}.}\label{tab:reasoning_domain}
\resizebox{\columnwidth}{!}{%
\begin{tabular}{ccc}\toprule
Discrete & Discrete + Time & Discrete + Continuous \\
\midrule
\multicolumn{1}{p{11cm}}{
GQE \citeps{gqe}, GQE w hash \citeps{gqe_bin}, CGA \citeps{cga}, TractOR \citeps{tractor2020}, Query2Box \citeps{q2b}, BetaE \citeps{betae}, EmQL \citeps{emql}, MPQE \citeps{mpqe}, Shv \citeps{sheaves}, Q2B Onto \citeps{q2b_onto}, RotatE-Box \citeps{regex}, BiQE \citeps{biqe}, HyPE \citeps{hype}, NewLook \citeps{newlook}, CQD \citeps{cqd}, PERM \citeps{perm}, ConE \citeps{cone}, LogicE \citeps{logic_e}, MLPMix \citeps{mlpmix}, FuzzQE \citeps{fuzz_qe}, GNN-QE \citeps{gnn_qe}, GNNQ \citeps{gnnq}, SMORE \citeps{smore}, KGTrans \citeps{kgtrans2022}, LinE \citeps{line2022}, Query2Particles \citeps{query2particles2022}, TAR \citeps{abin_abductive2022}, TeMP \citeps{temp2022}, FLEX \citeps{lin2022flex}, NodePiece-QE \citeps{galkin2022}, ENeSy \citeps{enesy}, GammaE \citeps{gammae}, NMP-QEM \citeps{nmp_qem}, StarQE \citeps{starqe}, QTO \citeps{qto}, SignalE 
\citeps{signale}, LMPNN \citeps{lmpnn}, NQE \citeps{nqe}, Var2Vec \citeps{var2vec}, $\text{CQD}^{\gA}$ \citeps{cqda}, Query2Geom \citeps{query2geom}, SQE \citeps{sqe}
}
& 
TFLEX \citeps{lin2022tflex} 
& None \\
\bottomrule
\end{tabular}
}
\end{table}

\subsection{Background Semantics}
\label{sec:bg_sem}

\begin{figure}[!ht]
	\centering
	\includegraphics[width=\linewidth]{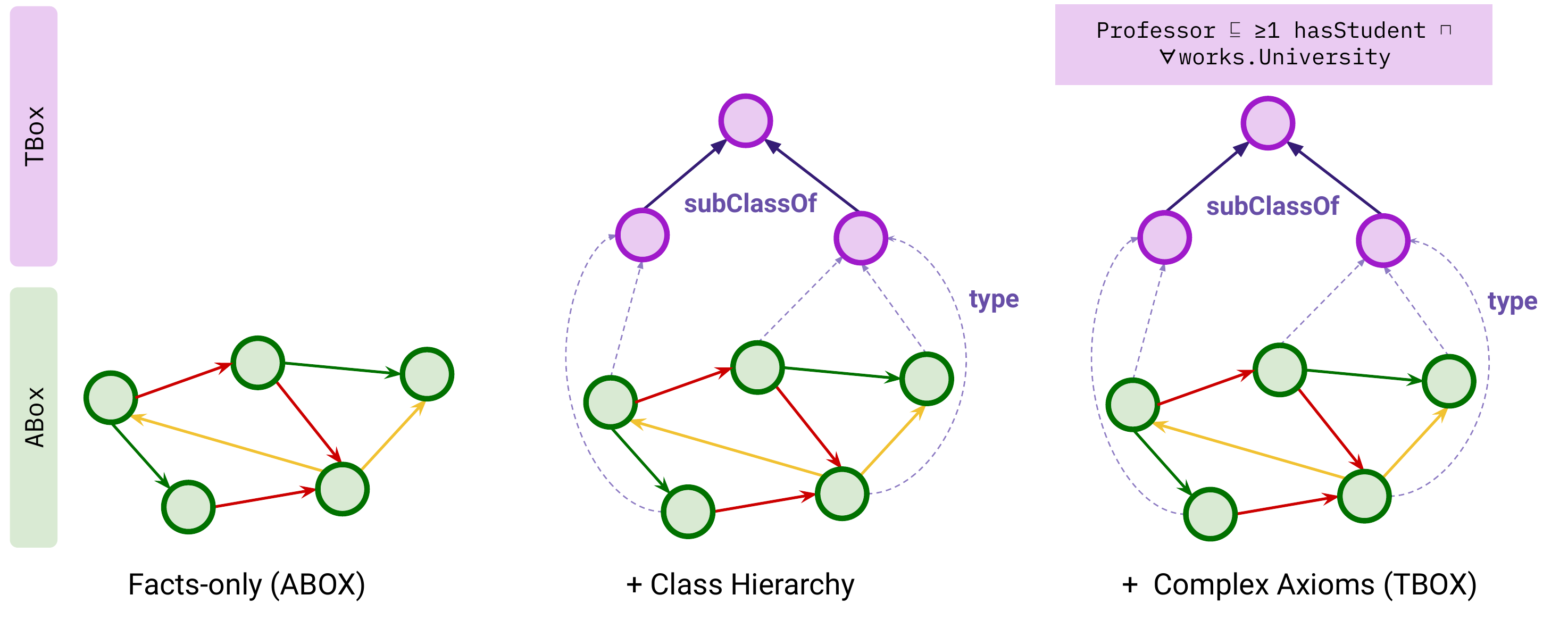}
	\caption{Background Semantics. \emph{Facts-only} are graphs that only have assertions (\abox) and have no higher-level schema. \emph{Class Hierarchy} introduces node types (classes) and hierarchical relationships between classes. Finally, \emph{Complex Axioms} add even more complex logical rules (\tbox) governed by certain OWL profiles, \eg, \emph{A professor is someone who has one or more students and works at university.}}
	\label{fig:bg_sem}
\end{figure}

Relational databases often contain a \emph{schema}, that is, a specification of how tables and columns are organized that gives a high-level overview of the database content.
In graphs databases, schemas exist as well and might describe, for instance, node types are common in the Labeled Property Graph (LPG) paradigm. 
RDF graphs, however, might provide an additional layer of logical consistence by employing standards based on \emph{Description Logics}~\citep{baader2003description}. 
As incorporating schema is crucial for designing effective query answering ML models, we introduce the \emph{Background Semantics} (\autoref{fig:bg_sem}) as the notion of additional schema information available on top of plain facts (statements).

\paragraph{Facts-only.} In the simplest case, there is no background schema such that a KG consists of statements (facts) only, that is, a KG follows Definition~\ref{def:kg}, $\gG = (\gE, \gR, \gS)$. 
In terms of description logics, a graph only has \emph{assertions} (\abox). 
Queries, depending on the Reasoning Domain (\autoref{sec:domain}), involve only entities, relations, and literals.
The original GQE~\citep{gqe} focused on facts-only graphs and the majority of subsequent query answering approaches (\autoref{tab:bg_semantics}) operate exclusively on schema-less KGs.

\paragraph{Class Hierarchy.} 
Classes of entities, or node types, are a natural addition to a facts-only graph as a basic schema. 
Extending Definition~\ref{def:kg} with a set of types $\gT$, a graph $\gG$ is defined as $\gG = (\gE, \gR, \gS, \gT)$. 
Each entity $e$ might have one or more associated classes $\textit{type}(e) = \{ t_1, \dots, t_k \} | t \in \gT$.
Both LPG and RDF graphs support classes albeit in RDF it is possible to specify hierarchical relationships between classes using RDF Schema (RDFS)~\citep{brickley2014rdf}\footnote{RDFS has more expressive means (\eg, a hierarchy of relations)  but we leave them to \emph{Complex Axioms}}.
Formally, types become explicit nodes in a graph, and statements $\gS$ can now have types as subjects or objects, $\gS \subset ((\gE \cup \gT) \times \gR \times (\gE \cup \gT))$, or other statement elements in graphs of other modalities.
Practically, edges involving types might be present physically in a KG or be considered as an additional supervised input to a particular model.

To date, we are aware of three query answering approaches (\autoref{tab:bg_semantics}) that incorporate entity types.
Contextual Graph Attention (\textsc{CGA}, \citet{cga}) only uses types for entity embedding initialization and requires each entity to have only one type. 
The queries are not conditioned by entity types and the answer set still includes entities only. 
That is, query constants, variables, and the answer set follow  $\con \subseteq \gE, \variable \subseteq \gE, \mu \subseteq \gE$. 

In Type-Aware Message Passing (\textsc{Temp}, \citet{temp2022}), type embeddings are used to enrich entity and relation representations that are later sent to a downstream query answering model. 
Each node might have several types. 
In the inductive scenario (we elaborate on inference scenarios in \autoref{sec:trans_ind}) with unseen nodes at inference time, type and relation embeddings are learned \emph{invariants} that transfer across training and inference entities.
Query-wise, constants, variables, and the answer set are still limited to entities only.

The \tbox\xspace and \abox\xspace Neural Reasoner (\textsc{TAR}, \citet{abin_abductive2022}) incorporates types and their hierarchy to improve predictions over entities as well as introduces a task of predicting types of answer entities, that is, $\mu \subseteq (\gE \cup \gT)$. 
Constants and variables are still entity-only such that using types as query variables is not allowed.
The class hierarchy in \textsc{TAR} is used in three auxiliary losses besides the original entity prediction, that is, \emph{concept retrieval} -- prediction of the answer set of types, \emph{subsumption} -- predicting which type is a subclass of another type, and \emph{instantiation} -- predicting a type for each entity.

A natural next step for the \emph{Class Hierarchy} family of approaches is to incorporate types in queries in the form of constants and variables, $\con \subseteq (\gE \cup \gT), \variable \subseteq (\gE \cup \gT)$.

\paragraph{Complex Axioms.}
Finally, a schema might contain not just a class hierarchy but a set of more complex axioms involving, for example, a hierarchy of relations, qualified restrictions on relations, or composite classes. 
Such a complex schema can now be treated as an \emph{ontology} $\gO$ that extends the definition of the graph $\gG = (\gE, \gR, \gS, \gO)$.
In \autoref{fig:bg_sem}, the axiom $\texttt{Professor} \sqsubseteq \: \geq1 \: \textit{hasStudent} \sqcap \forall\textit{works}.\texttt{University}$ describes that \emph{A professor is someone who has one or more students and works at university.}
In terms of description logics, a graph has an additional \emph{terminology} component (\tbox). 
The expressiveness of \tbox\xspace directly affects the complexity of symbolic reasoning engines up to exponential (\textsc{ExpTime}) for most expressive fragments.
To alleviate this issue, ontology languages (like Web Ontology Language, OWL) specify less expressive but computationally feasible \emph{profiles}, e.g., OWL 2 introduces three profiles \emph{OWL EL}, \emph{OWL QL}, \emph{OWL RL}~\citep{motik2009owl}. \emph{OWL-QL}, in turn, is loosely connected to the $\textit{DL-Lite}_{\gR}$ ontology language~\citep{artale2009dl}.

In graph representation learning, incorporating complex ontological axioms is non-trivial even for simple link prediction models~\citep{zhang2022_logic_emb_survey}.
In the query answering literature, the only attempt to include complex axioms is taken by \citet{q2b_onto}. 
In Q2B Onto (O2B), an extension of Query2Box (Q2B, \cite{q2b}), the set of considered complex axioms belongs to the $\textit{DL-Lite}_{\gR}$ fragment and supports the hierarchy of classes (\emph{subclasses}), the hierarchy of relations (\emph{subproperties}), as well as \emph{range} and \emph{domain} of relations.
The model architecture is not directly conditioned on the axioms and remains the original Query2Box. 
Instead, the axioms affect the graph structure, query sampling, and an auxiliary loss, that is, \emph{query rewriting} mechanisms are used to materialize more answers to original queries as if executed against the complete graph (\emph{deductive closure}) akin to data augmentation.
During optimization, an auxiliary regularization loss aims at including a specialized query box %
$q$ into the  more general version of this query $q'$.

Still, even the expensive procedure of incorporating complex axioms in query sampling in O2B benefits mostly the \emph{deductive} capabilities of query answering, that is, inferring answers that are already implied by the graph $\gG$ and ontology $\gO$, and does not improve the \emph{generalization} capabilities when missing edges cannot be inferred by ontological axioms. 
We elaborate on \emph{deductive}, \emph{generalization}, and other setups in \autoref{sec:datasets}. 

Another avenue for future work is a better understanding of theoretical expressiveness of Graph Neural Network (GNN) encoders when applied to multi-relational KGs. 
Initial works on non-relational graphs~\citep{Barcelo2020logic} map the expressiveness to the $\text{FOC}_2$ subset of FOL with two variables and counting quantifiers, and to $\text{FOC}_B$~\citep{luo2022on} for hypergraphs of maximum arity $B$. 
In relational graphs, \citet{rwl} quantified the expressiveness of relational GNNs in terms of the \emph{relational Weisfeiler-Leman} (RWL) test proving that RWL is more expressive than classical WL test~\citep{wl1968} and that common relational GNN architectures like R-GCN~\citep{rgcn} and CompGCN~\citep{compgcn} are bounded by 1-RWL.
Using RWL, \citet{huang2023theory} derive that the family of GNNs conditioned on the query node, such as Neural Bellman-Ford Networks~\citep{zhu2021neural}, are bounded by the asymmetric local 2-RWL and expressive as $\text{rGFO}^3_{\text{cnt}}$, restricted guarded first-order logic fragment with three variables and counting.
Concurrently, \citet{gao2023double} study KGs as double permutation equivariant structures (to permuting nodes and edge types) and map their expressiveness to universally quantified entity-relation (UQER) Horn clauses.
However, it is still an open question if GNNs can capture OWL-like axioms and leverage them as an inductive bias in complex query answering.

\begin{table}[!ht]
\centering
\caption{Complex Query Answering approaches categorized under \emph{Background Semantics }.}\label{tab:bg_semantics}
\resizebox{\columnwidth}{!}{%
\begin{tabular}{ccc}\toprule
Facts-only (ABOX) & + Class Hierarchy & + Complex Axioms (TBOX) \\
\midrule
\multicolumn{1}{p{10cm}}{
GQE \citeps{gqe}, GQE w hash \citeps{gqe_bin}, TractOR \citeps{tractor2020}, Query2Box \citeps{q2b}, BetaE \citeps{betae}, EmQL \citeps{emql}, Shv \citeps{sheaves}, RotatE-Box \citeps{regex}, MPQE \citeps{mpqe}, BiQE \citeps{biqe}, HyPE \citeps{hype}, NewLook \citeps{newlook}, CQD \citeps{cqd}, PERM \citeps{perm}, ConE \citeps{cone}, LogicE \citeps{logic_e}, MLPMix \citeps{mlpmix}, FuzzQE \citeps{fuzz_qe}, GNN-QE \citeps{gnn_qe}, GNNQ \citeps{gnnq}, SMORE \citeps{smore}, KGTrans \citeps{kgtrans2022}, LinE \citeps{line2022}, Query2Particles \citeps{query2particles2022}, FLEX \citeps{lin2022flex}, TFLEX \citeps{lin2022tflex}, NodePiece-QE \citeps{galkin2022}, ENeSy \citeps{enesy}, GammaE \citeps{gammae}, NMP-QEM \citeps{nmp_qem}, StarQE \citeps{starqe}, QTO \citeps{qto}, SignalE \citeps{signale}, LMPNN \citeps{lmpnn}, NQE \citeps{nqe}, Var2Vec \citeps{var2vec}, $\text{CQD}^{\gA}$ \citeps{cqda}, Query2Geom \citeps{query2geom}, SQE \citeps{sqe}
} 
& 
\multicolumn{1}{p{3cm}}{
CGA \citeps{cga}, TeMP \citeps{temp2022}, TAR \citeps{abin_abductive2022}
} 
& 
Q2B Onto \citeps{q2b_onto} \\
\bottomrule
\end{tabular}
}
\end{table}

\section{Modeling}
\label{sec:learning}

\begin{figure}[!ht]
	\centering
	\includegraphics[width=\linewidth]{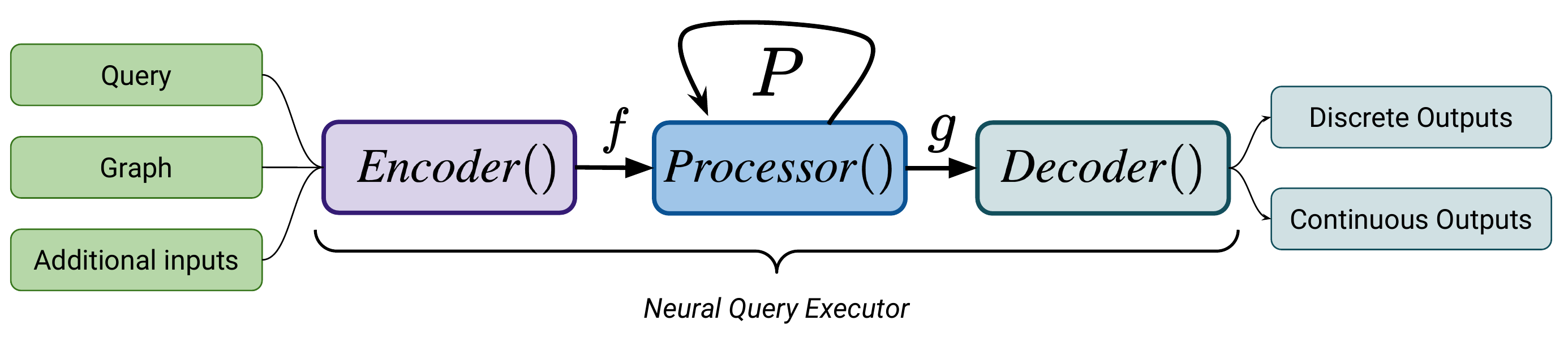}
	\caption{\emph{Neural Query Execution} through the \emph{Encoder}-\emph{Processor}-\emph{Decoder} modules. Encoder function $f$ builds representations of inputs (query, target graph, auxiliary data) in the latent space. Processor $P$ executes the query with its logical operators against the graph conditioned on other inputs. Decoder function $g$ builds requested outputs that might be discrete or continuous.}
	\label{fig:enc_proc_dec}
\end{figure}

In this section, we discuss the literature from the perspective of \emph{Modeling}. Following the common methodology~\citep{battaglia2018relational}, we segment the \emph{Modeling} methods through the lens of \emph{Encoder-Processor-Decoder} modules (illustrated in \autoref{fig:enc_proc_dec}).
(1) The \emph{Encoder} $\textsc{Enc}()$ takes an input query $q$, target graph $\gG$ with its entities and relations, and auxiliary inputs (\eg, node, edge, graph features) to build their representations in the latent space.
(2) The \emph{Processor} $P$ leverages the chosen inductive biases to process representations of the query with its logical operators in the latent or symbolic space.
(3) The \emph{Decoder} $\textsc{Dec}()$ takes the processed latents and builds desired outputs such as a distribution over discrete entities or regression predictions in case of continuous tasks.
Generally, encoder, processor, and decoder can be parameterized with a neural network $\theta$ or be non-parametric.
Finally, we analyze computational complexity of existing processors.

\subsection{Encoder}
\label{sec:trans_ind}

\begin{figure}[!ht]
	\centering
	\includegraphics[width=\linewidth]{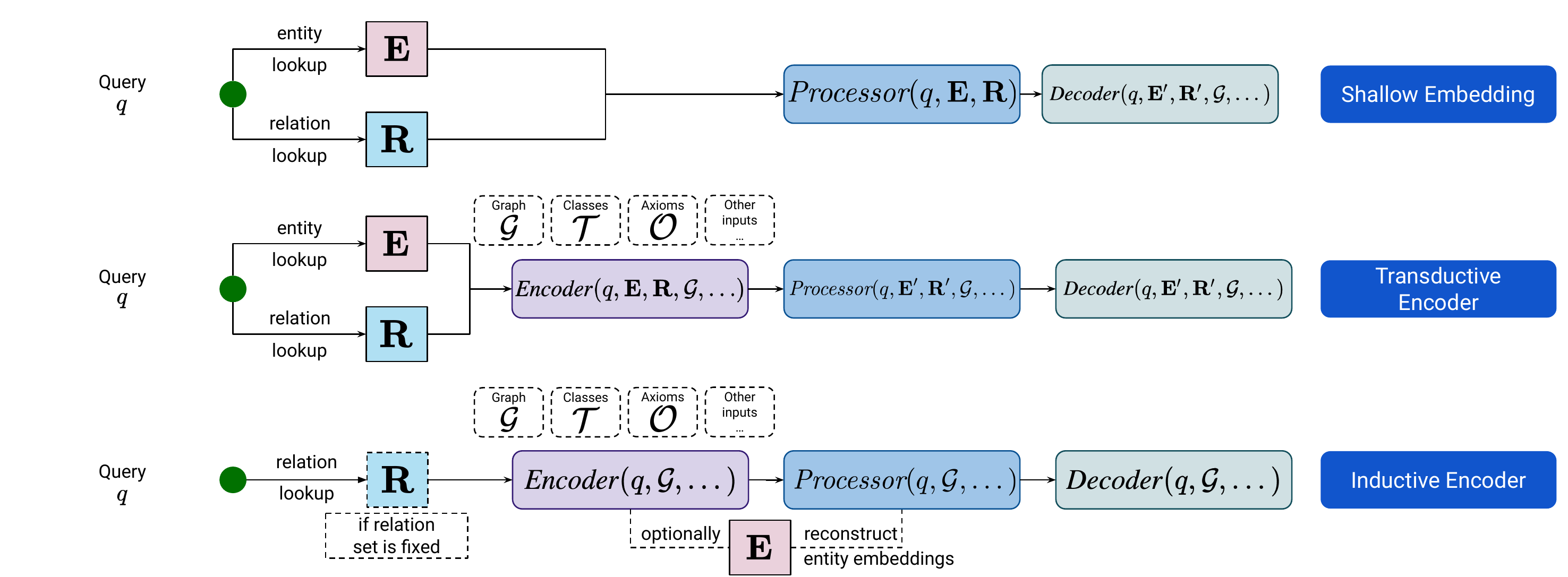}
	\caption{Categorization of \emph{Encoders}. Shallow encoders perform entity and relation embedding lookup and send them to the processor. Transductive encoders additionally enrich the  representations with query, graph, classes, or other latents. Inductive encoders do not need learnable entity embeddings.}
	\label{fig:model_encoders}
\end{figure}

We start the modeling section with encoders, \ie, how different methods encode and represent entities and relations from the KG. There are three different categories, \emph{Shallow Embedding}, \emph{Transductive Encoder}, and \emph{Inductive Encoder} each representing a different way of producing the neural representation of the entities/relations.
Different encoding methods are suitable in different inference setups (details in \Secref{sec:graph_inference}), and may further require different logical operator methods (details in \Secref{sec:models}).
\autoref{fig:model_encoders} illustrates the three common encoding approaches.

\begin{table}[!ht]
\centering
\caption{Complex Query Answering approaches categorized under \emph{Encoder}. 
}
\label{tab:encoder}
\resizebox{\columnwidth}{!}{%
\begin{tabular}{ccc}\toprule
Shallow Embedding & Transductive Encoder & Inductive Encoder \\
\midrule
\multicolumn{1}{p{10cm}}{
GQE \citeps{gqe}, GQE w hash \citeps{gqe_bin}, CGA \citeps{cga}, TractOR \citeps{tractor2020}, Query2Box \citeps{q2b}, BetaE \citeps{betae}, EmQL \citeps{emql}, Shv \citeps{sheaves}, Q2B Onto \citeps{q2b_onto}, RotatE-Box \citeps{regex}, HyPE \citeps{hype}, NewLook \citeps{newlook}, CQD \citeps{cqd}, PERM \citeps{perm}, ConE \citeps{cone}, LogicE \citeps{logic_e}, FuzzQE \citeps{fuzz_qe}, SMORE \citeps{smore}, LinE \citeps{line2022}, Query2Particles \citeps{query2particles2022}, TAR \citeps{abin_abductive2022}, FLEX \citeps{lin2022flex}, TFLEX \citeps{lin2022tflex}, GammaE \citeps{gammae}, NMP-QEM \citeps{nmp_qem}, QTO \citeps{qto}, SignalE \citeps{signale}, Var2Vec \citeps{var2vec}, $\text{CQD}^{\gA}$ \citeps{cqda}, Query2Geom \citeps{query2geom}
} 
&
\multicolumn{1}{p{4cm}}{
MPQE \citeps{mpqe}, BiQE \citeps{biqe}, KGTrans \citeps{kgtrans2022}, StarQE \citeps{starqe}, MLPMix \citeps{mlpmix}, ENeSy \citeps{enesy}, LMPNN \citeps{lmpnn}, NQE \citeps{nqe}, SQE \citeps{sqe}
} 
& 
\multicolumn{1}{p{4cm}}{
NodePiece-QE \citeps{galkin2022}, GNN-QE \citeps{gnn_qe}, GNNQ \citeps{gnnq}, TeMP \citeps{temp2022} 
} \\
\bottomrule
\end{tabular}
}
\end{table}

\paragraph{Shallow Embeddings.} The first line of approaches encodes each entity/relation on the graph as a low-dimensional vector, and thus we achieve an entity embedding matrix $\mE$ and a relation embedding matrix $\mR$. The shape of the entity embedding matrix is $|\gE|\times d$ ($|\gR|\times d$), where $d$ is the dimension of the embedding. Shallow embedding methods assume independence of the representation of all the nodes on the graph. This independence assumption gives the model much freedom, free parameters to learn. %
Such modeling origins from the KG completion literature, where the idea is to learn the entity and relation embedding matrices by optimizing a pre-defined distance/score function over all edges on the graph, \eg, a triplet fact $\texttt{dist}(\Em{e_s}, \Em{r}, \Em{e_o})$. The majority of query answering literature follows the same paradigm with various different embedding spaces and distance functions to learn the entity and relation embedding matrices.
Multiple embedding spaces have been proposed. For example, GQE~\citep{gqe} and Query2Box~\citep{q2b} embed into $\mathbb{R}^d$ (point vector in the Euclidean space); FuzzQE~\citep{fuzz_qe} embeds into the space of real numbers in range $[0,1]$ (fuzzy logic score); BetaE~\citep{betae} uses Beta distribution, a probabilistic embedding space; ConE~\cite{cone} on the other hand embeds entities as a point on a unit circle. Each design choice motivates the inductive bias for executing logical operators (as we show in \Secref{sec:models}).
Some approaches employ shallow entity and relation embeddings already pre-trained on a simple link prediction task and just apply on top of them a query answering decoder with non-parametric logical operators. 
For example, CQD~\citep{cqd}, LMPNN~\citep{lmpnn}, Var2Vec~\citep{var2vec}, and $\text{CQD}^{\gA}$~\citep{cqda} take pre-trained embeddings in the complex space $\mathbb{C}^d$ and apply non-parametric \emph{t-norms} and \emph{t-conorms} to model intersection and union, respectively. 
QTO~\citep{qto} goes even further and fully materializes scores of all possible triples in one $[0,1]^{|\gR| \times |\gE| \times |\gE|}$ matrix given pre-trained entity and relation embeddings at preprocessing stage.

Despite being the mainstream design choice, the downside of shallow methods is that (1) shallow embeddings do not use any inductive bias and prior knowledge of the entity or its neighboring structure since the parameters of all entities/relations are free parameters learned from scratch; (2) they are not applicable in the inductive inference setting since these methods do not have a representation/embedding for those unseen novel entities by design. 
One possible solution is to randomly initialize one embedding vector for a novel entity and finetune the embedding vector by sampling queries involving the novel entity (detailed in \autoref{sec:graph_training}) 
However, such a solution requires gradient steps during inference, rendering it not ideal.

\paragraph{Transductive Encoder.} Similar to shallow embedding methods, transductive encoder methods learn the same entity embedding matrix $\gE$. Besides, they learn an additional encoder $\textsc{Enc}_{\theta}(q, \mE, \mR, \dots)$ (parameterized with $\theta$) on top of the query $q$, entity and relation embedding matrices (and, optionally, other available inputs). The goal is to apply the encoder to the embeddings of entities in the query $q$ in order to capture dependencies between neighboring entities in the graph. 
Specifically, the additional encoder may take several rows of the feature matrix as input and further apply transformations. For example, BiQE~\citep{biqe} and kgTransformer~\citep{kgtrans2022} linearize a query graph $\gG_q$ into a sequence and apply a Transformer~\citep{transformer} encoder that attends to all other embeddings in the query and obtain the final representation of the \texttt{[MASK]} token as the target query. MPQE~\citep{mpqe} and StarQE~\citep{starqe} run a message passing GNN architecture on top of the query graph $\gG_q$ to enrich entity and relation embeddings and extract the final node representation as the query embedding. These methods share similar benefit and disadvantage of the shallow embeddings. Namely, there are many free parameters in the method to train. Unlike the shallow embeddings, the additional encoder leverages relational inductive bias between an entity and its neighboring entities or other entities in a query, allowing for a better learned entity representation and generalization capacity. However, since at its core the method is still based on the large look-up matrix of entity embeddings, it still exhibits the same downside that all such methods cannot be directly applied to an inductive setting where we may observe new entities.

\paragraph{Inductive Encoder.}
In order to address the aforementioned challenges of shallow embeddings and transductive encoders, inductive encoder methods aim to avoid learning an embedding matrix $\mE$ for a fixed number of entities. Instead, inductive representations are often calculated by leveraging certain \emph{invariances}, that is, the features that remain the same when transferred onto different graphs with new entities at inference time. 
As we describe in \autoref{sec:graph_inference}, inductive encoders might employ different invariances albeit the majority of inductive encoders rely on the assumption of the fixed set of relation types $\gR$.
Formally, following Definition~\ref{def:basic_graph_query}, given a complex query $\gQ = (\gE', \gR', \gS', \bar{\gS'})$ composed of entity and relation terms $\gE', \gR'$ (that, in turn, contain constants $\con$ and variables $\variable$), relation projections $R(a,b) \in \gS$ (and, optionally, in $\bar{\gS'}$), a target graph $\gG$ (and, optionally, other inputs), inductive encoders learn a conditional representation function $\textsc{Enc}_\theta(e|\gE', \gR', \gG, \dots)$ for each entity $e \in \gE$.
\citet{galkin2022} devise two families of inductive representations, \ie, (1) inductive \emph{node representations} and (2) inductive \emph{relational structure} representations. 

Inductive \textbf{node representation} approaches parameterize $\textsc{Enc}_\theta$ as a function of a fixed-size invariant vocabulary.
For instance, NodePiece-QE~\citep{galkin2022} employs the invariant vocabulary of relation types and parameterizes each entity through the set of incident relations.
TeMP~\citep{temp2022} employs the invariant vocabulary of entity types and class hierarchy and injects their representations into entity representations.
Inductive node representation approaches reconstruct embeddings of new entities and can be used as a drop-in replacement of shallow lookup tables paired with any \emph{processor} method, \eg, NodePiece-QE used CQD as the processor while TeMP was probed with GQE, Query2Box, BetaE, and LogicE processors.

Inductive \textbf{relational structure} representation methods parameterize $\textsc{Enc}_\theta$ as a function of the relative relational structure that only requires learning of relation embeddings and uses relations as invariants. 
Such methods often employ various \emph{labeling tricks}~\citep{labeling_trick} to label constants (anchor entities) of the input query $\gQ$ such that after the message passing procedure all other nodes would encode a graph structure relative to starting nodes.
In particular, GNN-QE~\citep{gnn_qe} labels anchor nodes with the embedding vector of the queried relations, \eg, for a projection query $(h, r, ?)$ a node $h$ will be initialized with the embedding of relation $r$, whereas all other nodes are initialized with the zero vector. In this way, GNN-QE learns only relation embeddings $\mR$ and GNN weights. 
GNNQ~\citep{gnnq} represents a query with its variables and relations as a hypergraph and learns a relational structure through applying graph convolutions on hyperedges. Hyperedges are parameterized with multi-hot feature vectors of participating relations, so the only learnable parameters are GNN weights.

Still, there exists a set of open problems for inductive models. 
As the majority of inductive methods rely on learning relation embeddings, they cannot be easily used in setups where at inference time KGs are updated with new, unseen relation types, that is, relations are not invariant. 
This fact might require exploration of novel invariances and featurization strategies~\citep{huang2022fewshot,gao2023double,chen2023generalizing}.
Inductive models are more expensive to train in terms of both time and memory than shallow models and cannot yet be easily extended to large-scale graphs.
We conjecture that inductive encoders will be in the focus of the future work in \clqa as generalization to unseen entities and graphs at inference time without re-training is crucial for updatability of \ngdb.
Furthermore, updatability might increase the role of \emph{continual learning}~\citep{cont_learning_thrun, cont_learning_ring} and amplify the negative effects of \emph{catastrophic forgetting}~\citep{mccloskey1989catastrophic} that have to be addressed by the encoders.
Larger inference graphs also present a major \emph{size generalization} issue~\citep{DBLP:conf/icml/YehudaiFMCM21,buffelli2022sizeshiftreg,zhou2022ood} when performance of GNNs trained on small graphs decreases when running inference on much larger graphs. The phenomenon has been  observed by \citet{galkin2022} in the inductive complex query answering setup.

\subsection{Processor}
\label{sec:models}

Having encoded the query and other available inputs, the \emph{Processor} $P$ executes the query in the latent (or symbolic) space against the input graph. 
Recall that a query $q$ is defined as $q(\gE', \gR', \gS, \bar{\gS})$ where $\gE'$ and $\gR'$ terms include constants $\con$ and variables $\variable$, statements in $\gS$ and $\bar{\gS}$ include relation projections $R(a,b)$, and logical operators $\textit{ops}$ over the variables.
We define \emph{Processor} $P$ as a collection of modules that perform relation projections $R(a,b)$ given constants $\con$ and logical operators $\textit{ops} \subseteq \{\wedge, \vee, \neg, \dots \}$ over variables $\variable$ (we elaborate on the logical operators in \autoref{sec:query_ops}). 
Depending on the chosen inductive biases and parameterization strategies behind those modules, we categorize \emph{Processors} into \emph{Neural} and \emph{Neuro-Symbolic} (\autoref{tab:processor1}).
Furthermore, we break down the Neuro-Symbolic processors into \emph{Geometric}, \emph{Probabilistic}, and \emph{Fuzzy Logic} (\autoref{tab:processor2}). 
Note that in this section we omit pure encoder approaches like TeMP~\citep{temp2022} and NodePiece-QE~\citep{galkin2022} that can be paired with any neural or neuro-symbolic processor.
To describe processor models more formally, we denote $\Em{e}$ as an entity vector, $\Em{r}$ as a relation vector, and $\Em{q}$ as the query embedding that is often a function of $\Em{e}$ and $\Em{r}$. We use $\gG_q$ as the query graph.

\begin{table}[ht]
\label{tab:processors_all}
    \caption{Categorization of Query Processors. \autoref{tab:processor1} provides a general view on Neural, Symbolic, Neuro-Symbolic methods as well as encoders that can be paired with any processor. \autoref{tab:processor2} further breaks down neuro-symbolic processors.}
    \begin{subtable}[h]{\columnwidth}
    \centering
    \caption{Complex Query Answering approaches categorized under the  \emph{Processor} type.}
        \resizebox{\linewidth}{!}{
            \begin{tabular}{cccc}
            \toprule
            Any Processor & Neural & Neuro-Symbolic \\
            \midrule
            \multicolumn{1}{p{4cm}}{TeMP \citeps{temp2022}, NodePiece-QE \citeps{galkin2022}} 
            &
            \multicolumn{1}{p{9cm}}{GQE \citeps{gqe}, GQE w/ hashing \citeps{gqe_bin}, CGA \citeps{cga}, BiQE \citeps{biqe}, MPQE \citeps{mpqe}, StarQE \citeps{starqe}, MLPMix \citeps{mlpmix},  Query2Particles \citeps{query2particles2022}, KGTrans \citeps{kgtrans2022}, RotatE-m, DistMult-m, ComplEx-m \citeps{smore}, GNNQ \citeps{gnnq}, SignalE \citeps{signale}, LMPNN \citeps{lmpnn}, SQE \citeps{sqe}} 
            &
            \multicolumn{1}{p{2cm}}{\autoref{tab:processor2}} \\ 
            \bottomrule
            \end{tabular}
        }
        \label{tab:processor1}
    \end{subtable}
    \\
    \begin{subtable}[h]{\columnwidth}
    \caption{Neuro-symbolic Processors}
        \resizebox{\linewidth}{!}{
            \begin{tabular}{ccc}\toprule
            Geometric {\footnotesize [Neuro-Symbolic]} & Probabilistic {\footnotesize[Neuro-Symbolic]} & Fuzzy Logic {\footnotesize[Neuro-Symbolic]} \\
            \midrule
            \multicolumn{1}{p{6.5cm}}{Query2Box \citeps{q2b}, Query2Onto \citeps{q2b_onto}, RotatE-Box \citeps{regex}, NewLook \citeps{newlook}, Knowledge Sheaves \citeps{sheaves}, HypE \citeps{hype}, ConE \citeps{cone}, Query2Geom \citeps{query2geom}} 
            &
            \multicolumn{1}{p{4.5cm}}{BetaE \citeps{betae}, PERM \citeps{perm}, LinE \citeps{line2022}, GammaE \citeps{gammae}, NMP-QEM \citeps{nmp_qem}} 
            &
            \multicolumn{1}{p{7cm}}{EmQL \citeps{emql}, TractOR~\citeps{tractor2020}, CQD \citeps{cqd}, LogicE \citeps{logic_e}, FuzzQE \citeps{fuzz_qe}, TAR \citeps{abin_abductive2022}, FLEX \citeps{lin2022flex}, TFLEX \citeps{lin2022tflex}, GNN-QE \citeps{gnn_qe}, ENeSy \citeps{enesy}, QTO \citeps{qto}, NQE \citeps{nqe}, Var2Vec \citeps{var2vec}, $\text{CQD}^{\gA}$ \citeps{cqda}} \\
            \bottomrule
            \end{tabular}
        }
        \label{tab:processor2}
     \end{subtable}
\end{table}

\begin{figure}[t]
	\centering
	\begin{subfigure}[b]{0.4\textwidth}
	    \centering
	    \includegraphics[width=\linewidth]{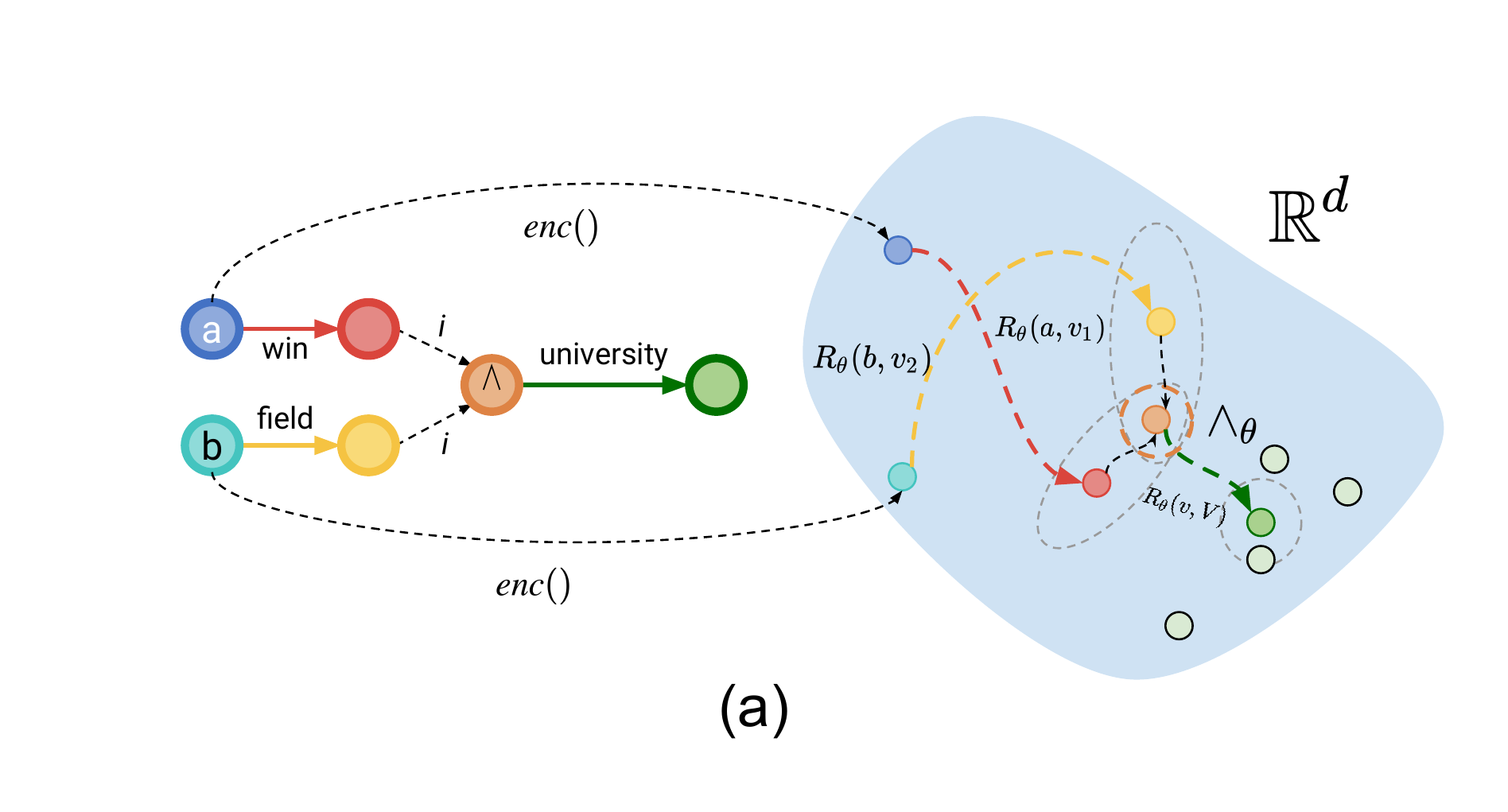}
	\end{subfigure}
	\begin{subfigure}[b]{0.55\textwidth}
	    \centering
	    \includegraphics[width=\linewidth]{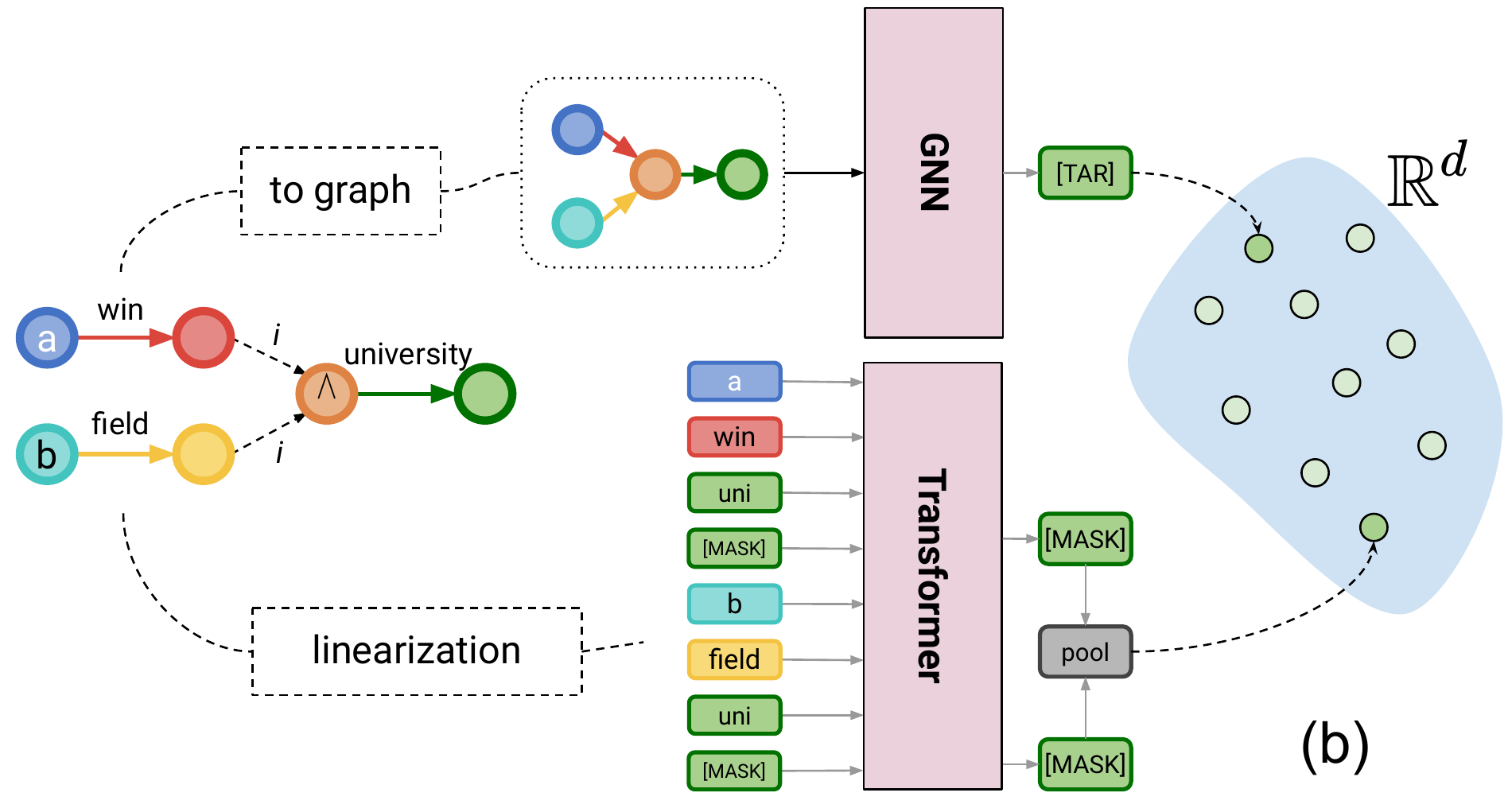}
	\end{subfigure}
	\caption{Neural Processors. (a) Relation projections $R_\theta(a,b)$ and logical operators (non-parametric or parameterized with $\theta$) are executed sequentially in the latent space; (b) a query is encoded to a graph or linearized to a sequence and passed through the encoder (GNN or Transformer, respectively). A pooled representation denotes the query embedding.}
	\label{fig:proc_neural}
\end{figure}

\paragraph{Neural Processors.}
Neural processors execute relation projections and logical operators directly in the latent space $\mathbb{R}^d$ parameterizing them with neural networks. 
To date, most existing purely neural approaches operate exclusively on the query graph $\gG_q$ only executing operators within a single query and do not condition the execution process on the full underlying graph structure $\gG$. 
Since the query processing is performed in the latent space with neural networks where Union ($\vee$) and Negation ($\neg$) are not well-defined, the majority of neural processors implement only relation projection $(R(a,b))$ and intersection $(\wedge)$ operators. 
We aggregate the characteristics of neural processors as to their embedding space, the way of executing relation projection, logical operators, and the final decoding distance function in \autoref{tab:proc_neural}.
We illustrate the difference between two families of neural processors (sequential execution and joint query encoding) in \autoref{fig:proc_neural}.

The original GQE~\citep{gqe} is the first example of the neural processor. 
That is, queries $\Em{q}$, entities $\Em{e}$, and relations $\Em{r}$ are vectors in $\mathbb{R}^d$.
Query embedding starts with embeddings of constants $\gC$ (anchor nodes $\Em{e}$) and they get progressively refined through relation projection and intersection, \ie, it is common to assume that query embedding at the initial step 0 is equivalent to embedding(s) of anchor node(s), $\Em{q}^{(0)}=\Em{e}$.
Relation projection is executed in the latent space with the translation function $\Em{q}+\Em{r}$, and intersection is modeled with the permutation-invariant DeepSet~\citep{deepsets} neural network.
Several follow-up works improved GQE to work with hashed binary vectors $\{+1, -1\}^d$~\citep{gqe_bin} or replaced DeepSet with self-attention and translation-based projection to a matrix-vector product~\citep{cga}.
Recently, \citet{smore} proposed DistMult-m, ComplEx-m, and RotatE-m, extensions of simple link prediction models for complex queries that, inspired by GQE, perform relation projection by the respective composition function and model the intersection operator with DeepSet and, optionally, L2 norm.

The other line of works apply neural encoders to whole query graphs $\gG_q$ without explicit execution of logical operators. Depending on the query graph representation, such encoders are often GNNs, Transformers, or MLPs. 
It is assumed that neural encoders can implicitly capture logical operators in the latent space during optimization.
For instance, MPQE~\citep{mpqe}, StarQE~\citep{starqe}, and LMPNN~\citep{lmpnn} represent queries as relational graphs (optionally, hyper-relational graphs for StarQE) where each edge is a relation projection and intersection is modeled as two incoming projections to the same variable node. 
All constants $\gC$ and known relation types are initialized from the respective embedding matrices. 
All variable nodes in all query graphs are initialized with the same learnable \texttt{[VAR]} feature vector while all target nodes are initialized with the same \texttt{[TAR]} vector.
Then, the query graph is passed through a GNN encoder (R-GCN~\citep{rgcn} for MPQE, StarE~\citep{stare} for StarQE, GIN~\citep{gin} for LMPNN), and the final state of the \texttt{[TAR]} target node is considered the final query embedding ready for decoding.
A recent LMPNN extends query graph encoding with an additional edge feature indicating whether a given projection $R(a,b)$ has a negation or not and derives a closed-form solution for the merged projection and negation operator for the ComplEx composition function.
A different approach is taken by GNNQ~\citep{gnnq} that frames query answering as a subgraph classification task.
That is, an input query is not directly executed over a given graph $\gG$, but, instead, the task is to classify whether a given precomputed subgraph $\gG' \subset \gG$ satisfies a given conjunctive query. 
For that, GNNQ first augments the graph with Datalog-derived triples and converts the subgraph to a hypergraph where only hyperedges are parameterized with learnable vectors. 
On the one hand, this strategy allows GNNQ to be inductive and not learn entity embeddings. On the other hand, GNNQ is limited to conjunctive queries only and extensions to union and negation queries are not defined.

A more exotic approach by \citet{sheaves} is based on the sheaf theory and algebraic topology~\citep{hansen2019sheaf}. There, a graph is represented as a cellular sheaf and conjunctive queries are modeled as chains of relations (0-cochains). A sheaf is induced over the query graph and relevant answers should be consistent with the induced sheaf and entity embeddings. The optimization problem is a harmonic extension of a 0-cochain using \emph{sheaf Laplacian} and \emph{Schur complement} of the sheaf Laplacian. Conceptually, this approach merges execution of projection and intersection operators as functions over topological structures. 

Considering Transformer encoders, BiQE~\citep{biqe}, kgTransformer~\citep{kgtrans2022}, and SQE~\citep{sqe} linearize a conjunctive query graph into a sequence of relational paths composed of entity constants $\gC$ and relation tokens. 
The order of tokens in paths and intersections of paths are marked with positional encodings.
The target node (present in many paths) is marked with the \texttt{[MASK]} token (optionally, kgTransformer also annotates existentially quantified variables with \texttt{[MASK]}). 
SQE does not model variables explicitly but instead relies on auxiliary \emph{bracket} tokens that separate branches of the computation graph.
Passing the sequence through the Transformer encoder, the final query embedding is the aggregated representation of the target node.
BiQE only supports conjunctive queries while kgTransformer converts queries with unions to the Disjunctive Normal Form (DNF) with post-processing of score distributions (we elaborate on query rewritings and normal forms in \autoref{sec:query_ops}).
SQE explicitly includes all operator tokens into the linearized sequence and thus supports negations.

Finally, MLPMix~\citep{mlpmix} sequentially executes operations of the query where projection, intersection, and negation operators are modeled as separate learnable MLPs. Union queries are converted to DNF such that they can be answered with projection and intersection operators with the final post-processing of scores as a union operator.
Similarly, Query2Particles~\citep{query2particles2022} represents each query as a set of vectors in the embedding space and models projection, intersection, and negation operators as attention over the set of particles followed by an MLP. Union is a concatenation of query particles.

\begin{table}[!ht]
\centering
\caption{Neural Processors. Most methods implement only Relation Projection and Intersection operators. The top part of models execute a query sequentially, the bottom part encode the whole query graph $\gG_q$. 
\label{tab:proc_neural} 
}
\resizebox{0.99\columnwidth}{!}{%
\begin{tabular}{ccccccc}
	\toprule
Model & Embedding Space & Relation Projection & Intersection & Union & Negation & Distance \\
\hline
GQE & $\Em{q}, \Em{e}, \Em{r}\in\RR^d$ & $\Em{q}+\Em{r}$ & $\texttt{DeepSet}(\{\Em{q_i}\})$ & - & - & $\|\Em{q}-\Em{e}\|$ \\\hline
GQE+hashing & $\Em{q}, \Em{e} \in \{\pm1\}^d, \Em{r} \in \RR^d$ & $\texttt{sgn}(\Em{q}+\Em{r})$ & $\texttt{sgn}(\texttt{DeepSet}(\{\Em{q_i}\}))$ & - & - & $-\textit{cos}(\Em{q},\Em{e})$ \\\hline
CGA & $\Em{q}, \Em{e}, \Em{r}\in\RR^d$ & $\Em{W}_r\Em{q}$ & $\texttt{SelfAttn}(\{\Em{q_i}\})$ & - & - & $-\textit{cos}(\Em{q},\Em{e})$ \\\hline
RotatE-m & $\Em{q}, \Em{e}, \Em{r}\in\CC^d$ & $\Em{q}\circ\Em{r}$ & $\texttt{DeepSet}(\{\Em{q_i}\})$ & - & - & $\|\Em{q}-\Em{r}\|$ \\\hline
DisMult-m & $\Em{q}, \Em{e}, \Em{r}\in\RR^d$ & $\texttt{L2Norm}(\Em{q}\circ\Em{r})$ & $\texttt{L2Norm}(\texttt{DeepSet}(\{\Em{q_i}\}))$ & - & - & $-\langle\Em{q},\Em{e}\rangle$ \\\hline
ComplEx-m & $\Em{q}, \Em{e}, \Em{r}\in\CC^d$ & $\texttt{L2Norm}(\Em{q}\circ\Em{r})$ & $\texttt{L2Norm}(\texttt{DeepSet}(\{\Em{q_i}\}))$ & - & - & $-\text{Re}(\langle\Em{q},\overline{\Em{e}}\rangle)$ \\ \hline
\multirow{2}{*}{Query2Particles} & \multirow{2}{*}{$\Em{q,e}\in\RR^{d\times K},\Em{r}\in\RR^{d}$} & $f(\Em{q}, \Em{r})$ & \multirow{2}{*}{$\texttt{MLP}(\texttt{Attn}([\Em{q_1}, \Em{q_2}]))$} & \multirow{2}{*}{$[\Em{q}^1, \dots, \Em{q}^N]$} & \multirow{2}{*}{$\texttt{MLP}(\texttt{Attn}(\Em{q}))$} & \multirow{2}{*}{$-\max_k \langle \Em{q}^k, \Em{e} \rangle $} \\
&&$f$ is neural gates &&& &\\\hline
\multirow{2}{*}{SignalE} & $\Em{q}, \Em{e}, \Em{r}=[\Em{z}, \Em{v}]$ &\multirow{2}{*}{$\Em{z}_q\circ\Em{z}_r$}&\multirow{2}{*}{$\texttt{SelfAttn}(\{{\Em{z}_q}_i\})$}& \multirow{2}{*}{DNF}&$(1-\Em{z}_q^{\text{amp}},$&$\beta\|\Em{v}_q-\Em{v}_e\|_2 + $ \\
&$\Em{z}\in\CC^d,\Em{v}=\texttt{IDFT}(\Em{z})\in\RR^d$&&&&$\Em{z}_q^{\text{phase}})$& $\|\Em{z}_q - \Em{z}_e\|_2$\\\hline
MLPMix & $\Em{q}, \Em{e}, \Em{r}\in\RR^d$ & $\texttt{MLP}_\texttt{mix}(\Em{q}, \Em{r})$ & $\texttt{MLP}_\texttt{mix}(\Em{q_1}, \Em{q_2})$ & DNF & $\texttt{MLP}(\Em{q})$ & $\|\Em{q} - \Em{e}\|$ \\ \hline \hline
MPQE & $\Em{q}, \Em{e}, \Em{r}\in\RR^d$ & $\Em{W}_r\Em{q}$ & $\texttt{RGCN}(\Em{q}, \gG_q)$ & - & - & $\textit{cos}(\Em{q}, \Em{e})$ \\\hline
StarQE & $\Em{q}, \Em{e}, \Em{r}\in\RR^d$ & $f(\Em{q}, g(\Em{r}, \Em{h}_{\text{quals}}))$ & $\texttt{StarE}(\Em{q}, \gG_q)$ & - & - & $\texttt{dot}(\Em{q},\Em{e})$ \\\hline
GNNQ & $\Em{q}, \Em{r}\in\RR^d$ & \multicolumn{2}{c}{Hypergraph $\texttt{ RGCN}(\gG')$}  & - & - & BCE \\ \hline
LMPNN & $\Em{q}, \Em{e}, \Em{r}\in\CC^d$ & $\rho(\Em{q}, \Em{r}, \text{dir}, \text{neg})$ & $\texttt{GIN}(\Em{q}, \gG_q)$ & DNF & $\rho(\Em{q}, \Em{r}, \text{dir}, \text{neg})$ & $\textit{cos}(\Em{q}, \Em{e})$ \\ \hline 
BiQE & $\Em{q}, \Em{e}, \Em{r}\in\RR^d$ & \multicolumn{2}{c}{Transformer$(\Em{q}, \text{linearized}\: \gG_q)$} & - & - & CE \\\hline 
kgTransformer & $\Em{q}, \Em{e}, \Em{r}\in\RR^d$ & \multicolumn{2}{c}{Transformer$(\Em{q}, \text{linearized}\: \gG_q)$} & DNF & - & $\texttt{dot}(\Em{q},\Em{e})$ \\\hline
SQE & $\Em{q}, \Em{e}, \Em{r}\in\RR^d$ & \multicolumn{4}{c}{LSTM / Transformer$(\Em{q}, \text{linearized}\: \gG_q \: \text{with operator tokens})$} & $\texttt{dot}(\Em{q},\Em{e})$ \\\hline
Sheaves & $\Em{q}, \Em{e} \in \RR^d, \mathcal{F}_r \in \RR^{d\times d}$ & \multicolumn{2}{c}{$f(\Em{q}, \gG_q \text{ as cochain})$} & - & - & $\|\mathcal{F}_r\Em{q}-\mathcal{F}_r\Em{e}\|$ \\
\bottomrule
\end{tabular}
}%
\end{table}

\paragraph{Neuro-Symbolic Processors.}
In contrast to purely neural and symbolic models, we define \emph{neuro-symbolic} processors as those who (1) explicitly design logic modules (or neural logical operators) that simulate the real logic/set operations, or rely on various kinds of fuzzy logic formulations to provide a probabilistic view of the query execution process, and (2) execute relation traversal in the latent space. The key difference between neuro-symbolic processors and the previous two is that neuro-symbolic processors explicitly model the logical operations with strong inductive bias so that the processing / execution is better aligned with the symbolic operation (\eg, by imposing restrictions on the embedding space) and more interpretable. We further segment these methods into the following categories.

\paragraph{Geometric Processors.}

\begin{figure}[!t]
	\centering
	\includegraphics[width=\linewidth]{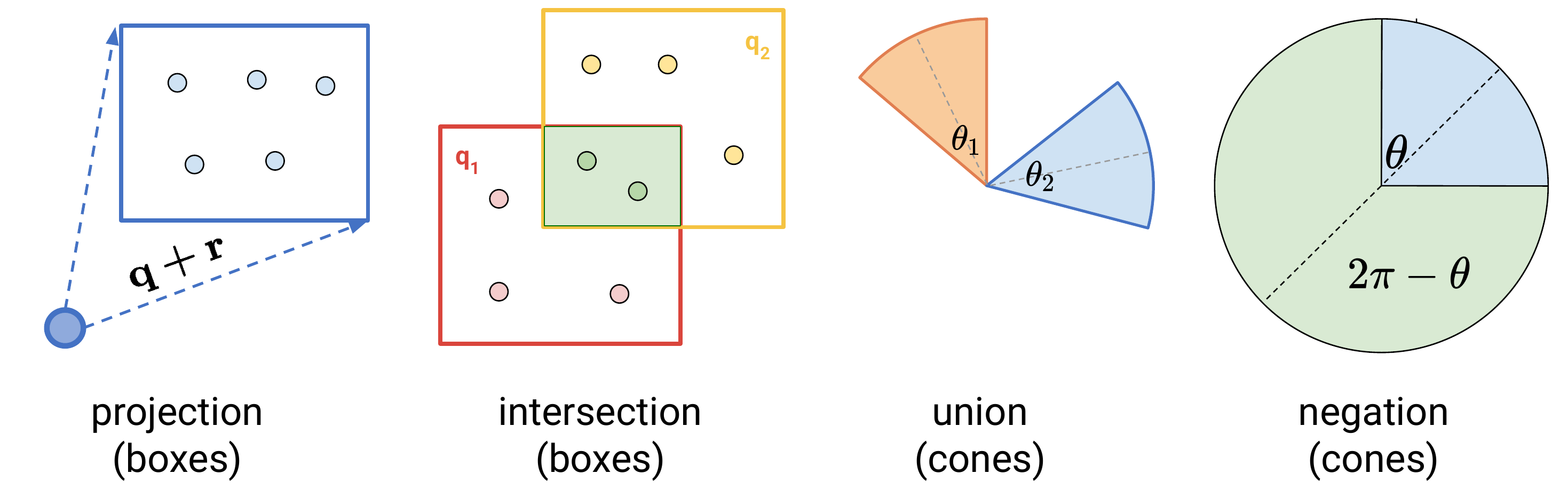}
	\caption{Geometric Processors and their inductive biases.}
	\label{fig:proc_geometric}
\end{figure}

Geometric processors design an entity/query embedding space with different geometric intuitions and further customize neuro-symbolic operators that directly simulate their logical counterparts with similar properties (as illustrated in \autoref{fig:proc_geometric}). 
We aggregate the characteristics of geometric models as to their embedding space and inductive biases for logical operators in \autoref{tab:proc_geometric}.

Query2Box~\citep{q2b} embeds queries $\Em{q}$ as hyper-rectangles (high-dimensional boxes) in the Euclidean space. To achieve that, entities $\Em{e}$ and relations $\Em{r}$ are embedded as points in the Euclidean space where each relation has an additional learnable offset vector $\Em{r_o}$ (entities' offsets are zeros). The projection operator is modeled as an element-wise summation $\Em{q} + \Em{r}$ of centers and offsets of the query and relation, that is, the initial box is obtained by projecting the original anchor node embedding $\Em{e}$ (with zero offset) with the relation embedding $\Em{r}$ and relation offset $\Em{r_o}$. Accordingly, an attention-based neuro-intersection operator is designed to simulate the set intersection of the query boxes in the Euclidean space. The operator is closed, permutation invariant and aligns well with the intuition that the size of the intersected set is smaller than that of all input sets. 
The union operator is achieved via DNF, that is, union is the final step of concatenating results of operand boxes. 
Several works extend Query2Box, \ie, Query2Onto~\citep{q2b_onto} attempts to model complex ontological axioms by materializing entailed triples and enforcing hierarchical relationship using inclusion of the box embeddings; RotatE-Box~\citep{regex} designs an additional rotation-based Kleene plus ($+$) operator denoting relational paths (we elaborate on the Kleene plus operator in \autoref{sec:query_ops}); NewLook~\citep{newlook} adds symbolic lookup from the adjacency tensor\footnote{Incorrect implementation led to the major test set leakage and incorrect reported results.} to the operators, modifies projection with MLPs and models the \emph{difference} operator as attention over centers and offsets (note that the difference operator is a particular case of the \emph{2in} negation query, we elaborate on that in \autoref{sec:query_ops}); Query2Geom~\citep{query2geom} replaces an attention-based intersection operator with a simple non-parametric closed-form geometric intersection of boxes. 

\begin{table}[t]
\centering
\caption{Geometric Neuro-Symbolic Processors. $\odot$ denotes element-wise multiplication, $\oplus_c$ denotes M\"{o}bius addition, $\odot_c$ denotes M\"{o}bius scalar product. \texttt{DS} denotes DeepSets neural network. NewLook only partially implements negation as the difference operator.}
\label{tab:proc_geometric}
\resizebox{0.99\columnwidth}{!}{%
\begin{tabular}{ccccccc}
	\toprule
Model & Embedding Space & Relation Projection & Intersection & Union & Negation & Distance \\
\hline
Query2Box & \multirow{2}{*}{$\Em{q, r}\in\RR^{2d}$, $\Em{e}\in\RR^d$} & \multirow{2}{*}{$\Em{q}+\Em{r}$} & $\Em{q}_c=\texttt{Attn}(\{\Em{q_c^i}\})$ & \multirow{2}{*}{DNF} & \multirow{2}{*}{-} & \multirow{2}{*}{$\text{d}_\text{out}+\alpha\text{d}_\text{in}$}\\
Query2Onto &&&$\Em{q_o}=\text{min}(\{\Em{q_o^i}\})\odot\sigma(\texttt{DS}(\{\Em{q_o^i}\}))$&&\\\hline
\multirow{2}{*}{Query2Geom} & \multirow{2}{*}{$\Em{q, r}\in\RR^{2d}$, $\Em{e}\in\RR^d$} & \multirow{2}{*}{$\Em{q}+\Em{r}$} & $\Em{q}_c=\frac{1}{2}((\Em{q_c^i}+\Em{q_o^i}) + (\Em{q_c^j}-\Em{q_o^j}))$ & \multirow{2}{*}{DNF} & \multirow{2}{*}{-} & \multirow{2}{*}{$\text{d}_\text{out}+\alpha\text{d}_\text{in}$} \\ 
& & & $\Em{q}_o = \Em{q_c} - (\Em{q_c^i}-\Em{q_o^i})$ & & & \\\hline
\multirow{2}{*}{RotatE-Box} & \multirow{2}{*}{$\Em{q, r}\in\CC^{2d},\Em{e}\in\CC^d$} & \multirow{2}{*}{$(\Em{q_c}\circ \Em{r_c},\Em{q_o}+\Em{r_o})$} & \multirow{2}{*}{-} & DNF /  & \multirow{2}{*}{-} & \multirow{2}{*}{$\text{d}_\text{out}+\alpha\text{d}_\text{in}$} \\
&&&&$\texttt{DS}(\{\Em{q}^i\})$&\\\hline
\multirow{2}{*}{NewLook} & $\Em{q, r}\in\RR^{2d},\Em{e}\in\RR^d$ & $\texttt{MLP} [ \texttt{MLP}(\Em{q_c} + \Em{r_c})\: \| $ & $\Em{q_c}=\texttt{Attn}(\{\Em{q_c^i}, \Em{x^i} \})$ & \multirow{2}{*}{DNF}   &  $\texttt{Attn}(\{\Em{q_c^i}\})$ & \multirow{2}{*}{$\text{d}_\text{out}+\alpha\text{d}_\text{in}$} \\
& $\Em{x} \in \gT^{|\gR| \times |\gE| \times |\gE|} $ & $\quad \texttt{MLP}(\Em{r_o})\: \| \: \Em{x}_t ] $  & $\Em{q_o}=\text{min}(\{\Em{q_o^i}\})\odot\sigma(\texttt{DS}(\{\Em{q_o^i}\}))$  &  & $\texttt{Attn}(\{\Em{q_o^i}, \Em{x^i}\})$ & \\\hline
\multirow{2}{*}{HypE} & \multirow{2}{*}{$\Em{q, r}\in\RR^{2d}$, $\Em{e}\in\RR^d$} & \multirow{2}{*}{$\Em{q} \oplus_c \Em{r}$} & $\Em{q_c}=\texttt{Attn}(\{\Em{q_c^i}\})$ & \multirow{2}{*}{DNF} & \multirow{2}{*}{-} & \multirow{2}{*}{$\text{d}_\text{out}+\alpha\text{d}_\text{in}$}\\
 &&&$\Em{q_o}=\text{min}(\{\Em{q_o^i}\})\odot_c\sigma(\texttt{DS}(\{\Em{q_o^i}\}))$&&\\\hline
\multirow{4}{*}{ConE} & $\Em{q,r}=(\vtheta_\text{ax}, \vtheta_\text{ap})$ & \multirow{2}{*}{$g(\texttt{MLP}(\Em{q}+\Em{r}))$} & \multirow{2}{*}{$\vtheta_\text{ax}=\texttt{SemanticAvg}(\Em{q}_1, \dots, \Em{q}_n)$} & \multirow{4}{*}{DNF/DM} & \multirow{2}{*}{$\vtheta_\text{ax}=\vtheta_\text{ax}\pm\pi$} & \multirow{4}{*}{$\text{d}_\text{out}+\alpha \text{d}_\text{in}$} \\&  $\Em{e}=(\vtheta_\text{ax},\vzero)$ & & & & & \\ & $\vtheta_\text{ax}\in[-\pi,\pi)^d$  & \multirow{2}{*}{$g$ gates $\vtheta_\text{ax}$ and $\vtheta_\text{ap}$} &\multirow{2}{*}{$\vtheta_\text{ap}=\texttt{CardMin}(\Em{q}_1, \dots, \Em{q}_n)$}& &   \multirow{2}{*}{$\vtheta_\text{ap}=2\pi - \vtheta_\text{ap}$} & \\
& $\vtheta_\text{ap}\in[0,2\pi]^d$ & & & & & \\\bottomrule
\end{tabular}
}%
\end{table}

HypE~\citep{hype} extends the idea of Query2Box and embeds a query as a hyperboloid (two parallel pairs of arc-aligned horocycles) in a Poincar\'e hyperball to better capture the hierarchical information. A similar attention-based neuro-intersection operator is designed for the hyperboloid embeddings with the goal to shrink the limits with DeepSets. 

ConE~\citep{cone}, on the other hand, embeds queries on the surface of a set of unit circles. Each query is represented as a cone section and the benefit is that in most cases the intersection of cones is still a cone, and the negation/complement of a cone is also a cone thanks to the angular space bounded by $2\pi$. Based on this intuition, they design geometric neuro-intersection and negation operators.

To sum up, the geometric-based processors are often designed with a strong geometric prior such that 
properties of the logical/set operations can be better simulated or satisfied.

\paragraph{Probabilistic Processors.}

\begin{figure}[!t]
	\centering
	\includegraphics[width=\linewidth]{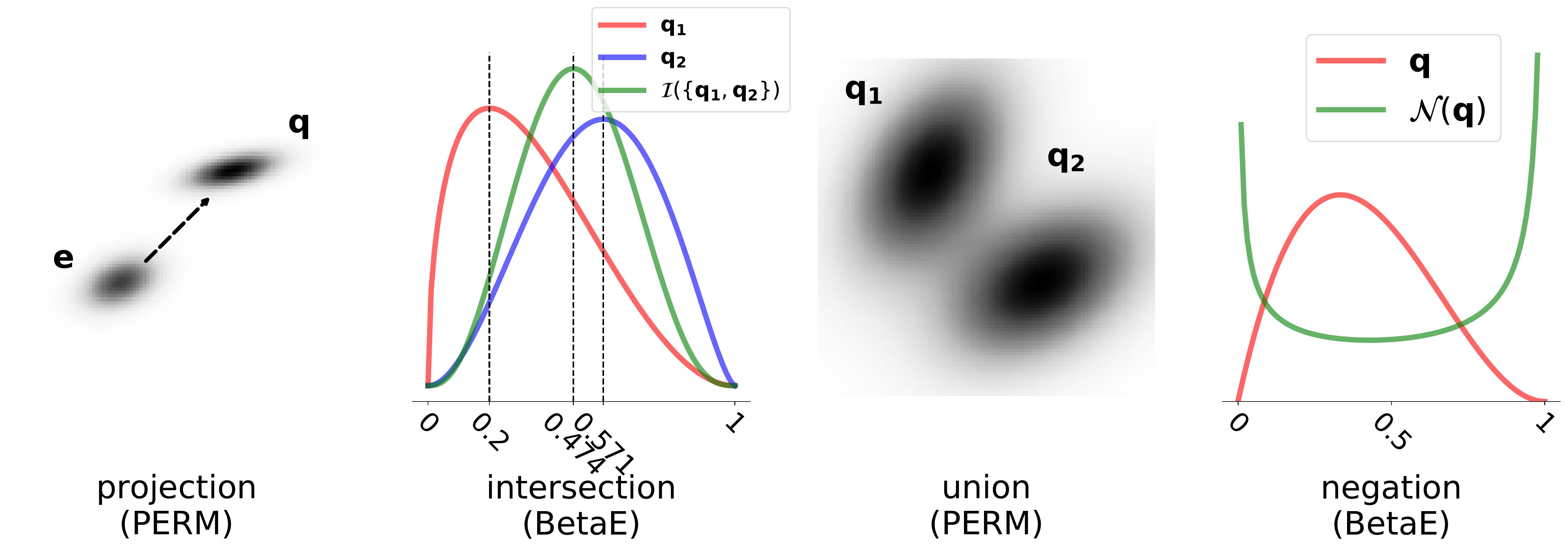}
	\caption{Probabilistic Processors and their inductive biases.}
	\label{fig:proc_probabilistic}
\end{figure}

Instead of a geometric embedding space, probabilistic processors aim to model the query and the logic/set operations in a probabilistic space. Such methods are similar to the geometric based methods in that they also require a probabilistic space on which one can design a neuro-symbolic logic/set operator that aligns well with the real one. Some examples of implementing logical operators in a probabilistic space are illustrated in \autoref{fig:proc_probabilistic}. The aggregated characteristics are presented in \autoref{tab:proc_prob}.

BetaE~\citep{betae} builds upon the Beta distribution and embeds entities/queries as high-dimensional Beta distribution with learnable parameters. The benefit is that one can design a parameterized neuro-intersection operator over two Beta embeddings where the output is still a Beta embedding with more concentrated density function. A neuro-negation operator can be designed by simply taking the reciprocal of the parameters in order to flip the density. 
PERM~\citep{perm} looks at the Gaussian distribution space and embeds queries as a multivariate Gaussian distribution. Since the product of Gaussian probability density functions (PDFs) is still a Gaussian PDF, the neuro-intersection operator accordingly calculates the parameters of the Gaussian embedding of the intersected set. 
NMP-QEM~\citep{nmp_qem} develops this idea further and represents a query as a mixture of Gaussians where logical operators are modeled with MLP or attention over distribution parameters. 
LinE~\citep{line2022} transforms the Beta distribution into a discrete sequence of values. A similar neuro-negation operator is introduced by taking the reciprocal as BetaE while designing a new neuro-intersection/union operator by taking element-wise min/max. %
GammaE~\citep{gammae} replaces Beta distribution with Gamma distribution as entity and query embedding space. 
Parameterizing logical operators with operations over mixtures of Gamma distributions, union and negation become closed, do not need DNF or DM transformations, and can be executed sequentially along the query computation graph.
Overall, probabilistic processors are similar to geometric processors since they are all inspired by certain properties of the probability and geometry used for embeddings and customize neuro-logic operators.

\begin{table}[!ht]
\centering
\caption{Probabilistic Neuro-Symbolic Processors.\label{tab:proc_prob}}
\resizebox{0.99\columnwidth}{!}{%
\begin{tabular}{ccccccc}
	\toprule
Model & Embedding Space & Relation Projection & Intersection & Union & Negation & Distance \\
\hline
BetaE & $\Em{q},\Em{e}\in\RR^{2d}$ & $\texttt{MLP}_r(\Em{q})$ & $\Em{q}=[(\sum w_i\alpha_i,\sum w_i \beta_i)]$ & DNF/DM & $\frac{1}{\Em{q}}$ & $\texttt{KL}(\Em{e};\Em{q})$\\\hline
\multirow{2}{*}{PERM} & \multirow{2}{*}{$\Em{q,e,r}=\gN(\vmu,\Sigma)$} & $\gN(\vmu_q+\vmu_r,$ & $\Sigma_q^{-1}=\Sigma_{q_1}^{-1}+\Sigma_{q_2}^{-1}$ & 
\multirow{2}{*}{$\texttt{Attn}(\{\Em{q}_i\})$} 
& \multirow{2}{*}{-} &  $(\vmu_q-\vmu_e)^T\Sigma_q^{-1}$ \\
 & & $ (\Sigma_q^{-1}+\Sigma_r^{-1})^{-1})$ & $\vmu_q=\Sigma_q(\Sigma_{q_2}^{-1}\vmu_1 + \Sigma_{q_1}^{-1}\vmu_2)$ & & &$(\vmu_q-\vmu_e)$ \\\hline
LinE & $\Em{q},\Em{e} \in\RR^{k\times h}$ & $\texttt{MLP}_r(\Em{q})$ & $\text{min}\{q_1, q_2\}$ & $\text{max}\{q_1, q_2\}$ & $[\frac{1}{p_1^q},\dots,\frac{1}{p_k^q}]^h$ & $\|\Em{q} - \Em{e}\|^2_2$ \\ \hline
GammaE & $\Em{q},\Em{e}\in\RR^{2d}$ & $\texttt{MLP}_r(\Em{q})$ & $\Em{q}=[(\sum w_i\alpha_i,\sum w_i \beta_i)]$ & $\texttt{Attn}(\{\Em{q}_i\})$ & $[\frac{1}{\Em{\alpha}}, \Em{\beta}]$ & $\texttt{KL}(\Em{e};\Em{q})$ \\ \hline
\multirow{2}{*}{NMP-QEM} & $\Em{e}\in\RR^{d}$ & \multirow{2}{*}{$\texttt{MLP}_r(\Em{q})$} & \multirow{2}{*}{$\texttt{Attn}(\{\Em{q}_i\})$} & \multirow{2}{*}{DNF} & \multirow{2}{*}{$\texttt{MLP}(\Em{q})$} & \multirow{2}{*}{$\| \Em{e} - \sum_{i=1}^K \omega_{i}^q \vmu_{i}^q \|$} \\
& $\Em{q} = \sum_{i=1}^K \omega_i \gN(\vmu_i,\Sigma_i)$ & & & & & \\
 \bottomrule
\end{tabular}
}%
\end{table}

\paragraph{Fuzzy-Logic Processors.}

\begin{figure}[t]
	\centering
	\includegraphics[width=\linewidth]{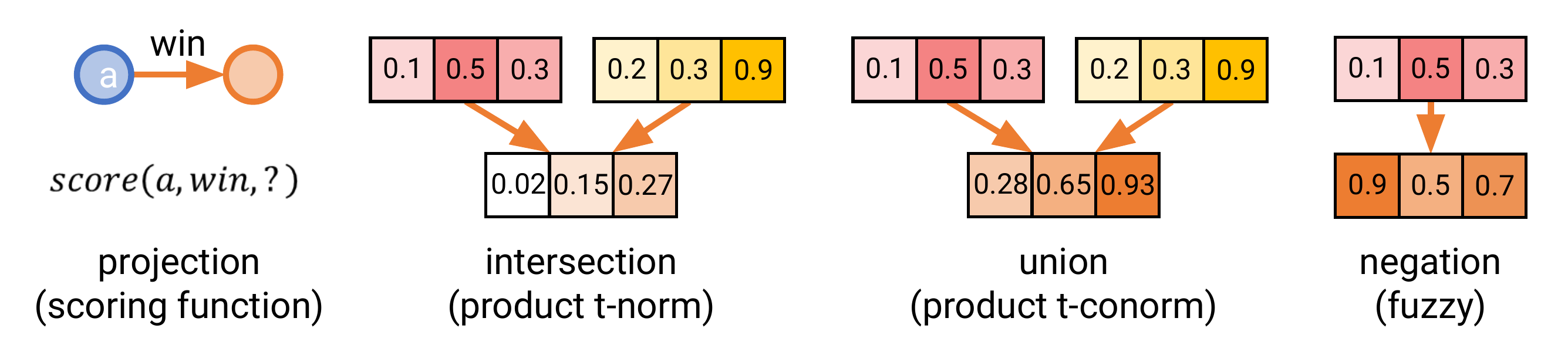}
	\caption{Fuzzy-Logic Processors and fuzzy logical operators.}
	\label{fig:proc_fuzzy}
\end{figure}

Unlike the methods above, fuzzy-logic processors directly model all logical operations using existing fuzzy logic theory~\citep{tnorm,vankrieken_fuzzy} where intersection can be expressed via \emph{t-norms}
and union via corresponding \emph{t-conorms} (\autoref{sec:tnorms}). 
In such a way, fuzzy-logic processors avoid the need to manually design or learn neural logical operators as in the previous two processors but rather directly use established fuzzy operators commonly expressed as differentiable, element-wise algebraic operators over vectors (\autoref{fig:proc_fuzzy}). While intersection, union, and negation are non-parametric, the projection operator might still be parameterized with a neural network. The aggregated characteristics of fuzzy-logic processors are presented in \autoref{tab:proc_fuzzy}. Generally, fuzzy processors aim to combine execution in \emph{embedding space} (vectors) with \emph{entity space} (symbols). The described methods are different in designing such a combination.

One of the first fuzzy processors is EmQL~\citep{emql} that imbues entity and relation embeddings with a count-min sketch~\citep{count_min_sketch}. There, projection, intersection, and union are performed both in the embedding space and in the symbolic sketch space, \eg, intersection is modeled as an element-wise multiplication and union is an element-wise summation of two sketches.
CQD~\citep{cqd} scores each atomic formula in a query with a pretrained neural link predictor and uses t-norms and t-conorms to compute the final score of a query directly in the \emph{embedding space}. CQD does not train any neural logical operators and only requires pretraining the entity and relation embeddings with one-hop links such that the projection operator is equivalent to \emph{top-k} results of the chosen scoring function, \eg, ComplEx~\citep{lacroix2018complex}. 
The idea was then extended in several directions: Query Tree Optimization (QTO)~\citep{qto} added a look-up from the materialized tensor of scores of all possible triples $\Em{M}\in [0,1]^{|\gR|\times |\gE| \times |\gE|}$ to the relation projection step; $\text{CQD}^{\gA}$~\citep{cqda} and Var2Vec~\citep{var2vec} learn an additional linear transformation of the entity-relation concatenation $\Em{W}[\Em{q},\Em{r}]$.

LogicE~\citep{logic_e} designs logic embeddings for each entity with a list of \emph{lower bound -- upper bound} pairs in range $[0,1]$, which can be interpreted as a uniform distribution between the lower and upper bound. LogicE executes negation and conjunction with continuous t-norms over the lower and upper bounds. FuzzQE~\citep{fuzz_qe} and TAR~\citep{abin_abductive2022} embed query to a high-dimensional fuzzy space $[0,1]^d$ and similarly use G\"odel t-norm and \L{}ukasiewicz t-norm to model disjunction, conjunction and negation.
FuzzQE models relation projection as a relation-specific MLP whereas TAR uses a geometric translation (element-wise sum). FLEX~\citep{lin2022flex} and TFLEX~\citep{lin2022tflex} embed a query as a mixture of feature and logic embedding. For the logic part, both methods use the real logical operations in vector logic~\citep{mizraji2008vector}. TFLEX adds a temporal module conditioning logical operators on the time embedding.

GNN-QE~\citep{gnn_qe} models the likelihood of all entities for each relation projection step with a graph neural network NBFNet~\citep{zhu2021neural}. It further adopts product logic to directly model the set operations (intersection, union, and negation) over the fuzzy set obtained after a relation projection. GNN-QE employs a node labeling technique where a starting node is initialized with the relation vector (while other nodes are initialized with zeros). This allows GNN-QE to be inductive and not rely on trainable entity embeddings.  
ENeSy~\citep{enesy}, on the other hand, maintains both vector and symbolic representations for queries, entities, and relations (where symbolic relations are encoded into $\Em{M}_r$ sparse adjacency matrices). Logical operators are executed first in the neural space, \eg, relation projection is RotatE composition function~\citep{sun2018rotate},  and then get intertwined with symbolic representations. Logical operators in the symbolic space employ a generalized version of the product logic and corresponding t-(co)norms. 

In summary, fuzzy-logic processors directly rely on established fuzzy logic formalisms to perform all the logical operations in the query and avoid manually designing and learning neural operators in (possibly) unbounded embedding space.
The fuzzy logic space is continuous but bounded within $[0, 1]$ -- this is both the advantage and weakness of such processors. The bounded space is beneficial for closed logical operators as their output values still belong to the same bounded space.
On the other hand, most of the known t-norms (and corresponding t-conorms) still lead to vanishing gradients and only the Product logic norms are stable~\citep{vankrieken_fuzzy,badreddine2022logic}.
Another caveat is designing an effective and differentiable interaction mechanism between the fuzzy space $[0,1]^d$ and unbounded embedding space $\RR^d$ (or $\CC^d$) where relation representations are often initialized from. 
That is, re-scaling and squashing of vector values when processing a computation graph might lead to noisy gradients and unstable training which is observed, for instance, by GNN-QE that has to turn off gradients from all but last projection step.

\begin{table}[t]
\centering
\caption{Fuzzy Neuro-Symbolic Processors. 
\label{tab:proc_fuzzy}}
\resizebox{0.99\columnwidth}{!}{%
\begin{tabular}{ccccccc}
	\toprule
Model & Embedding Space & Relation Projection & Intersection & Union & Negation & Distance \\
\hline
\multirow{2}{*}{EmQL} & $\Em{e,r}\in\RR^{d}, \Em{q}\in\RR^{3d}$ & \multirow{2}{*}{$\texttt{MIPS}(\Em{q}, [\Em{r}, \Em{e}_h, \Em{e}_t])$} & $(\Em{q_1} + \Em{q_2}) / 2$ & $(\Em{q_1} + \Em{q_2}) / 2$ & \multirow{2}{*}{-} & \multirow{2}{*}{$\texttt{dot}(\Em{q},\Em{e})$} \\ 
& $\Em{b}$ count-min sketch & & $\Em{b_1} \odot \Em{b_2}$ & $\Em{b_1} + \Em{b_2}$ & & \\\hline
\multirow{2}{*}{CQD} & \multirow{2}{*}{$\Em{q,e, r}\in\CC^{d}$} & $\min_{\Em{q}_2} d(\Em{q}_1, \Em{r}, \Em{q}_2)$ or & Product: $\Em{q_1}\cdot\Em{q_2}$ & $\Em{q_1}+\Em{q_2}-\Em{q_1}\cdot\Em{q_2}$ & \multirow{2}{*}{-} & \multirow{2}{*}{$\texttt{dist}(\Em{q}, \Em{e})$} \\
 & & $\text{top}_k (d(\Em{q}_1, \Em{r}, \Em{e_k}))$ & G\"odel: $\text{min}(\Em{q_1}, \Em{q_2})$ &$\text{max}(\Em{q_1}, \Em{q_2})$&  & \\ \hline
\multirow{2}{*}{$\text{CQD}^{\gA}$} & $\Em{q,e, r}\in\CC^{d}$ & $\theta=\Em{W}[\Em{q}_1, \Em{r}]$ & Product: $\Em{q_1}\cdot\Em{q_2}$ & $\Em{q_1}+\Em{q_2}-\Em{q_1}\cdot\Em{q_2}$ & $\Em{1} - \Em{q}$ & \multirow{2}{*}{$\texttt{dist}(\Em{q}, \Em{e})$} \\
& $\Em{W} \in \RR^{2 \times 2d}$ & $\text{top}_k [\rho_{\theta}(d(\Em{q}_1, \Em{r}, \Em{e_k}))]$ & G\"odel: $\text{min}(\Em{q_1}, \Em{q_2})$ &$\text{max}(\Em{q_1}, \Em{q_2})$& $(1+\cos(\pi \Em{q}))/2$ & \\ \hline
\multirow{2}{*}{Var2Vec} & $\Em{q,e, r}\in\CC^{d}$ & 
$\Em{q}_2 = \Em{W}[\Em{q}_1, \Em{r}]$
& Product: $\Em{q_1}\cdot\Em{q_2}$ & $\Em{q_1}+\Em{q_2}-\Em{q_1}\cdot\Em{q_2}$ & \multirow{2}{*}{$1-\Em{q}$} & \multirow{2}{*}{$\texttt{dist}(\Em{q}, \Em{e})$} \\
 & $\Em{W} \in \RR^{d \times 2d}$ & $d(\Em{q}_1, \Em{r}, \Em{q}_2)$ & G\"odel: $\text{min}(\Em{q_1}, \Em{q_2})$ &$\text{max}(\Em{q_1}, \Em{q_2})$&  & \\ \hline
 \multirow{2}{*}{QTO} & $\Em{q,e, r}\in\CC^{d}$ & $d(\Em{q}_1, \Em{r}, \Em{q}_2)$ & \multirow{2}{*}{$\Em{q_1}\cdot\Em{q_2}$} & \multirow{2}{*}{$\Em{q_1}+\Em{q_2}-\Em{q_1}\cdot\Em{q_2}$} & $d(\Em{q}_1, \Em{r}, \Em{q}_2)$ & \multirow{2}{*}{$\texttt{dist}(\Em{q}, \Em{e})$} \\ 
 & $\Em{M}\in [0,1]^{|\gR|\times|\gE|\times|\gE|}$ & $\text{row}_{q1}(\Em{M}_r)$ & & & $\text{row}_{q1}(\Em{1} - \Em{M}_r)$ &  \\ \hline
\multirow{3}{*}{LogicE} & $\Em{q,e}=([l_i,u_i]$,  & $\sigma(\max(0,\max(0,$ & $([\top(l_i^{(1)}, \dots, l_i^{(n)}),\quad$ & \multirow{3}{*}{DM} & $([1-l_i, $ &
\multirow{3}{*}{
$\|\Em{q}-\Em{e}\|_1$
} 
\\
& $l_i, u_i \in [0,1])_{i=1}^d$ & $[\Em{r},\Em{q}]\rmF_1)\rmF_2)\rmF_3)$ & $\:\top(u_i^{(1)}, \dots, u_i^{(n)})])$ & & $ \quad 1-u_i])$ & \\
& $\Em{r} \in \RR^d$ & & $i=1 \dots d$ & & $i=1 \dots d$ & \\\hline
\multirow{2}{*}{FuzzQE} & \multirow{2}{*}{$\Em{q,e}\in[0,1]^d$} & \multirow{2}{*}{$\sigma(\texttt{MLP}_r(\Em{q}))$} & Product: $\Em{q_1}\cdot\Em{q_2}$ & $\Em{q_1}+\Em{q_2}-\Em{q_1}\cdot\Em{q_2}$ & $\Em{1}-\Em{q}$ & \multirow{2}{*}{
$\texttt{dot}(\Em{q}, \Em{e})$
} 
\\
& & & G\"odel: $\text{min}(\Em{q_1}, \Em{q_2})$ &$\text{max}(\Em{q_1}, \Em{q_2})$& - & \\\hline
TAR & $\Em{q,e,r}\in\RR^d$ & $\Em{q}+\Em{r}$ & 
$\texttt{Attn}(\Em{q}_1, \Em{q}_2)$ 
& $\max(\Em{q}_1, \Em{q}_2)$ & $\Em{1}-\Em{q}$ & $\|\Em{q}-\Em{e}\|$ \\\hline
GNN-QE &  $\Em{q, r}\in\RR^{|\gE|}$ & $\sigma(\texttt{GNN}(\Em{q},\gG))$ &  $\Em{q_1}\cdot\Em{q_2}$ & $\Em{q_1}+\Em{q_2}-\Em{q_1}\cdot\Em{q_2}$ & $\Em{1}-\Em{q}$ & $\texttt{BCE}(\Em{q})$ \\\hline
\multirow{4}{*}{FLEX} & $\Em{q}=(\Em{\theta}_f, \Em{\theta}_l)$ & $g(\texttt{MLP}([\Em{\theta}_f + \Em{\theta}_{f,r};$ & \multirow{2}{*}{$\Em{\theta}_f=\sum_i \mathbf{a}_i\Em{\theta}_{q_i,f}$} & $\Em{\theta}_f=\sum_i \mathbf{a}_i\Em{\theta}_{q_i,f}$ & $ \Em{\theta}_f = L \cdot  \text{tanh}($ & 
\\
& $\Em{\theta}_f\in[-L,L]^d$  & $\quad\quad\quad\: \Em{\theta}_l + \Em{\theta}_{l,r}]))$ & & $\Em{\theta}_l=\sum_i \Em{\theta}_{q_i,l} - \quad\quad$ & $\texttt{MLP}([\Em{\theta}_f;\Em{\theta}_l]))$ & $\|\Em{\theta}^e_{f} - \Em{\theta}^q_{f}\|_1 +$ \\
 & $\Em{e}=(\Em{\theta}_f,\Em{0})$ & \multirow{2}{*}{$g$ gates $\Em{\theta}_f$, $\Em{\theta}_e$.} & \multirow{2}{*}{$\Em{\theta}_l=\Pi_i \Em{\theta}_{q_i,l}$} & $ \quad\quad\sum_{1\leq i<j\leq n}\Em{\theta}_{q_i,l}\Em{\theta}_{q_j,l} + $ &  \multirow{2}{*}{$ \Em{\theta}_l = 1 - \Em{\theta}_l$} & $\quad \Em{\theta}^q_{l}$ \\
 & $\Em{\theta}_l\in[0,1]^d$ & & & $\quad\quad\dots + (-1)^{n-1}\Pi_i \Em{\theta}_{q_i,l}$ & & \\\hline
\multirow{4}{*}{TFLEX} & $\Em{q,r}=(\Em{q}_f^e,\Em{q}_l^e,\Em{q}_f^t,\Em{q}_l^t)$ & $\text{Entity:} \: g(\texttt{MLP}($ & $\sum_i \mathbf{\alpha}\Em{q}_{i,f}^e, \bigotimes_i(\{\Em{q}_{i,l}^e\}),$ & $\sum_i \mathbf{\alpha}\Em{q}_{i,f}^e, \bigoplus_i(\{\Em{q}_{i,l}^e\}),$ & $ f_\text{not}^e(\Em{q}_f^e),$ & $ \|\Em{e}_f^e - \Em{q}_f^e\|_1+$ 
\\
& $\Em{q}_f^e,\Em{q}_f^t, \Em{e}_f^e\in\RR^d$ & $\Em{q} + \Em{r} + \Em{t}))$ &  $\sum_i \mathbf{\beta}\Em{q}_{i,f}^t, \bigotimes_i(\{\Em{q}_{i,l}^t\})$ & $ \sum_i \mathbf{\beta}\Em{q}_{i,f}^t, \bigotimes_i(\{\Em{q}_{i,l}^t\})$ &  $\quad\ominus(\Em{q}_l^e),\Em{q}_f^t, \Em{q}_l^t$ & $\quad\Em{q}_f^e\|_1+\Em{q}_l^e$ \\ \cmidrule{3-7}
 & $\Em{e}=(\Em{e}_f^e,\Em{0},\Em{0},\Em{0})$ & $\text{Time:} \: g(\texttt{MLP}($ & $\sum_i \mathbf{\alpha}\Em{q}_{i,f}^e, \bigotimes_i(\{\Em{q}_{i,l}^e\}),$ & $\sum_i \mathbf{\alpha}\Em{q}_{i,f}^e, \bigoplus_i(\{\Em{q}_{i,l}^e\}),$ & $\Em{q}_f^e, \Em{q}_l^e,$ & $\|\Em{t}_f^t - \Em{q}_f^t\|_1+$ 
 \\ 
 & $\Em{q}_l^e,\Em{q}_l^t\in[0,1]^d$ & $\Em{q_1} + \Em{r} + \Em{q_2}))$ & $ \sum_i \mathbf{\beta}\Em{q}_{i,f}^t, \bigotimes_i(\{\Em{q}_{i,l}^t\})$ & $ \sum_i \mathbf{\beta}\Em{q}_{i,f}^t, \bigotimes_i(\{\Em{q}_{i,l}^t\})$ & $ f_\text{not}^t(\Em{q}_f^t),\ominus(\Em{q}_l^t)$ & $\quad\Em{q}_l^t$ \\ \hline
\multirow{3}{*}{ENeSy} & $\Em{q, e, r}\in\CC^{d}$ & Neural: $\Em{q} \circ \Em{r}$ & \multirow{3}{*}{$g(\Em{p_1} \cdot \Em{p_2})$} & \multirow{3}{*}{$g(\Em{p_1} + \Em{p_2} - \Em{p_1}\cdot \Em{p_2})$} & \multirow{3}{*}{$g(\frac{\alpha}{|\gE|} -\Em{p})$} & \multirow{3}{*}{$\|\Em{q}-\Em{e}\|$} \\
 & $\Em{p}_q, \Em{p}_e \in \{0,1\}^{|\gE|}$  & Symb: $g(\Em{p}_q \Em{M}_r)^\top$ &  &  &  & \\
 & $\Em{M}_r \in \{0,1\}^{|\gE|\times |\gE|}$  & $g = \Em{x}/\text{sum}(\Em{x})$ & & & & \\ \hline
NQE &  $\Em{q, e, r}\in [0,1]^d$ & $\sigma(\texttt{Trf}(\Em{q},\gG_q))$ &  $\Em{q_1}\cdot\Em{q_2}$ & $\Em{q_1}+\Em{q_2}-\Em{q_1}\cdot\Em{q_2}$ & $\Em{1}-\Em{q}$ & $\texttt{dot}(\Em{q}, \Em{e})$ \\
 \bottomrule
\end{tabular}
}%
\end{table}

\subsection{Decoder}
\label{sec:decoder}
The goal of decoding is to obtain the final set of answers or a ranking of all the entities. It is the final step of the query answering task after processing. Here we categorize the methods into two buckets: \emph{non-parametric} and \emph{parametric}. Parametric methods require a parameterized method to score an entity (or predict a regression target from the processed latents) while non-parametric methods can directly measure the similarity (or distance) between a pair of query and entity on the graph. 
Most of the methods belong to the non-parametric category as shown in the \emph{Distance} column of processor tables \autoref{tab:proc_neural}, \autoref{tab:proc_geometric}, \autoref{tab:proc_prob}, \autoref{tab:proc_fuzzy}. For instance, geometric models~\citep{q2b, q2b_onto, regex, hype, cone} pre-define a distance function between the representation of the query and that of an entity. Commonly employed distance functions are L1~\citep{gqe, smore, sheaves, mlpmix, abin_abductive2022, enesy,nmp_qem},  L2~\citep{line2022, fuzz_qe}, or their variations~\citep{logic_e, lin2022flex, lin2022tflex}, cosine similarity~\citep{gqe_bin, cga, mpqe, lmpnn}, dot product~\citep{emql, smore, starqe,kgtrans2022,query2particles2022,nqe}, 
or naturally model the likelihood of all the entities without the need of a distance function~\citep{cqd,biqe,gnn_qe,gnnq,qto,cqda,var2vec}. 
Probabilistic models often employ KL divergence~\citep{betae, gammae} or Mahalanobis distance~\citep{perm}.

One important direction (orthogonal to the distance function) that current methods largely ignore is how to perform efficient answer entity retrieval over extremely large graphs with billions of entities. A scalable and approximate nearest neighbor (ANN) search algorithm is necessary. Existing frameworks including FAISS~\citep{faiss} or ScaNN~\citep{scann} provide scalable implementations of ANN. However, ANN is limited to L1, L2 and cosine distance and mostly optimized for CPUs. It is still an open research problem how to design efficient scalable ANN search algorithms for more complex distance functions such as KL divergences so that we can retrieve with much better efficiency for different \clqa methods with different distance functions (preferably, using GPUs).

We conjecture that parametric decoders are to gain more traction in numerical tasks on top of plain entity retrieval for query answering. Such tasks might involve numerical and categorical features on node-, edge-, and graph levels, \eg, training a regressor to predict numerical values for node attributes like \emph{age}, \emph{length}, etc. Besides a parametric decoder gives new opportunities to generalize to inductive settings where we may have unseen entities during evaluation. 
SE-KGE~\citep{se_kge} takes a step in this direction by predicting geospatial coordinates of query targets.

\subsection{Computation Complexity}
\label{sec:complexity}

Here we analyze the time complexity of different query reasoning models categorized by different operations, including relation projection, intersection, union, negation and answer retrieval after obtaining the representation of the query. We list the asymptotic complexity in \autoref{tab:complexity-stepwise} for methods that perform stepwise encoding of the query graph, and \autoref{tab:complexity-together} for methods (mostly GNN-based) that encode the whole query graph simultaneously including projection and other logic operations.

\begin{table}[!ht]
\centering
\small
\caption{Time complexity of each operation on $\gG=(\gE,\gR,\gS)$. Across all methods, we denote the embedding dimension and hidden dimension of MLPs as $d$, number of layers of MLPs/GNNs as $l$, number of branches in an intersection operation as $i$, number of branches in a union operation as $u$.
\label{tab:complexity-stepwise}
}
\resizebox{\columnwidth}{!}{%
\begin{tabular}{ccccccc}
	\toprule
Model & Projection & Intersection & Negation & Union & Answer Retrieval & Definitions \\
\hline
GQE & \multirow{5}{*}{$\gO(d)$} & \multirow{5}{*}{$\gO(ild^2)$} & \multirow{5}{*}{-} & \multirow{5}{*}{-} & \multirow{5}{*}{$\gO(|\gE|d)$} & \multirow{5}{*}{-} \\
GQE+hashing &&&&&& \\
RotatE-m &  &  &  &  & & \\
Distmult-m &  &  &  &  & & \\
ComplEx-m &  &  &  &  & & \\
\hline
CGA & $\gO(d^2)$ & $\gO(id+d^2)$ & - & - & $\gO(|\gE|d)$ & - \\\hline
Query2Particles & $\gO(Kd^2)$ & $\gO(iKd^2)$ & $\gO(Kd^2)$ & DNF & $\gO(|\gE|dK)$ & $K:$ \#particles. \\\hline
SignalE & $\gO(d)$ & $\gO(ild^2)$ & $\gO(d)$ & DNF & $\gO(|\gE|d)$ & - \\\hline
MLPMix & $\gO(ld^2)$ & $\gO(ild^2)$ & $\gO(ld^2)$ & DNF & $\gO(|\gE|d)$ & - \\\hline
Query2Box & \multirow{2}{*}{$\gO(d)$} & \multirow{2}{*}{$\gO(ild^2)$} & \multirow{2}{*}{-} & \multirow{2}{*}{DNF} & \multirow{2}{*}{$\gO(|\gE|d)$}& \multirow{2}{*}{-} \\
Query2Onto &&&&&& \\
\hline
Query2Geom & $\gO(d)$ & $\gO(d)$ & - & DNF & $\gO(|\gE|d)$  & -\\\hline
RotatE-Box & $\gO(d)$ & - & - & DNF / $\gO(uld^2)$ & $\gO(|\gE|d)$  & -\\\hline
NewLook & $\gO(|\gE|d+ld^2)$ & $\gO(i(|\gE|+ld^2))$ & $\gO(ld^2)$ & DNF & $\gO(|\gE|d)$  & -\\\hline
HypE & $\gO(d)$ & $\gO(ild^2)$ & - & DNF & $\gO(|\gE|d)$ & - \\\hline
ConE/BetaE & $\gO(ld^2)$ & $\gO(ild^2)$ & $\gO(d)$ & DNF & $\gO(|\gE|d)$ & - \\\hline
PERM & $\gO(d^2)$ & $\gO(id^3)$ & - & DNF & $\gO(|\gE|d^2)$ & - \\\hline
LinE & $\gO(ld^2)$ & $\gO(d)$ & $\gO(d)$ & $\gO(d)$ & $\gO(|\gE|d)$ & - \\\hline
GammaE & $\gO(ld^2)$ & $\gO(ild^2)$ & $\gO(d)$ & $\gO(uld^2)$ & $\gO(|\gE|d)$ & - \\\hline
NMP-QEM & $\gO(Kld^2)$ & $\gO(iK^2d)$ & $\gO(Kld^2)$ & DNF & $\gO(|\gE|dK)$ & $K:$ \# centers \\\hline
EmQL & $\gO(|\gE|d)$ & $\gO(d)$ & - & $\gO(d)$ & $\gO(|\gE|d)$ & - \\\hline
CQD-CO & $\gO(|\gE|d)$ & Opt & - & Opt & $\gO(|\gE|d)$ & - \\\hline
CQD-Beam & $\gO(|\gE|d)$ & $\gO(|\gE|kd)$ & - & $\gO(|\gE|kd)$ & $\gO(|\gE|d)$ & - \\\hline
\multirow{4}{*}{QTO} &  & \multirow{4}{*}{$\gO(|\gE|)$} & \multirow{4}{*}{$\gO(|\gE|)$} & \multirow{4}{*}{$\gO(|\gE|)$} & \multirow{4}{*}{$\gO(|\gE|)$} & $T^*(v):$ the  \\
&$\gO(\max_k|T^*(v_k)$&&&&&maximum truth value \\
&$>0||\gE|)$&&&&& for the subquery  \\
&&&&&& rooted at node $v$.\\
\hline
LogicE & $\gO(ld^2)$ & $\gO(d)$ & $\gO(d)$ & $\gO(d)$ & $\gO(|\gE|d)$ & - \\\hline
FuzzQE & $\gO(d^2)$ & $\gO(d)$ & $\gO(d)$ & $\gO(d)$ & $\gO(|\gE|d)$ & - \\\hline
TAR & $\gO(d)$ & $\gO(d)$ & $\gO(d)$ & $\gO(d)$ & $\gO(|\gE|d)$ & - \\\hline
GNN-QE & $\gO(|\gE|d^2+|\gS|d)$ & $\gO(|\gE|)$ & $\gO(|\gE|)$ & $\gO(|\gE|)$ & $\gO(|\gE|)$ & - \\\hline
FLEX & $\gO(ld^2)$ & $\gO(ild^2)$ & $\gO(ld^2)$ & $\gO(uld^2+2^ud)$ & $\gO(|\gE|d)$ & - \\\hline
TFLEX & $\gO(ld^2)$ & $\gO(ild^2)$ & $\gO(ld^2)$ & $\gO(uld^2)$ & $\gO(|\gE|d)$ & - \\\hline
\bottomrule
\end{tabular}
}%
\end{table}
\begin{table}[!ht]
\centering
\caption{Time complexity of answering a query for methods that directly encode the query graph.
Besides the operators defined in \autoref{tab:complexity-stepwise}, we denote $n_q$ as average degree of the query graph $\gG_q=(\gE_q, \gR_q, \gS_q)$. 
\label{tab:complexity-together} 
}
\begin{tabular}{ccccccc}
	\toprule
Model & Projection & Intersection & Negation & Union & Answer Retrieval \\
\hline
MPQE / GNNQ & \multicolumn{2}{c}{$\gO(d^2n_ql|\gE_q|)$} & - & - & $\gO(|\gE|d)$ &  \\\hline
StarQE & \multicolumn{2}{c}{$\gO(d^2n_ql|\gE_q|+|\gS_q||qp|d)$} & - & - & $\gO(|\gE|d)$ & \\\hline
LMPNN & \multicolumn{3}{c}{$\gO(d^2n_ql|\gE_q|)$} & DNF & $\gO(|\gE|d)$ & \\\hline
BiQE & \multicolumn{2}{c}{$\gO((|\gE_q|d^2+|\gE_q|^2d)l)$} & - & - & $\gO(|\gE|d)$\\\hline
kgTransformer & \multicolumn{3}{c}{$\gO((d^2n_q|\gE_q|+|\gE_q|n_q^2d)l)$} & - & $\gO(|\gE|d)$ \\\hline
SQE &  \multicolumn{4}{c}{$\gO((|\gE_q|d^2+|\gE_q|^2d)l)$} & $\gO(|\gE|d)$ \\
\bottomrule
\end{tabular}
\end{table}

\section{Queries}
\label{sec:queries}

The third direction to segment the methods is from the queries point of view. Under the queries category, we have three subcategories: \emph{Query Operators}, \emph{Query Patterns}, and \emph{Projected Variables}. For query operators, methods have different operator expressiveness, which means the set of query operators one model is able to handle including existential quantification ($\exists$), conjunction ($\wedge$), disjunction ($\vee$), negation ($\neg$), Kleene plus (+), filter and various aggregation operators. For query patterns, we refer to the structure/pattern of the (optimized) query plan, ranging from paths and trees to arbitrary directed acyclic graphs (DAGs) and cyclic patterns. 
As to projected variables (by \emph{projected} we refer to target variables that have to be bound to particular graph elements like entity or relation), queries might have a different number (zero or more) of target variables.
We are interested in the complexity of such projections as binding of two and more variables involves relational algebra~\citep{codd1970relational} and might result in a Cartesian product of all retrieved answers.

\subsection{Query Operators}
\label{sec:query_ops}

Different query processors have different expressiveness in terms of operators a method can handle. Throughout all the works, we compiled \autoref{tab:proc_query_ops} that classifies all methods based on the supported operators.

\begin{table}[ht]
    \centering
    \caption{Query answering processors and supported query operators. Processors supporting unions $\vee$ also support projection and intersection ($\wedge$). Models supporting negation ($\neg$) also support unions, projections, and intersections. }
    \label{tab:proc_query_ops}
    \resizebox{\linewidth}{!}{
        \begin{tabular}{ccccc}
        \toprule
        Projection and Intersection ($\wedge$) & Union ($\vee$) & Negation ($\neg$) & Kleene $+$ & Filter \& Aggr \\
        \midrule
        \makecell[tl]
        {GQE \citeps{gqe} \\ GQE hashed  \citeps{gqe_bin} \\ CGA  \citeps{cga} \\ TractOR  \citeps{tractor2020} \\ MPQE  \citeps{mpqe} \\ BiQE  \citeps{biqe} \\ Sheaves  \citeps{sheaves} \\ StarQE  \citeps{starqe} \\ RotatE-m, DistMult-m, \\ ComplEx-m  \citeps{smore} \\ GNNQ  \citeps{gnnq}} 
        &
        \makecell[tl]{Query2Box  \citeps{q2b} \\ EmQL  \citeps{emql} \\ Query2Onto \citeps{q2b_onto} \\ HypE  \citeps{hype} \\ NewLook  \citeps{newlook} \\ PERM  \citeps{perm} \\ CQD  \citeps{cqd} \\ kgTransformer  \citeps{kgtrans2022} \\ Query2Geom \citeps{query2geom}} 
        &
        \makecell[tl]{BetaE  \citeps{betae} \\ ConE  \citeps{cone} \\ LogicE  \citeps{logic_e} \\ MLPMix  \citeps{mlpmix} \\  Query2Particles  \citeps{query2particles2022} \\ LinE  \citeps{line2022} \\ GammaE  \citeps{gammae} \\ NMP-QEM \citeps{nmp_qem} \\ FuzzQE  \citeps{fuzz_qe} \\ TAR  \citeps{abin_abductive2022} \\ GNN-QE  \citeps{gnn_qe} \\ FLEX  \citeps{lin2022flex} \\ TFLEX  \citeps{lin2022tflex} \\ ENeSy  \citeps{enesy} \\ QTO \citeps{qto} \\ SignalE \citeps{signale} \\ LMPNN  \citeps{lmpnn} \\ NQE \citeps{nqe} \\ Var2Vec \citeps{var2vec} \\ $\text{CQD}^{\gA}$ \citeps{cqda} \\ SQE \citeps{sqe}}   
        &
        \makecell[tl]{RotatE-Box \\ \citeps{regex}} 
        & 
        \makecell[tc]{None} \\
        \bottomrule
        \end{tabular}
    }
\end{table}

We start from simplest conjunctive queries that involve only existential quantification ($\exists$) and conjunction ($\wedge$) and gradually increase the complexity of supported operators.

\begin{figure}[t]
	\centering
    \includegraphics[width=\textwidth]{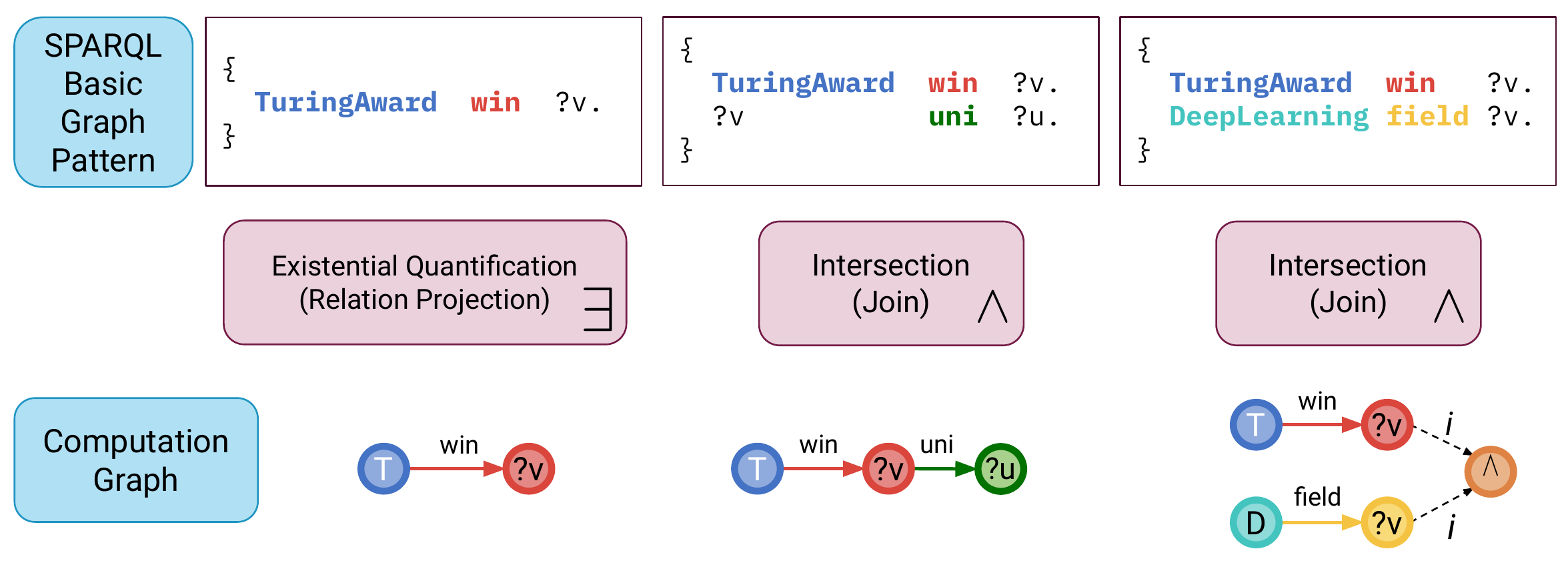}
	\caption{Query operators (relation projection and intersection), corresponding SPARQL basic graph patterns (BGP), and their computation graphs. Relation Projection (left) corresponds to a triple pattern. 
    }
	\label{fig:query_ops1}
\end{figure}

\paragraph{Existential Quantification ($\exists$).}
When $\exists$ appears in a query, this means that there exists at least one existentially quantified variable. For example, given a query ``At what universities do the Turing Award winners work?'' and its logical form $q = U_?\:.\: \exists V\::\:\textit{win}(\texttt{TuringAward}, V)\wedge \textit{university}(V, U_?)$, here $V$ is the existentially quantified variable. Query processors model existential quantification by using a relation projection operator. 
Mapping to query languages like SPARQL, relation projection is equivalent to a \emph{triple pattern} with one variable, \eg, \texttt{\{TuringAward win ?v\}} (\autoref{fig:query_ops1}).
Generally, as introduced in \autoref{sec:models}, query embedding methods embed a query in a bottom-up fashion, starting with the embedding of the anchors (leaf nodes) and gradually traversing the query tree up to the root. In such a way, query embedding methods (\eg, geometric or probabilistic) explicitly obtain an embedding for the existentially quantified variables. The embeddings/representations of these variables are calculated by a relation projection function implemented as shallow vector operations~\citep{gqe,q2b,hype,cqd,smore} or deep neural nets~\citep{betae,cone,query2particles2022,mlpmix}. 
Another set of methods based on GNNs and Transformers directly assigns a learnable initial embedding for the existentially quantified variables. These embeddings are then updated through several message passing layers over the query plan~\citep{mpqe,starqe,gnnq,lmpnn} or attention layers over the serialized query plan~\citep{biqe,kgtrans2022}.

The major drawback of existing neural query processors is the assumption of at least one anchor entity in the query from which the answering process starts and relation projections can be executed. 
It remains an open challenge and an avenue for future work to support queries without anchor entities, \eg, $q_1 = U_? . \exists v_1, v_2 : \textit{win}(v_1, v_2) \wedge \textit{university}(v_2, U_?)$, and queries where relations are variables, \eg, $q_2 = r_? : r_?(\texttt{TuringAward}, \texttt{Bengio})$, that can be framed as the relation prediction task.

\paragraph{Universal Quantification ($\forall$).}
A universally quantified variable $\forall x.P(x)$ means that a logical formula $P(x)$ holds for all possible $x$. 
Usually, the universal quantifier does not appear in facts-only ABox graphs without types and complex axioms (see \autoref{sec:bg_sem}) as it would imply that some entity is connected to all other entities. 
For example, $\forall V. \textit{win}(\texttt{TuringAward}, V)$ without other constraints implies that all entities are connected to \texttt{TuringAward} by the \textit{win} relation (which does not occur in practice). 
However, universal quantifiers are more useful when paired with the class hierarchy (unary relations), \eg, $\exists \textit{?paper}, \forall r \in \texttt{Researcher}: \texttt{Researcher}(r) \wedge \textit{authored}(r, \textit{?paper})$ means that the \textit{authored} relation projection would be applied only to entities of class \texttt{Researcher}.

Currently, existing \clqa approaches do not support universal quantifiers explicitly nor the datasets include queries with the $\forall$ quantifier. 
Still, using the basic identity  $\forall x. P(x) \equiv \neg (\exists x. \neg P(x))$ it is possible to model the universal quantifier by any approach supporting existential quantification $\exists$ and negation $\neg$.
By default, we assume the closed-world assumption in {\ngdb}s.
We leave the implications of universal quantification pertaining to the open-world assumption (OWA) out of the scope of this work.

\paragraph{Conjunction ($\wedge$).}
We elaborate on the differences of those query patterns in the following \autoref{sec:query_pat} and emphasize our focus on more complex intersection queries going beyond simpler path-like queries.

Query processors, as described in \autoref{sec:models}, employ different parameterizations of conjunctions as permutation-invariant set functions. 
A family of neural processors~\citep{gqe, gqe_bin, smore} often resort to the DeepSet architecture~\citep{deepsets} that first projects each set element independently and then pools representations together with a permutation-invariant function (\eg, sum, mean) followed by an MLP. Alternatively,~\citep{cga, query2particles2022} self-attention, can serve as a replacement of the DeepSet where set elements are weighted with the attention operation. 
The other family of neural processors combine projection and intersection by processing the whole query graph with GNNs~\citep{mpqe, starqe, gnnq, lmpnn} or with the Transformer over linearized query sequences~\citep{biqe, kgtrans2022}. 
Geometric processors~\citep{q2b, newlook, hype, cone} implement conjunction as the attention-based average of centroids and offsets of respective geometric objects (boxes, hyperboloids, or cones).
Probabilistic processors~\citep{betae, perm, line2022, gammae} implement intersection as a weighted sum of parametric distributions that represent queries and variables.

Fuzzy-logic processors~\citep{cqd,logic_e, fuzz_qe, gnn_qe} commonly resort to \emph{t-norms}, generalized versions of conjunctions in the continuous $[0,1]$ space
and corresponding \emph{t-conorms} for modeling unions (\autoref{sec:tnorms}). 
Often, due to the absence of a principal study, the choice of the fuzzy logic is a hyperparameter. 
We posit that such a study is an important avenue for future works in fuzzy processors.
More exotic neuro-symbolic methods for modeling conjunctions include element-wise product of count-min sketches~\citep{emql} or as a weighted sum in the \emph{feature logic}~\citep{lin2022flex, lin2022tflex}.
Finally, some processors~\citep{abin_abductive2022, enesy} perform conjunctions both in the embedding and symbolic space with neural and fuzzy operators.

We note that certain neural query processors that embed queries directly~\citep{gqe,gqe_bin, cga,tractor2020, smore} or via GNN/Transformer encoder over a query graph~\citep{mpqe, biqe, sheaves, starqe, cbr_subg2022, gnnq} support only projections and intersections, that is, their extensions to more complex logical operators are non-trivial and might require changing the underlying assumptions of modeling entities, variables, and queries. 
In some cases, support of unions might be enabled when re-writing a query to the disjunctive normal form (discussed below).

\paragraph{Disjunction ($\vee$).}

Query processors implement the disjunction operator in several ways. However, modeling disjunction is notoriously hard since it requires modeling any powerset of entities\footnote{Here the most rigorous description should be powerset of sets of ``isomorphic'' entities on the graph. By ``isomorphic'' entities, it refer to an entity set where each elemen} on the graph in a vector space.
Before delving into details about different ways of modeling disjunction, we first refer the readers to the Theorem 1 in Query2Box~\citep{q2b}. The theorem proves that we need the VC dimension of the function class of the distance function to be around the number of entities on the graph.

The theorem shows that in order to accurately model \emph{any} EPFO query with the existing framework, the complexity of the distance function measured by the VC dimension needs to be as large as the number of KG entities. This implies that if we use common distance functions based on hyper-plane, Euclidean sphere, or axis-aligned rectangle,\footnote{For the detailed VC dimensions of these function classes, see \cite{vapnik2013nature}. Crucially, their VC dimensions are all linear with respect to the number of parameters $d$.} their parameter dimensionality needs to be $\Theta(M)$, which is $\Theta(|\gE|)$ for real KGs we are interested in. In other words, the dimensionality of the logical query embeddings needs to be $\Theta(|\gE|)$, which is not low-dimensional; thus not scalable to large KGs and not generalizable in the presence of unobserved KG edges.

The first idea proposed in Query2Box~\citep{q2b} is that given a model has defined a distance function between a query representation and the entity representation, then a query can be transformed (or re-written) into its equivalent \emph{disjunctive normal form} (DNF), \ie, a disjunction of conjunctive queries. For example, we can safely convert a query $(A\vee B)\wedge (C\vee D)$ to $((A\wedge C) \vee (A\wedge D) \vee (B \wedge C) \vee (B \wedge D))$, where $A,B,C,D$ are atomic formulas. In such a way, we only need to process disjunction $\vee$ at the very last step. For models that have defined a distance function between the query representation and entity representation $d(\Em{q}, \Em{e})$ (such as geometric processors~\citep{q2b, q2b_onto, regex, hype, cone, query2geom} and some neural processors~\citep{kgtrans2022, mlpmix, lmpnn}, the idea of using DNF to handle disjunction is to (1) embed each atomic formula / conjunctive query in the DNF into a vector $\Em{q_i}$, (2) calculate the distance between the representation/embedding of each atomic formula / conjunctive query and the entity $d(\Em{q_i}, \Em{e})$, (3) take the minimum of the distances $\textit{min}_i(d(\Em{q_i}, \Em{e}))$. 
The intuition is that since disjunction models the union operation, as long as the node is close to one atomic formula / conjunctive query, it should be close to the whole query.
Potentially, many neural processors with the defined distance function and originally supporting only intersection and projection can be extended to supporting unions with DNF. 
One notable downside of this modeling is that it is exponentially expensive (to the number of disjunctions) in the worst case when converting a query to its DNF.

Another category of mostly probabilistic processors~\citep{perm, gammae} proposes a neural disjunction operator implemented with the permutation-invariant attention over the input set with the \emph{closure} assumption that the result of attention weighting  union remains in the same probabilistic space as its inputs.  %
Such models design a more black-box framework to handle the disjunction operation under the strong closture assumption that might not be true in all cases. 

The third way of modeling disjunction is based on the De Morgan's laws~\citep{betae}. According to the De Morgan's laws (DM), the disjunction is equivalent to the negation of the conjunction of the negation of the statements making up the disjunction, \ie, $A\vee B = \neg (\neg A \wedge \neg B)$. For methods that can handle the negation operator (detailed in the following paragraph), they model disjunction by using three negation operations and one conjunction operation. 
DM conversion was explicitly probed in probabilistic~\citep{betae}, geometric~\citep{cone}, and fuzzy~\citep{logic_e} processors.

Finally, most fuzzy-logic~\citep{cqd, fuzz_qe, abin_abductive2022, gnn_qe, qto, cqda, var2vec} processors employ \emph{t-conorms}, generalized versions of disjunctions in the continuous $[0,1]$ space (\autoref{sec:tnorms}).
More exotic versions of neuro-symbolic disjunctions include element-wise summation of count-min sketches~\citep{emql}, feature logic operations~\citep{lin2022flex,lin2022tflex}, as well as performing a union in both embedding and symbolic spaces~\citep{abin_abductive2022,enesy} with fuzzy operators.

\begin{figure}[!t]
	\centering
    \includegraphics[width=\textwidth]{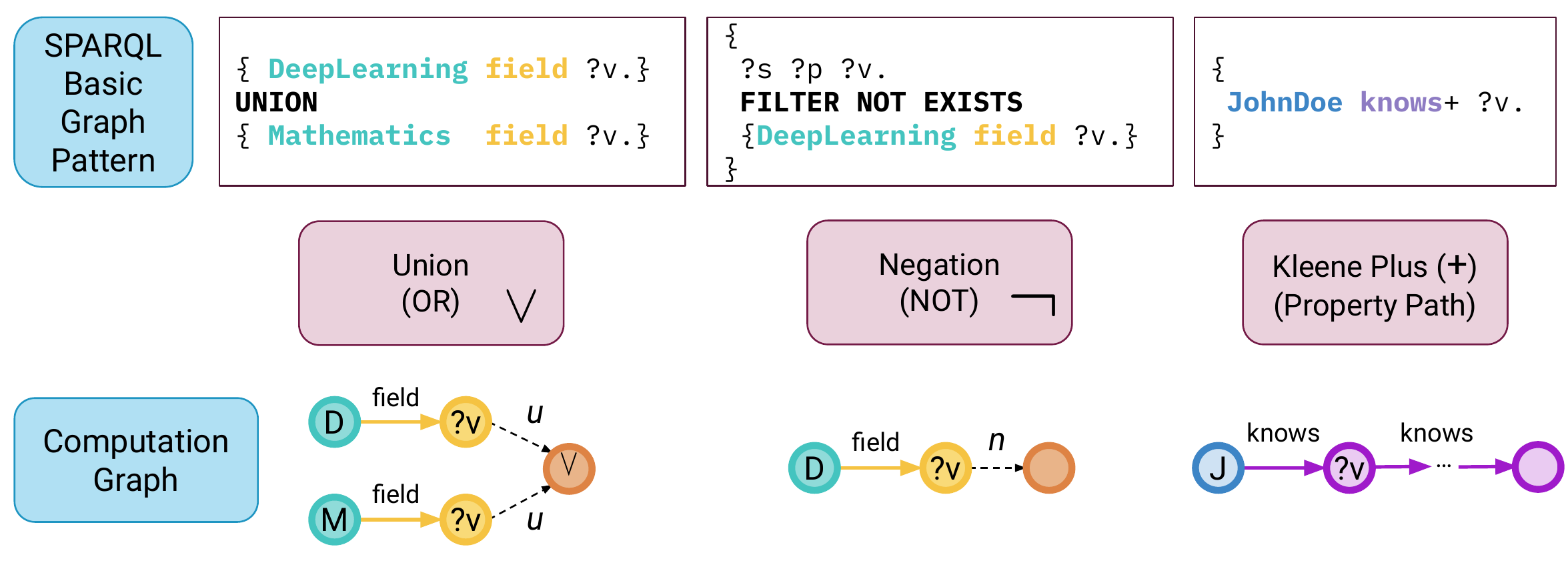}
	\caption{Query operators (union, negation, Kleene plus), corresponding SPARQL basic graph patterns (BGP), and their computation graphs.}
	\label{fig:query_ops2}
\end{figure}

\paragraph{Negation ($\neg$).}
For negation operation, the goal is to model the complement set, \ie, the answers $\gA_q$ to a query $q=V_?: \neg r(v, V_?)$ are the exact complement of the answers $\gA_{q'}$ to query $q'=V_?:r(v,V_?)$: $\gA_q = \gV / \gA_{q'}$. 
Correspondingly, negation in SPARQL can be implemented with \texttt{FILTER NOT EXISTS} or \texttt{MINUS} clauses. 
For example (\autoref{fig:query_ops2}), a logical formula with negation $\neg\textit{field}(\texttt{DeepLearning, V})$ is equivalent to the SPARQL BGP \texttt{\{?s ?p ?v. FILTER NOT EXISTS \{DeepLearning field ?v\}\}} where \texttt{\{?s ?p ?v\}} models the \emph{universe set} ($\Em{1}$) of all facts that gets filtered by the triple pattern.

Modeling the \emph{universe set} ($\Em{1}$) and its complement is the key problem when designing a negation operator in neural query processors, \eg, an arbitrary real $\RR$ or complex $\CC$ space is unbounded such that $\Em{1}$ is not defined.
For that reason, many neural processors do not support the negation operator. 
Still, there exist several approaches to handle negation. 

The first line of works~\citep{query2particles2022, mlpmix, nmp_qem} designs a purely neural MLP-based negation operator over the query representation avoiding the universe set altogether. 
Similarly, a token of the negation operator can be included into the linearized query representation~\citep{sqe} to be encoded with Transformer or recurrent network.
A step aside from purely neural operators is taken by GNN-based processors~\citep{lmpnn} that treat a negation edge as a new edge type during message passing over the query computation graph.

The second line is customized to different embedding spaces and aims to simulate the calculation of the universe and complement in the embedding space, \eg, using geometric cones~\citep{cone}, parameters are angles $\Em{\theta}$ such that the space (and, hence, $\Em{1}$) is bounded to $2\pi$ and the complement is straight $2\pi - \Em{\theta}$. 
Probabilistic methods~\citep{betae, line2022, gammae} naturally represent negation as an inverse of distribution parameters.

Thirdly, fuzzy logic processors explicitly model the universe set $\Em{1}$ and the complement %
over the same real valued logic space. 
For instance, LogicE~\citep{logic_e} and FuzzQE~\citep{fuzz_qe} restrict the query embedding space to the range $[0,1]^d$ where each query $\Em{q} \in [0,1]^d$ is a vector. This way, the universe $\Em{1}$ is represented with a vector of all ones (in the embedding space $\Em{1}^d$) and negation is simply $\Em{1} - \Em{q}$.
TAR~\citep{abin_abductive2022} and GNN-QE~\citep{gnn_qe} operate over fuzzy sets where each entity has a corresponding scalar $\Em{q} \in [0,1]$ in the bounded range. Therefore, the universe $\Em{1}$ can still be a vector of all ones (in the entity space $\Em{1}^{|\gE|}$) and negation is $\Em{1} - \Em{q}$. 
ENeSy~\citep{enesy} defines the universe as the uniform distribution over the entity space with each element weighting $\frac{\alpha}{|\gE|}$ ($\alpha$ is a hyperparameter). 
$\text{CQD}^{\gA}$~\citep{cqda} employs a strict cosine fuzzy negation $\frac{1}{2}(1+\cos(\pi \Em{q}))$ over scalar scores $\Em{q}$.
More exotic processors~\citep{lin2022flex, lin2022tflex} employ feature logic for modeling negation.
We also note that the \emph{difference} operator introduced in \citet{newlook} is in fact a common intersection-negation (\emph{2in}) query pattern used in all standard benchmarks (\autoref{sec:datasets}).

\paragraph{Kleene Plus (+) and Property Paths.}
Kleene Plus is an operator that applies compositionally and recursively to any regular expression (RegEx) that denotes \emph{one or more} occurrence of the specified pattern. 
Regular expressions exhibit a direct connection to \emph{property paths} in SPARQL.
We defined a very basic regular graph query in \autoref{def:regular_path_query}, here we generalize that further to property paths.
To define property paths more formally, given a set of relations $\gR$ and operators $\{+, *, ?, !, \hat{},  /, | \}$, a property path $p$ can be obtained from the recursive grammar 
$p::= r \: | \: p^+ \:|\: p^* \:|\: p? \:|\: !p \: |\: \hat{p} \:|\: p_1 / p_2 \:|\: ``p_1|p_2" $
Here, $r$ is any element of $\gR$, $+$ is a Kleene Plus denoting \emph{one or more} occurrences, $*$ is a Kleene Star denoting \emph{zero or more} occurrences, $?$  denotes \emph{zero or one} occurences, $!$ denotes negation of the relation or path, $\hat{p}$ 
traverses an edge of type $p$ in the opposite direction,
$p_1/p_2$ is a sequence of relations (corresponds to \emph{relation projection}), and $p_1 | p_2$ denotes an alternative path of $p_1$ or $p_2$ (corresponds to a \emph{union} operation).
For example (\autoref{fig:query_ops2}), an expression with Kleene plus $\textit{knows}(\texttt{JohnDoe}, V)^+$ can be represented as a SPARQL property path \texttt{\{JohnDoe knows+ ?v.\}}.

Property paths are non-trivial to model for neural query processors due to compositionalilty and recursive nature. To the best of our knowledge, RotatE-Box~\citep{regex} is the only geometric processor that handles the Kleene Plus operator implementing the subset of operators $\{ +, /, | \}$.
RotatE-Box provides two ways to handle Kleene Plus. The first method is to define a $\Em{r^+}$ embedding for each relation $r\in\gR$, note this is independent and separate from the regular relation embedding for $r$; another way is to use a trainable matrix to transform the relation embedding $\Em{r}$ to the $\Em{r^+}$ embedding. 
Note the two methods do not support Kleene Plus over paths.
RotatE-Box also implements relation projection (as a rotation in the complex space) and union (with DeepSets or DNF) but does not support the intersection operator.
 
We hypothesize that better support of the property paths vocabulary might be one of main focuses in future neural query processors.
Particularly for Kleene Plus, some unresolved issues include supporting
idempotence ($(r^+)^+=r^+$) and infinite union of sets ($r^+=r | (r/r) | (r/r/r)\dots$).

\begin{figure}[!t]
	\centering
    \includegraphics[width=\textwidth]{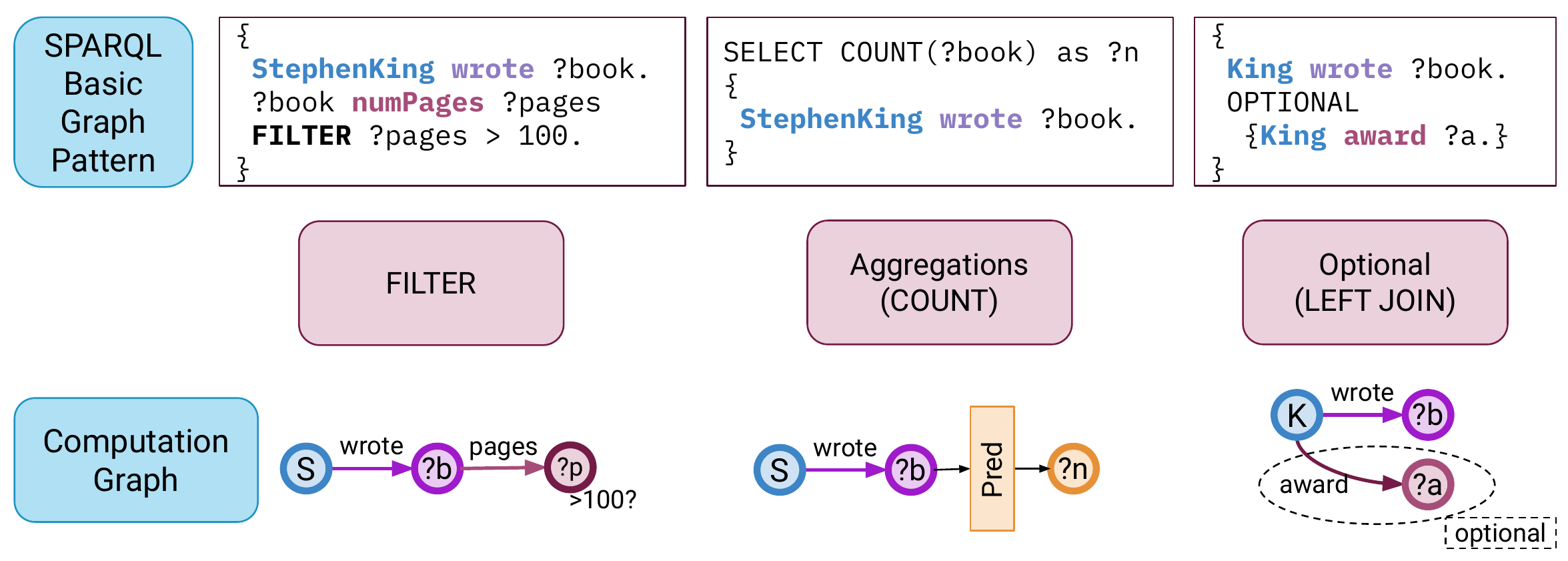}
	\caption{Query operators (\texttt{Filter}, \texttt{Count} Aggregation, \texttt{Optional}), corresponding SPARQL basic graph patterns (BGP), and their computation graphs.}
	\label{fig:query_ops3}
\end{figure}

\paragraph{Filter.}
Filter is an operation that can be inserted in a SPARQL query. %
It takes any expression of boolean type as input and aims to filter the results based on the boolean value, \ie, only the results rendered \texttt{True} under the expression will be returned. The boolean expression can thus be seen as a condition that the answers to the query should follow. For the filter operator, we can do filter on values/literals/attributes, \eg, $\texttt{Filter}(V_\text{date} \geq ``2000-01-01" \&\& V_\text{date} \leq ``2000-12-31")$ means we would like to filter dates not in the year 2000; $\texttt{Filter}(\texttt{LANG}(V_\text{book}) = ``en")$ means we would like to filter books not written in English, $\texttt{Filter}(?\text{pages} > 100)$ means returning the books that have more than 100 pages (as illustrated in \autoref{fig:query_ops3}). To the best of our knowledge, there does not exist a reasoning model that claims to handle Filters, which leaves room for future work on this direction.

We envision several possibilities to support filtering in neural query engines: (1) the simplest option used by  \citet{thorne2021acl,thorne2021vldb} in natural language engines is to defer filtering to the postprocessing stage when the set of candidate nodes is identified and their attributes can be extracted by a lookup. 
(2) Filtering often implies reasoning over literal values and numerical node attributes, that is, processors supporting continuous values (as described in \autoref{sec:modality}) might be able to perform filtering in the latent space by attaching, for instance, a parametric regressor decoder (\autoref{sec:decoder}) when predicting $?\text{pages} > 100$.

\paragraph{Aggregation.}
Aggregation is a set of operators in SPARQL queries including \texttt{COUNT} (return the number of elements), \texttt{MIN}, \texttt{MAX}, \texttt{SUM}, \texttt{AVG} (return the minimum / maximum / sum / average value of all elements), \texttt{SAMPLE} (return any sample from the set). 
For example (\autoref{fig:query_ops3}), given a triple pattern \texttt{\{StephenKing wrote ?book.\}}, the clause $\texttt{COUNT (?book) as ?n)}$ returns the total number of books written by \texttt{StephenKing}.

Most aggregation operators require reasoning over sets of numerical values/literals. Such symbolic operations have long been considered a challenge for neural models~\citep{hendrycksmath2021}. How to design a better representation for numerical values / literals requires remains an open question.
Some neural query processors~\citep{q2b, betae, cone, gnn_qe}, however, have the means to estimate the cardinality of the answer set (including predicted hard answers) that directly corresponds to the \texttt{COUNT} aggregation over the target projected variable (assumed to be an entity, not a literal).
For example, GNN-QE~\citep{gnn_qe} returns a fuzzy set, \ie, a scalar likelihood value for each entity, that, after thresholding, 
has low mean absolute percentage error (MAPE) of the number of ground truth answers.
Answer cardinality estimation is thus obtained as a byproduct of the neural query processor without tailored predictors.
Alternatively, when models cannot predict the exact count, Spearman's rank correlation is a surrogate metric to evaluate the correlation between model predictions and the exact count.
Spearman's rank correlation and MAPE of the number of ground truth answers are common metrics to evaluate the performance of neural query processors and we elaborate on the metrics in \autoref{sec:metrics}.

\paragraph{Optional and Solution Modifiers.}
SPARQL offers many features yet to be incorporated into neural query engines to extend their expressiveness.
Some of those common features include the \texttt{OPTIONAL} clause that is essentially a \texttt{LEFT JOIN} operator.
For example (\autoref{fig:query_ops3}), given a triple pattern \texttt{\{King wrote ?book.\}} that returns books, the optional clause \texttt{\{King wrote ?book. OPTIONAL \{King award ?a.\}\}} enriches the answer set with any existing awards received by \texttt{King}. 
Importantly, if there are no bindings to the optional clause, the query still returns the values of \texttt{?book}.
In the query's computation graph, the optional clause corresponds to the optional branch of a relation projection.
A particular challenge for neural query processors operating on incomplete graphs is that the absence of the queried edge in the graph does not mean that there are no bindings -- instead, the edge might be missing and might be predicted during query processing.

Solution modifiers, \eg, \texttt{GROUP BY}, \texttt{ORDER BY}, \texttt{LIMIT}, apply further postprocessing of  projected (returned) results and are particularly important when projecting several variables in the query. 
So far, all existing neural query processors are tailored for only one return variable. We elaborate on this matter in \autoref{sec:vars}.

\paragraph{A General Note on Incompleteness.}
Finally, we would like to stress out that neural query engines performing all the described operators (\textbf{Projection}, \textbf{Intersection}, \textbf{Union}, \textbf{Negation}, \textbf{Property Paths}, \textbf{Filters}, \textbf{Aggregations}, \textbf{Optionals}, and \textbf{Modifiers}) assume the underlying graph is incomplete and queries might have some missing answers to be predicted, hence, all the operators should incorporate predicted \emph{hard answers} in addition to \emph{easy answers} reachable by graph traversal as in symbolic graph databases.
Evaluation of query performance with those operators in light of incompleteness is still an open challenge (we elaborate on that in \autoref{sec:metrics}), \eg, having an \emph{Optional} clause, it might be unclear when there is no true answer (even predicted ones are in fact false) or a model is not able to predict them.

\subsection{Query Patterns}
\label{sec:query_pat}

Here, we introduce several types of query patterns commonly used in practical tasks and sort them in the increasing order of complexity. Starting with chain-like \emph{Path} queries known in the literature for years, we move to \emph{Tree-Structured} queries (the main supported pattern in modern \clqa systems). Then, we overview \emph{DAG} and \emph{cyclic} patterns which currently are not supported by any neural query answering system and represent a solid avenue for future work.

\begin{figure}[!t]
	\centering
    \includegraphics[width=\textwidth]{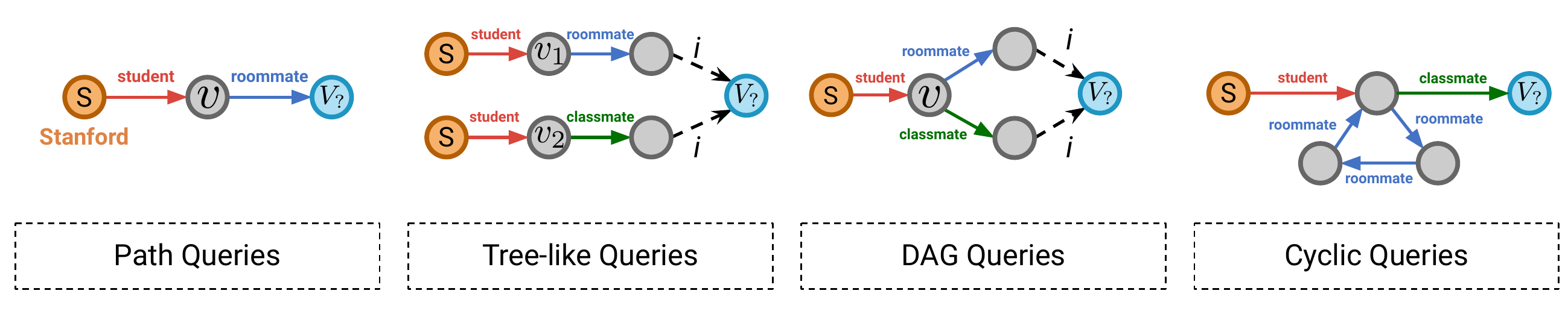}
	\caption{Query patterns: path, tree-like, DAG, and cyclic queries. A DAG query has two branches from the intermediate variable, a cyclic query contains a 3-cycle. Existing neural query processors support path and tree-like patterns.}
	\label{fig:query_pats1}
\end{figure}

\begin{figure}[!t]
	\centering
    \includegraphics[width=\textwidth]{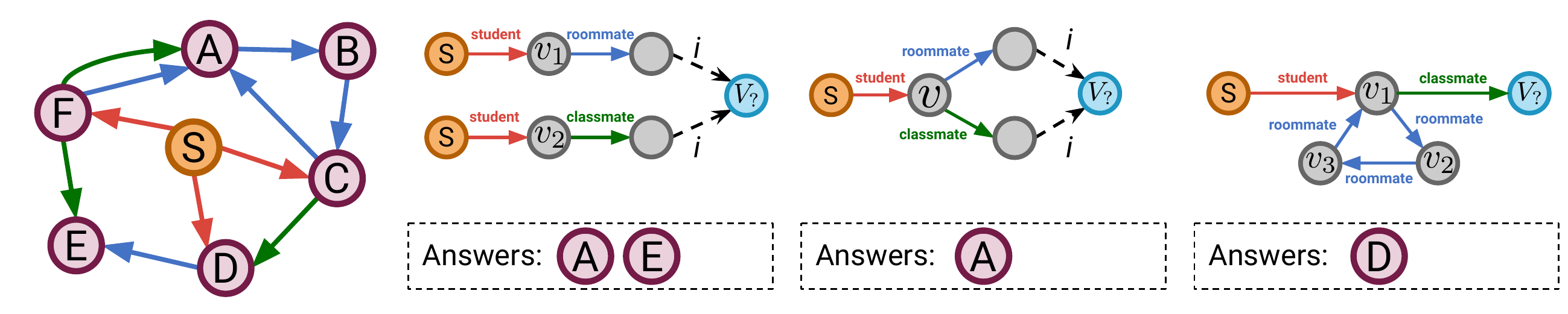}
	\caption{Answers to example tree-like, DAG, and cyclic query patterns given a toy graph. Note the difference in the answer set to the tree-like and DAG queries -- in the DAG query, a variable $v$ must have two outgoing edges from the same node.}
	\label{fig:query_pats2}
\end{figure}

\paragraph{Path Queries.} As introduced in \autoref{sec:multihop_vs_complex}, previous literature starts with path queries (\emph{aka} multi-hop queries), where the goal is simply to go beyond one-hop queries such as $q=V_? . r(v, V_?)$, where $r\in\gR, v\in\gV$ and $V_?$ represents the answer variable. 
As shown in \autoref{fig:query_patterns} and \autoref{fig:query_pats1}, there is no logical operator such as branch intersection or union involved. 
Therefore, in order to answer such a query, we simply find or infer the neighbors of the entity $v$ with relation $r$. 
Path queries are a natural extension of one-hop queries.
Formally, we denote a path query as follows. $q_\text{path}=V_? . \exists V_1,\dots,V_{k-1} : r_1(v, V_1)\wedge r_2(V_1, V_2) \wedge \dots \wedge r_k(V_{k-1}, V_?)$, where $r_i \in \gR, \forall i\in[1,k]$, $v\in\gV$, $V_i$ are all existentially quantified variables. We denote a $k$-hop path query if it has $k$ atomic formulas. 
The query plan of a $k$-hop path query is a chain of length $k$ starting from the anchor entity. 
For example (\autoref{fig:query_pats1}), a 2-hop path query is $V_?.\exists v:\textit{student}(\texttt{Stanford}, v) \wedge \textit{roommate}(v, V_?)$ where \texttt{Stanford} is the starting anchor node, $v$ is a tail variable of the first projection $\textit{student}$ and at the same time is the head variable of the second projection $\textit{roommate}$ thus forming a chain.

As shown in the definition, in order to handle path queries, it is necessary to develop a method to handle existential quantification $\exists$ and conjunction $\wedge$ operators. 
Several query reasoning methods~\citep{guu-2015-traversing, das-2017-chains} aim to answer the path queries with sequence models using either chainable KG embeddings (\eg, TransE) in \citet{guu-2015-traversing} or LSTM in \citet{das-2017-chains}. These methods initiated one of the first efforts that use embeddings and neural methods to answer multi-hop path queries.
We acknowledge the efforts in this domain but emphasize their limitations in terms of query expressiveness and, therefore, focus our attention in this work on more expressive query answering methods that operate on tree-like and more complex patterns.

\paragraph{Tree-Structured Queries.} Path queries only have one anchor entity and one answer variable. 
Such queries have limited expressiveness and are far away from real-world query complexity seen in the logs~\citep{wikidata_query_logs} of real-world KGs like Wikidata.
One direct extension to increase the expressiveness and complexity is to support tree-structured (tree-like) queries.
Tree-like queries may have multiple anchor entities, and different branches (from different anchors) will merge at the final single answer node, thus forming a tree structured query plan. 
Such merge can be achieved by intersection, union, or negation operators. For example, as shown in \autoref{fig:clqa_1}, the query plan of ``At what universities do the Turing Award winners in the field of Deep Learning work?'' is not a path but a tree. 
Alternatively, the example in \autoref{fig:query_pats1} depicts a query $q = V_?, \exists v_1, v_2 : \textit{student}(\texttt{Stanford}, v_1) \wedge \textit{roommate}(v_1, V_?) \wedge \textit{student}(\texttt{Stanford}, v_2) \wedge \textit{classmate}(v_2, V_?)$ that consists of two branches of 2-hop path queries joined by the intersection operator at the end.

Tree-like queries pose more challenges to the previous models that are only able to handle path (multi-hop) queries since a sequence model no longer applies to tree-structured execution plans with logical operators. 
In light of the challenges, neural and neuro-symbolic query processors (described in \autoref{sec:learning}) %
are designed to execute more complex query patterns.
These processors design neural set/logic operators and do a bottom-up traversal of the tree up to the single root node. 

\paragraph{Arbitrary DAGs.}
Based on tree-structured queries, one can further increase the complexity of the query pattern to arbitrary directed acyclic graphs (DAGs). The key difference between the two types of queries is that for DAG-structured queries, one variable node in the query plan (that represents a set of entities) may be split and routed to different reasoning paths, while the number of branches/reasoning paths in the query plan always decreases from the anchor nodes to the answer node. We show one example in \autoref{fig:query_pats1} and in \autoref{fig:query_pats2}. 
Consider the tree-like query from the previous paragraph $q_1=V_?, \exists v_1, v_2 : \textit{student}(\texttt{Stanford}, v_1) \wedge \textit{roommate}(v_1, V_?) \wedge \textit{student}(\texttt{Stanford}, v_2) \wedge \textit{classmate}(v_2, V_?)$ and the DAG query $q_2=V_?, \exists v : \textit{student}(\texttt{Stanford}, v) \wedge \textit{roommate}(v, V_?) \wedge \textit{classmate}(v, V_?)$.
The two queries search for $V_?$ who are \emph{roommate} and \emph{classmate} with \texttt{Stanford} students.
However, the answer sets of the two queries are different (illustrated in \autoref{fig:query_pats2}).
That is, the answer to the DAG query $V_{q_2} = \{\texttt{A}\}$ is the subset of the answers to the tree-like query $V_{q_1} = \{\texttt{A,E}\}$ because the answers to $q_2$ have to be both \textit{roommate} and \textit{classmate} with the \textbf{same} Stanford student in the intermediate variable $v$. 
On the other hand, the two branches of the tree-like query $q_1$ are independent such that intermediate variables $v_1$ and $v_2$ need not be the same entities, hence, the query has more valid intermediate answers and more correct answers.
To the best of our knowledge, there still does not exist a neural query processor that can faithfully handle any DAG query.
Although BiQE~\citep{biqe} claims to support DAG queries, the mined dataset consists of tree-like queries. 
Nevertheless, we hypothesize that, potentially, processors with message passing or Transformer architectures that consider the entire query graph structure $\gG_q$ may be capable of handling DAG queries and leave this question for future work.

\paragraph{Cyclic Queries.}
Cyclic queries are more complex than DAG-structured queries. 
A cycle in a query naturally entails no particular order to traverse the query plan.
An example of the cyclic query is illustrated in \autoref{fig:query_pats1} and \autoref{fig:query_pats2}: $q = V_?, \exists v_1, v_2, v_3 : \textit{student}(\texttt{Stanford}, v_1) \wedge \textit{roommate}(v_1, v_2) \wedge \textit{roommate}(v_2, v_3) \wedge \textit{roommate}(v_3, v_1) \wedge \textit{classmate}(v_1, V_?)$. 
In $q$, three variables form a triangle cycle $\textit{roommate}(v_1, v_2) \wedge \textit{roommate}(v_2, v_3) \wedge \textit{roommate}(v_3, V_?)$.
Given a graph in \autoref{fig:query_pats2}, the cycle starts and ends at node \texttt{C}, hence the only correct answer is obtained after performing the \textit{classmate} relation projection from \texttt{C} ending in \texttt{D}, $V_{q} = \{\texttt{D}\}$. 

Reasoning methods and query processors that assume a particular traversal or node ordering on the query plan, therefore, cannot faithfully answer cyclic queries. 
It remains an open question how to effectively model a query with cyclic structures.
Moreover, cyclic structures often appear when processing queries with regular expressions and \emph{property paths} (\autoref{sec:query_ops}). 
We posit that supporting cycles might be a necessary condition to fully enable property paths in neural query engines.

\subsection{Projected Variables}
\label{sec:vars}

\begin{figure}[!t]
	\centering
    \includegraphics[width=\textwidth]{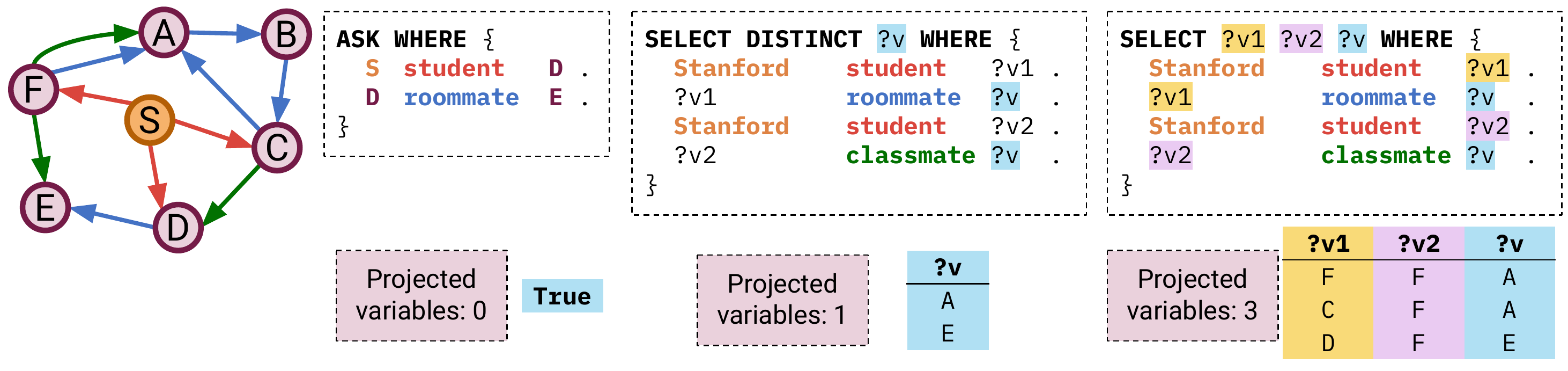}
	\caption{Projected variables of the tree-like query from \autoref{fig:query_pats2}. Current neural query processors support the single-variable \texttt{DISTINCT} mode (center) whereas queries might have zero return variables akin to a subgraph matching Boolean \texttt{ASK} query (left) or multiple projected variables (right) that imply returning intermediate answers and form output tuples.}
	\label{fig:query_projected_vars}
\end{figure}

By \emph{projected variables} we understand target query variables that have to be bound to particular graph elements such as entity, relation, or literals.
For example, a query in \autoref{fig:clqa_1} $q = V_?. \exists v : \textit{win}(\texttt{TuringAward}, v)\wedge \textit{field}(\texttt{DeepLearning}, v) \wedge \textit{university}(v, V_?)$ has one projected variable $V_?$ that can be bound to three answer nodes in the graph, $V_? = \{\texttt{UofT}, \texttt{UdeM}, \texttt{NYU}\}$.
In the SPARQL literature~\citep{sparql2013}, the \texttt{SELECT} query specifies which existentially quantified variables to project as final answers.
The pairs of projected variables and answers form \emph{bindings} as the result of the \texttt{SELECT} query.
Generally, queries might have zero, one, or multiple projected variables, and we align our categorization with this notion. 
Examples of such queries and their possible answers are provided in \autoref{fig:query_projected_vars}.
Currently, most neural query processors focus on the setting where queries have only one answer variable -- the leaf node of the computation graph, as shown in \autoref{fig:query_projected_vars} (center). 

\paragraph{Zero Projected Variables.}
Queries with zero projected variables do not return any bindings but rather probe the graph on the presence of a certain subgraph or relational pattern where the answer is Boolean \texttt{True} or \texttt{False}.
In SPARQL, the equivalent of zero-variable queries is the \texttt{ASK} clause.
Zero-variable queries might have all entities and relations instantiated with constants, \eg, $q = \textit{student}(\texttt{S}, \texttt{D}) \wedge \textit{roommate}(\texttt{D}, \texttt{E})$ as in \autoref{fig:query_projected_vars} (left) is equivalent to the SPARQL query \texttt{ASK WHERE \{S student D. D roommate E.\}}. 
The query probes whether a graph contains a particular subgraph (path) induced by the constants. Such a path exists, so the answer is $q = \{\texttt{True}\}$. 

Alternatively, zero-variable queries might have existentially quantified variables that are never projected (up to the cases where all subjects, predicates, or objects are variables). 
For example, a query $q_1 = \exists v_1, v_2, v_3 : \textit{student}(v_1, v_2) \wedge \textit{roommate}(v_2, v_3)$ probes whether there exist any nodes forming a relational path $v_1 \xrightarrow{\textit{student}} v_2 \xrightarrow{\textit{roommate}} v_3$.
In a general case, a query $q_2 = \exists p, s, o : p(s, o)$ asks if a graph contains at least one edge.

We note that in the main considered setting with incomplete graphs and missing edges zero-variable queries are still non-trivial to answer.
Particularly, a subfield of \emph{neural subgraph matching}~\citep{rex2020neural,huang2022fewshot} implies having incomplete graphs. 
We hypothesize such approaches might be found useful for neural query processors to support answering zero-variable queries.

\paragraph{One Projected Variable.}
Queries with one projected variable return bindings for one (of possibly many) existentially quantified variable. 
In SPARQL, the projected variable is specified in the \texttt{SELECT} clause, \eg, \texttt{SELECT DISTINCT ?v} in \autoref{fig:query_projected_vars} (center).
Although SPARQL allows projecting variables from any part of a query, most neural query engines covered in \autoref{sec:learning} follow the task formulation of GQE~\citep{gqe} and allow the projected target variable to be only the \textbf{leaf node} of the query computation graph. 
This limitation is illustrated in \autoref{fig:query_projected_vars} (center) where the target variable $?v$ is the leaf node of the query graph and has two bindings $v=\{\texttt{A}, \texttt{E}\}$.

It is worth noting that existing neural query processors are designed to return a \emph{unique set} of answers to the input query, \ie, it corresponds to the \texttt{SELECT DISTINCT} clause in SPARQL.
In contrast, the default \texttt{SELECT} returns \emph{multisets} with possible duplicates.
For example, the same query in \autoref{fig:query_projected_vars} (center) without \texttt{DISTINCT} would have bindings $v=\{\texttt{A},\texttt{A},\texttt{E}\}$ as there exist two matching graph patterns ending in \texttt{A}. 
Implementing non-\texttt{DISTINCT} query answering remains an open challenge.

Most neural query processors have a notion of intermediate variables and model their distribution in the entity space.
For instance, having a defined distance function, geometric processors~\citep{q2b,hype} can find nearest entities as intermediate variables.
Similarly, fuzzy-logic processors operating on fuzzy sets~\citep{abin_abductive2022,gnn_qe} already maintain a scalar distribution over all entities after each execution step.
Finally, GNN-based~\citep{mpqe,starqe} and Transformer-based~\citep{kgtrans2022} processors  explicitly include intermediate variables as nodes in the query graph (or tokens in the query sequence) and can therefore decode  their representations to the entity space.
The main drawback of all those methods is the lack of filtering mechanisms for the sets of intermediate variables \emph{after} the leaf node has been identified.
That is, in order to filter and project only those intermediate variables that lead to the final answer, some notion of \emph{backward pass} is required. 
The first step in this direction is taken by  QTO~\citep{qto} that runs the pruning backward pass after reaching the answer leaf node.

\paragraph{Multiple Projected Variables.}
The most general and complex case for queries is to have multiple projected variables as illustrated in \autoref{fig:query_projected_vars} (right).
In SPARQL, all projected variables are specified in the \texttt{SELECT} clause (with the possibility to project all variables in the query via \texttt{SELECT *}).
In the logical form, a query has several target variables $q = ?v_1, ?v_2, ?v : \textit{student}(\texttt{Stanford}, ?v_1) \wedge \textit{roommate}(?v_1, ?v) \wedge \textit{student}(\texttt{Stanford}, ?v_2) \wedge \textit{classmate}(?v_2, ?v)$ such that the output bindings are organized in \emph{tuples}.
For example, one possible answer tuple is $\{?v_1: \texttt{F}, ?v_2: \texttt{F}, ?v: \texttt{A}\}$ denotes particular nodes (variable bindings) that satisfy the query pattern.

As shown in the previous paragraph about one-variable queries, some neural query processors have the means to keep track of the intermediate variables.
However, none of them have the means to construct answer tuples with variables bindings and it remains an open challenge how to incorporate multiple projected variables into such processors.
Furthermore, some common caveats to be taken into account include (1) dealing with unbound variables that often emerge, for example, in \texttt{OPTIONAL} queries covered in \autoref{sec:query_ops}, where answer tuples might contain an empty value ($\emptyset$ or \texttt{NULL}) for some variables; (2) the growing complexity issue where the answer set might potentially be polynomially large depending on the number of projected variables.

\section{Datasets and Metrics}
\label{sec:datasets}

\begin{figure}[!t]
	\centering
    \includegraphics[width=0.7\textwidth]{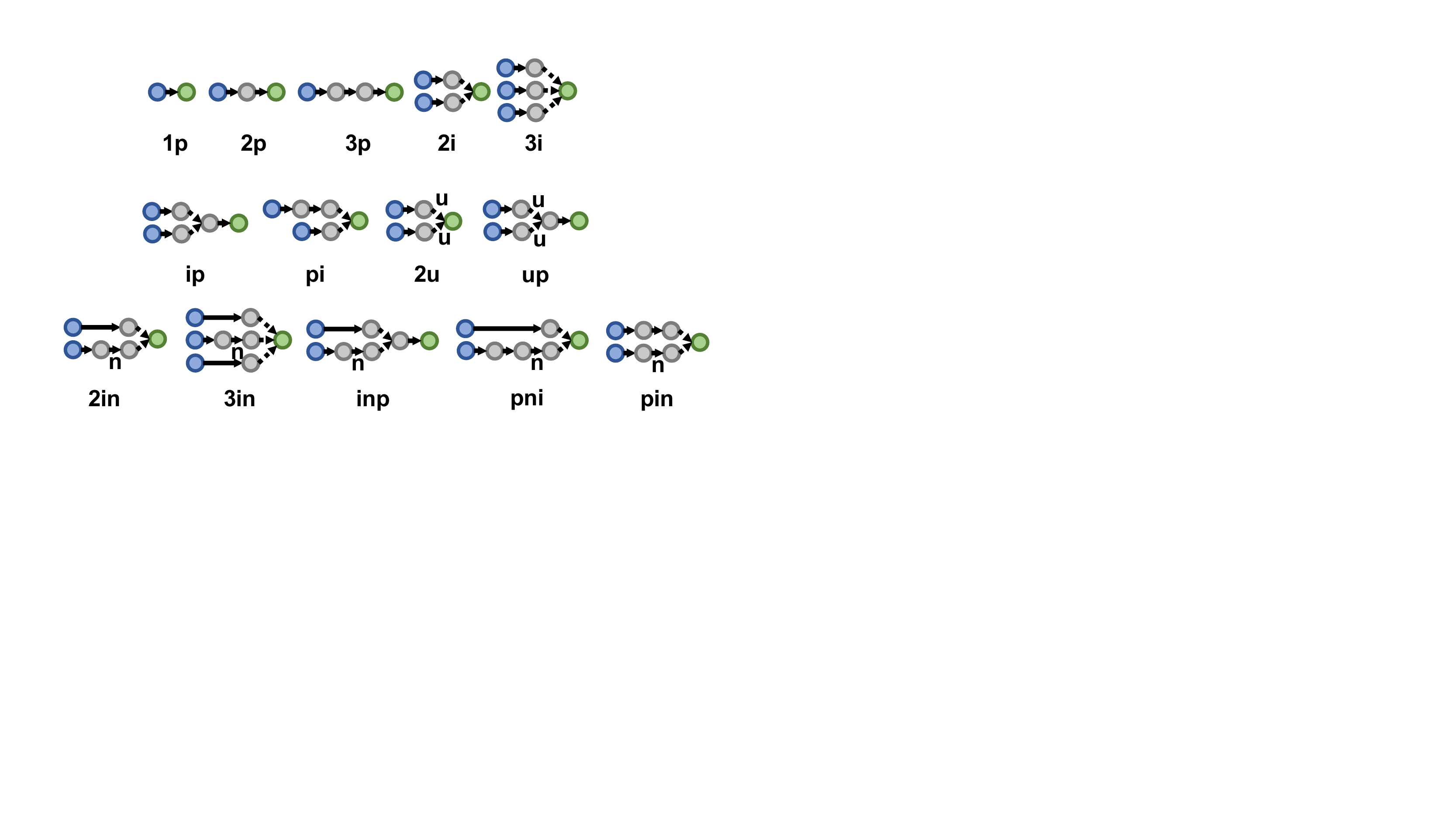}
	\caption{Standard query patterns with names, where \emph{p} is projection, \emph{i} is intersection, \emph{u} is union, \emph{n} is negation. In a pattern, blue node represents a non-variable entity, grey node represents a variable node, and the green node represents the answer node. In a typical training protocol, models are trained on 10 patterns (first and third rows) and evaluated on all patterns. In the hardest generalization case, models are only trained on \emph{1p} queries. Some datasets further modify the patterns with additional features like qualifiers or temporal timestamps.}
	\label{fig:query_graph}
\end{figure}

\subsection{Evaluation Setup}
\label{sec:evaluation}

Multiple datasets have been proposed for evaluation of query reasoning models. Here we introduce the common setup for \clqa task. Given a knowledge graph $\gG=(\gE,\gR,\gS)$, the standard practice is to split $\gG$ into a training graph $\gtrain$, a validation graph $\gval$ and a test graph $\gtest$ (simulating the unobserved complete graph $\hat{\gG}$ from \autoref{sec:prelim}). The standard experiment protocol is to train a query reasoning model only on the training graph $\gtrain$, and evaluate the model on answering queries over the validation graph $\gval$ and the test graph $\gtest$. Given a query $q$, denote the answers of this query on training, validation and test graph as $\dqtr$, $\dqv$ and $\dqte$. During evaluation, queries may have missing answers, \eg, a validation query $q$ may have answers $\dqv$ that are not in $\dqtr$, a test query $q$ may have answers $\dqte$ that are not in $\dqv$. 
The overall goal of \clqa task is to find these missing answers.
The details of typical training queries, training protocol, inference and evaluation metrics are introduced in \autoref{sec:dataset_query_types}, \autoref{sec:graph_training}, \autoref{sec:graph_inference}, and \autoref{sec:metrics}, respectively.

\subsection{Query Types}
\label{sec:dataset_query_types}

The standard set of graph queries used in many datasets includes 14 types: \emph{1p/2p/3p/2i/3i/ip/pi/2u/up/2in/3in/inp/pni/pin} where \emph{p} denotes relation projection, \emph{i} is intersection, \emph{u} is union, \emph{n} is negation, and a number denotes the number of hops for projection queries or number of branches to be merged by a logical operator. 
\autoref{fig:query_graph} illustrates common query patterns. 
For example, \emph{3p} is a chain-like query of three consecutive relation projections, \emph{2i} is an intersection of two relation projections, \emph{3in} is an intersection of three relation projections where one of the branches contains negation, \emph{up} is a union of two relation projections followed by another projection.
The original GQE by \citet{gqe} introduced 7 query patterns with projection and intersection \emph{1p/2p/3p/2i/3i/ip/pi}, Query2Box~\citep{q2b} added union queries \emph{2u/up}, and BetaE~\citep{betae} added five types with negation.

Subsequent works modified the standard set of query types in several ways, \eg, hyper-relational queries~\citep{starqe,nqe} with entity-relation qualifiers on relation projections, or temporal operators on edges~\citep{lin2022tflex}. 
New query patterns include queries with regular expressions of relations (property paths)~\citep{regex}, more tree-like queries~\citep{biqe}, and more combinations of projections, intersections, and unions~\citep{wang2021benchmarking, gnnq}.
We summarize existing query answering datasets and their properties in \autoref{tab:datasets1} covering supported query operators, inference setups, and additional features like temporal timestamps, class hierarchies, or complex ontological axioms.

Commonly, query datasets are sampled from different KGs to study model performance under different graph distributions, for example, BetaE datasets include sets of queries from denser Freebase~\citep{bollacker2008freebase} with average node degree of 18 and sparser WordNet~\citep{miller1998wordnet} and NELL~\citep{nell} with average node degree of 2. 
Hyper-relational datasets WD50K~\citep{starqe} and WD50K-NFOL~\citep{nqe} were sampled from Wikidata~\citep{wikidata} where qualifiers are natural.
TAR datasets with class hierarchy~\citep{abin_abductive2022} were sampled from YAGO 4~\citep{yago} and DBpedia~\citep{lehmann2015dbpedia} where class hierarchies are well-curated.
Q2B Onto datasets with ontological axioms~\citep{q2b_onto} were sampled from LUBM~\citep{guo2005lubm} and NELL. 
Temporal TFLEX datasets~\citep{lin2022tflex} were sampled from ICEWS~\citep{icews_dataset} and GDELT~\citep{leetaru2013gdelt} that maintain event information.
InductiveQE datasets~\citep{galkin2022} were sampled from Freebase and Wikidata, while inductive GNNQ datasets~\citep{gnnq} were sampled from the WatDiv benchmark~\citep{watdiv} and Freebase.

\subsection{Training}
\label{sec:graph_training}
Query reasoning methods are trained on the given $\gtrain$ with different objectives/losses and different datasets. Following the standard protocol, methods are trained on 10 query patterns \emph{1p/2p/3p/2i/3i/2in/3in/inp/pni/pin} and evaluated on all 14 patterns including generalization to unseen \emph{ip/pi/2u/up} patterns. That is, the training protocol assumes that models trained on atomic logical operators would learn to compositionally generalize to patterns using several operators such as \emph{ip} and \emph{pi} queries that use both intersection and projection. 

We summarize different training objectives in \autoref{tab:loss}. Most methods that learn a representation of the queries and entities on the graph optimize a contrastive loss, \ie, minimizing the distance between the representation of a query $q$ and its positive answers $e$ while maximizing that between the representation of a query and negative answers $e'$. Various objectives include: (1) max-margin loss (first column in \autoref{tab:loss}) with the goal that the distance of negative answers should be larger than that of positive answers at least by the margin $\gamma$. Such loss is often of the form as the equation below.
$$
\ell = \max(0, \gamma - \texttt{dist}(\Em{q}, \Em{e}) + \texttt{dist}(\Em{q}, \Em{e'}));
$$
(2) LogSigmoid loss (second column in \autoref{tab:loss}) with a similar goal that pushes the distance of negatives up and vice versa. Often the loss also includes a margin term and the gradient will gradually decrease when the margin is satisfied.
$$
\ell = -\log\sigma (\gamma - \texttt{dist}(\Em{q}, \Em{e})) - \sum \frac{1}{k} \log\sigma(\texttt{dist}(\Em{q}, \Em{e'}) - \gamma),
$$
where $k$ is the number of negative answers.
Other methods (third column in \autoref{tab:loss}) that directly model a logit vector over all the nodes on the graph may optimize a cross entropy loss instead of a contrastive loss.
Besides, methods such as the two variants of CQD~\citep{cqd} or QTO~\citep{qto} only optimize the link prediction loss since they do not learn a representation of the query.

\begin{table}[!ht]
\centering
\caption{Complex Query Answering approaches categorized under \emph{Loss}. 
}
\label{tab:loss}
\resizebox{\columnwidth}{!}{%
\begin{tabular}{ccc}\toprule
Max Margin & LogSigmoid~\citep{sun2018rotate} & Cross Entropy \\
\midrule
\multicolumn{1}{p{4cm}}{
GQE \citeps{gqe}, GQE w hash \citeps{gqe_bin}, CGA \citeps{cga}, MPQE \citeps{mpqe}, HyPE \citeps{hype}, Shv \citeps{sheaves}
} 
&
\multicolumn{1}{p{7.5cm}}{
Query2Box \citeps{q2b}, BetaE \citeps{betae}, RotatE-Box \citeps{regex}, ConE \citeps{cone}, NewLook \citeps{newlook}, Q2B Onto \citeps{q2b_onto}, PERM \citeps{perm}, LogicE \citeps{logic_e}, MLPMix \citeps{mlpmix}, FuzzQE \citeps{fuzz_qe}, FLEX \citeps{lin2022flex}, TFLEX \citeps{lin2022tflex}, SMORE \citeps{smore}, LinE \citeps{line2022}, GammaE \citeps{gammae}, NMP-QEM \citeps{nmp_qem}, ENeSy \citeps{enesy}, RoMA \citeps{roma}, SignalE \citeps{signale}, Query2Geom \citeps{query2geom}
} 
& 
\multicolumn{1}{p{5cm}}{
BiQE \citeps{biqe}, NodePiece-QE \citeps{galkin2022}, GNN-QE \citeps{gnn_qe}, GNNQ \citeps{gnnq}, KGTrans \citeps{kgtrans2022}, Query2Particles \citeps{query2particles2022}, EmQL \citeps{emql}, LMPNN \citeps{lmpnn}, StarQE \citeps{starqe}, NQE \citeps{nqe}, $\text{CQD}^{\gA}$ \citeps{cqda}, SQE \citeps{sqe}
} \\
\bottomrule
\end{tabular}
}
\end{table}

Almost all the datasets including GQE~\citep{gqe}, Q2B~\citep{q2b}, BetaE~\citep{betae}, RegEx~\citep{regex}, BiQE~\citep{biqe}, Query2Onto~\citep{q2b_onto}, TAR~\citep{abin_abductive2022}, StarQE~\citep{starqe}, GNNQ~\citep{gnnq}, TeMP~\citep{temp2022}, TFLEX~\citep{lin2022tflex}, InductiveQE~\citep{galkin2022}, SQE~\citep{sqe} provide a set of training queries of given structures sampled from the $\gtrain$. The benefit is that during training, methods do not need to sample queries online. However, it often means that only a portion of information is utilized from the $\gtrain$ since exponentially more multi-hop queries exist on $\gtrain$ and the dataset can never pre-generate all offline. SMORE~\citep{smore} proposes a bidirectional online query sampler such that methods can directly do online sampling efficiently without the need to pre-generate a training set offline. Alternatively, methods that do not have parameterized ways to handle logical operations, \eg, CQD~\citep{cqd}, only require one-hop edges to train the overall system.

\subsection{Inference}
\label{sec:graph_inference}

\begin{table}[t]
\centering
\caption{
Existing logical query answering datasets classified along supported \textbf{query operators}, \textbf{inference} scenarios, and \textbf{domain} properties. 
}
\label{tab:datasets1}
\begin{adjustbox}{width=\textwidth}
\begin{tabular}{llcccccccccccccc}\toprule
& &\multicolumn{5}{c}{\textbf{Query Operators}} &\multicolumn{2}{c}{\textbf{Inference}} &\multicolumn{3}{c}{\textbf{Domain}} & & & \\\cmidrule(l){3-7} \cmidrule(l){8-9} \cmidrule(l){10-12}
Source & Dataset &\rot{Conjunctive} &\rot{Union} &\rot{Negation} &\rot{Kleene Plus} &\rot{Filter + Agg} &\rot{Transductive} &\rot{Inductive} &\rot{Discrete} &\rot{+ Timestamps} &\rot{+ Continuous} &\rot{Types} &\rot{Rules} &\rot{Qualifiers} \\\midrule
\citet{gqe} & GQE datasets  &\OK & & & & &\OK & &\OK & & & & & \\
\citet{q2b} & Q2B datasets &\OK &\OK & & & &\OK & &\OK & & & & & \\
\citet{betae} & BetaE datasets &\OK &\OK &\OK & & &\OK & &\OK & & & & & \\
\citet{regex} & Regex queries &\OK &\OK & & \OK & &\OK & &\OK & & & & & \\
\citet{biqe} & DAG queries &\OK & & & & & \OK & & \OK & & & & & \\
\citet{wang2021benchmarking} & EFO-1 queries &\OK &\OK &\OK & & &\OK & &\OK & & & & & \\
\citet{q2b_onto} & LUBM/NELL (type) &\OK &\OK & & & &\OK & &\OK & & & &\OK & \\
\citet{abin_abductive2022} & TAR datasets &\OK &\OK & & & &\OK & &\OK & & &\OK & & \\
\citet{smore} & SMORE datasets &\OK &\OK & & & &\OK & &\OK & & & & & \\
\citet{starqe} & WD50K dataset &\OK & & & & &\OK & &\OK & & & & &\OK \\
\citet{temp2022} & TeMP dataset &\OK &\OK & & & & &\OK &\OK & & &\OK & & \\
\citet{gnnq} & GNNQ dataset &\OK & & & & & &\OK &\OK & & & & & \\
\citet{lin2022tflex} & TFLEX dataset &\OK &\OK &\OK & & &\OK & &\OK &\OK & & & & \\
\citet{galkin2022} & InductiveQE dataset &\OK &\OK &\OK & & & &\OK &\OK & & & & & \\
\citet{nqe} & WD50K-NFOL dataset &\OK &\OK &\OK & & & \OK & &\OK & & & & & \OK \\
\citet{sqe} & SQE dataset & \OK & \OK & \OK & & & \OK & & \OK & & & & & \\
\bottomrule
\end{tabular}
\end{adjustbox}
\end{table}

\begin{figure}[!ht]
	\centering
	\includegraphics[width=\linewidth]{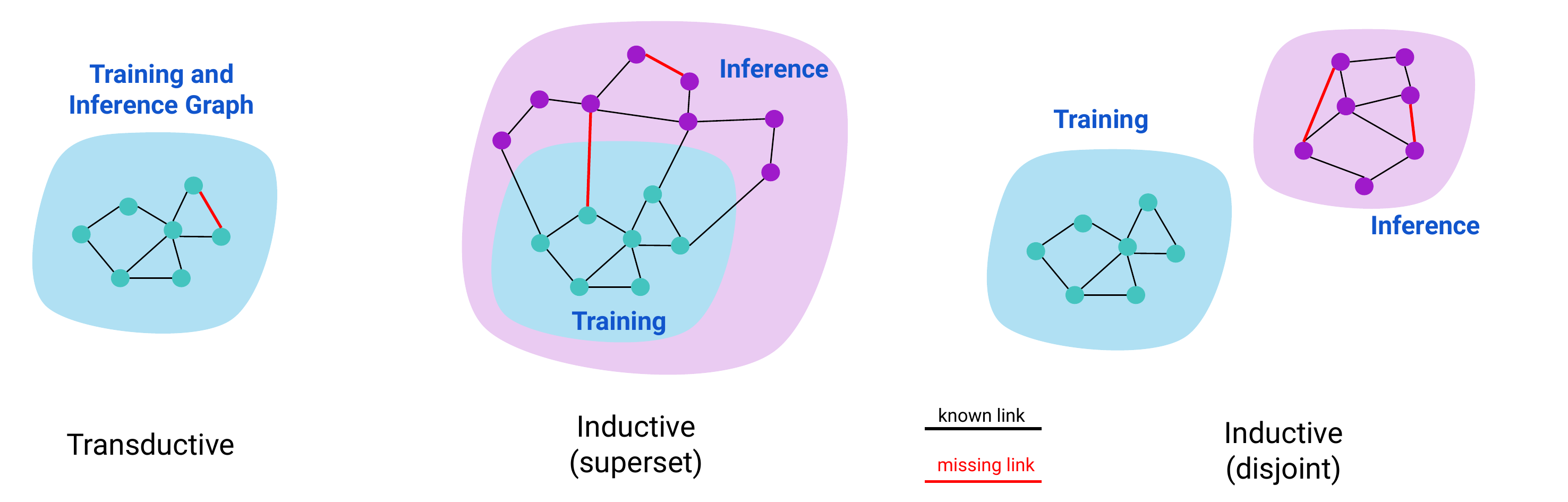}
	\caption{Inference Scenarios. In the \emph{Transductive} case, training and inference graphs are the same and share the same nodes ($\einf = \etrain$). \emph{Inductive} cases can be split into \emph{superset} where the inference graph extends the training one ($\etrain \subseteq \einf$, missing links cover both seen and unseen nodes) and \emph{disjoint}  where the inference graph is disconnected ($\etrain \cap \einf = \emptyset$, missing links are among unseen nodes).}
	\label{fig:graph_inference}
\end{figure}

By \emph{Inference} we understand testing scenarios on which a trained query answering model will be deployed and evaluated. 
Following the literature, we distinguish \emph{Transductive} and \emph{Inductive} inference (\autoref{fig:graph_inference}). In the transductive case, inference is performed on the graph with the same set of nodes and relation types as in training but with different edges. Any other scenario when either the number of nodes or relation types of an inference graph is different from that of the training is deemed inductive. 
The inference scenario plays a major role in designing query answering models, that is, transductive models can learn a shallow entity embedding matrix thanks to the fixed entity set whereas inductive models have to rely on other \emph{invariances} available in the underlying graph in order to generalize to unseen entity/relation types. 
We discuss many transductive and inductive models in \autoref{sec:models}. 
Below, we categorize existing datasets from the \emph{Inferece} perspective.
The overview of existing \clqa datasets is presented in \autoref{tab:datasets1} through the lens of supported query operators, inference scenario, graph domain, and other features like types or qualifiers.

\paragraph{Transductive Inference.}
Formally, given a training graph $\gtrain = (\etrain, \rtrain, \strain)$, the transductive inference graph $\ginf$~\footnote{Below we use $\ginf$ to refer to the graphs we use during inference, it can be $\gval$ or $\gtest$ without loss of generalization.} contains the same set of entities and relation types, that is, $\etrain = \einf$ and $\rtrain=\rinf$, while the edge set on $\gtrain$ is a subset of that on the inference graph $\ginf$, \ie, $\strain\subset\sinf$.
In this setup, query answering is performed on the same nodes and edges seen during training. 
From the entity set perspective, the prediction pattern is \emph{seen-to-seen} -- missing links are predicted between known entities. 

Traditionally, KG link prediction focused more on the transductive task. In \clqa, therefore, the majority of existing datasets (\autoref{tab:inference}) follow the transductive scenario. Starting from simple triple-based graphs with fixed query patterns in GQE datasets~\citep{gqe}, Query2Box datasets~\citep{q2b}, and BetaE datasets~\citep{betae} that became de-facto standard benchmarks for query answering approaches, newer datasets include  regex queries~\citep{regex}, wider set of query patterns~\citep{biqe, wang2021benchmarking, sqe}, entity type information~\citep{abin_abductive2022}, ontological axioms~\citep{q2b_onto}, hyper-relational queries with qualifiers~\citep{starqe,nqe}, temporal queries~\citep{lin2022tflex}, or very large graphs up to 100M nodes~\citep{smore}.  

\paragraph{Inductive Inference.} 
Formally, given a training graph $\gtrain = (\etrain, \rtrain, \strain)$, the inductive inference graph $\ginf =(\einf, \rinf, \sinf) $ is different from the training graph in either the entity set or the relation set or both. 
The nature of this difference explains several subtypes of inductive inference. 
First, the set of relations might or might not be shared at inference time, that is, $\rinf \subseteq \rtrain$ or $|\rinf \setminus \rtrain| > 0$. 
Most of the literature on inductive link prediction~\citep{teru2020inductive,zhu2021neural,galkin2022nodepiece} in KGs assumes the set of relations is shared whereas the setup where new relations appear at inference time is still highly non-trivial~\citep{huang2022fewshot,gao2023double,chen2023generalizing}.

On the other hand, the inference graph might be either a superset of the training graph after adding new nodes and edges, $\etrain \subseteq \einf$, or a disjoint graph with completely new entities as a disconnected component, $\einf \cap \etrain = \emptyset$ as illustrated in \autoref{fig:graph_inference}.
From the node set perspective, the superset inductive inference case might contain both \emph{unseen-to-seen} and \emph{unseen-to-unseen} missing links whereas in the disjoint inference graph only \emph{unseen-to-unseen} links are naturally appearing.

In \clqa, inductive reasoning is still an emerging area as it has a direct impact on the space of possible variables $\gV$, constants $\gC$, and answers $A$ that might now include entities unseen at training time. 
Several most recent works started to explore inductive query answering (\autoref{tab:inference}). 
InductiveQE datasets~\citep{galkin2022} focus on the inductive superset case where a training graph can be extended with up to 500\% new unseen nodes. Test queries start from unseen constants and answering therefore requires reasoning over both seen and unseen nodes. Similarly, training queries can have many new correct answers when answered against the extended inference graph.
GNNQ datasets~\citep{gnnq} focus on the disjoint inductive inference case where constants, variables, and answers all belong to a new entity set.
TeMP datasets~\citep{temp2022} focus on the disjoint inductive inference as well but offer to leverage an additional class hierarchy as a learnable \emph{invariant}. That is, the set of classes at training and inference time does not change.

As stated in \autoref{sec:ngdb}, inductive inference is crucial for NGDBs to enable running models over updatable graphs without retraining. 
We conjecture that inductive datasets and models are likely to be the major contribution area in the future work.

\begin{table}[!ht]
\centering
\caption{Complex Query Answering datasets categorized under \emph{Inference}.}\label{tab:inference}
\begin{tabular}{cc}\toprule
Transductive & Inductive \\
\midrule
\makecell[tl]
{GQE datasets~\citeps{gqe} \\ Query2Box datasets~\citeps{q2b} \\ BetaE datasets~\citeps{betae} \\ Regex datasets~\citeps{regex} \\ BiQE dataset~\citeps{biqe} \\ Query2Onto datasets~\citeps{q2b_onto} \\  EFO-1 dataset~\citeps{wang2021benchmarking} \\ TAR datasets~\citeps{abin_abductive2022} \\ SMORE datasets~\citeps{smore} \\ StarQE dataset~\citeps{starqe} \\ TFLEX dataset~\citeps{lin2022tflex} \\ WD50K-NFOL dataset~\citeps{nqe} \\ SQE dataset~\citeps{sqe}} 
& 
\makecell[tl]
{InductiveQE datasets~\citeps{galkin2022} \\ TeMP datasets (type-aware)~\citeps{temp2022} \\ GNNQ dataset~\citeps{gnnq}} \\
\bottomrule
\end{tabular}
\end{table}

\subsection{Metrics}
\label{sec:metrics}

Several metrics have been proposed to evaluate the performance of query reasoning models that can be broadly classified into \textbf{generalization}, \textbf{entailment}, and \textbf{query representation quality} metrics.

\paragraph{Generalization Metrics.}
Since the aim of query reasoning models is to perform reasoning over massive incomplete graphs, 
most metrics are designed to evaluate models' \emph{generalization} capabilities in discovering missing answers, \ie, $\dqte \backslash \dqv$ for a given test query $q$.
As one of the first works in the field, GQE~\citep{gqe} proposes ROC-AUC and average percentile rank (APR). The idea is that for a given test query $q$, GQE calculates a score for all its missing answers $e\in\dqte\backslash\dqv$ and the negatives $e'\notin \dqte$. The model's performance is the ROC-AUC score and APR, where they rank a missing answer against at most 1000 randomly sampled negatives of the same entity type. Besides GQE, GQE+hashing~\citep{gqe_bin}, CGA~\citep{cga} and TractOR~\citep{tractor2020} use the same evaluation metrics.

However, the above metrics do not reflect the real world setting where we often have orders of magnitude more negatives than the missing answers. Instead of ROC-AUC or APR, Query2Box~\citep{q2b} proposes ranking-based metrics, such as mean reciprocal rank (MRR) and hits@$k$. Given a test query $q$, for each missing answer $e\in\dqte\backslash\dqv$, we rank it against all the other negatives $e'\notin \dqte$. Given the ranking $r$, MRR is calculated as $\frac{1}{r}$ and hits@$k$ is $1[r\leq k]$. This has been the most used metrics for the task.
Note that the final rankings are computed only for the \emph{hard} answers that require predicting at least one missing link. 
Rankings for \emph{easy} answers reachable by edge traversal are usually discarded.

\paragraph{Representation Quality Metrics.}
Besides evaluating model's capability of finding missing answers, another aspect is to evaluate the quality of the learned query representation for all models. BetaE~\citep{betae} proposes to evaluate whether the learned query representation can model the cardinality of a query's answer set, and view this as a proxy of the quality of the query representation. For models with a sense of ``volume'' (\eg, differential entropy for Beta embeddings), the goal is to measure the Spearman's rank correlation coefficient and Pearson's correlation coefficient between the ``volume'' of a query (calculated from the query representation) and the cardinality of the answer set. BetaE also proposed to evaluate an ability to model queries without answers using ROC-AUC. 

\paragraph{Entailment Metrics.}
The other evaluation protocol is about whether a model is also able to discover the existing answers, \eg, $\dqv$ for test queries, that does not require inferring missing links but focuses on memorizing the graph structure (\emph{easy} answers in the common terminology). 
This is referred to as faithfulness (or \emph{entailment}) in EmQL~\citep{emql}. 
Natural for database querying tasks, it is expected that query answering models first recover \emph{easy} answers already existing in the graph (reachable by edge traversal) and then enrich the answer set with predicted \emph{hard} answers inferred with link prediction. 
A natural metric is therefore an ability to rank easy answers higher than hard answers -- this was studied by InductiveQE~\citep{galkin2022} that proposed to use ROC-AUC as the main metric for this task.

Still, we would argue that existing metrics might not fully capture the nature of neural query answering and new metrics might be needed.
For example, some under-explored but potentially useful metrics include (1) studying \emph{reasonable} answers (in between easy and hard answers) that can be deduced by symbolic reasoners using a higher-level graph schema (ontology)~\citep{q2b_onto}. 
A caveat in computing reasonable answers is a potentially infinite processing time of symbolic reasoners that have to be limited by time or expressiveness in order to complete in a finite time. Hence, the set of reasonable answers might still be incomplete; (2) evaluation in light of the \emph{open-world assumption} (OWA) stating that unknown triples in the graph might not necessarily be false (as postulated by the standard \emph{closed-world assumption} used a lot in link prediction). 
Practically, OWA means that even the test set might be incomplete and some high-rank predictions deemed incorrect by the test set might in fact be correct in the (possibly unobservable) complete graph. 
Initial experiments of \citet{yang2022rethinking} with OWA evaluation of link prediction explain the saturation of ranking metrics (\eg, MRR) on common datasets by the performance of neural link predictors able to predict the answers from the complete graph missed in the test set.
For example, saturated MRR of 0.4 on the test set might correspond to MRR of 0.9 on the true complete graph.
Studying OWA evaluation in the query answering task in both transductive and inductive setups is a solid avenue for future work.

\section{Applications}
\label{sec:apps}

In addition to graph database use cases, the framework of complex query answering is applied in a variety of graph-conditioned machine learning tasks.
For example, SE-KGE~\citep{se_kge} applies GQE to answer geospatial queries conditioned on numerical $\{x,y\}$ coordinates. 
The coordinates encoder fuses numerical representations with entity embeddings such that the prediction task is still entity ranking.

In case-based reasoning, CBR-SUBG~\citep{cbr_subg2022} is a method for question answering over KGs based on subgraph extraction and encoding. 
As a byproduct, CBR-SUBG is capable of answering conjunctive queries with projections and intersections.
However, due to a non-standard evaluation protocol and custom synthetic dataset, its performance cannot be directly compared to \clqa models. 
Similarly, \citet{wang2023unifying} merge a Query2Box-like model with a pre-trained language model to improve question answering performance.
LEGO~\citep{lego} also applies \clqa models for KG question answering. The idea is to simultaneously parse a natural language question as a query step and execute the step in the latent space with \clqa models.

LogiRec~\citep{wu2022highorder} frames product recommendation as a complex logical query such that source products are root nodes, combinations of multiple products form intersections, and non-similar products to be filtered out form negations. 
LogiRec employs BetaE as a query engine.
Similarly, \cite{syed2022context} design an explainable recommender system based on Query2Box. 
Given a logical query, they first use Query2Box to generate a set of candidates and rerank them using neural collaborate filtering \citep{he2017neural}.  

\section{Summary and Future Opportunities}
\label{sec:future}

We presented a summary of the literature of the \longclqa (\clqa) task, which is at the core of our proposed neural query engine and neural graph database (NGDB) concepts. {\ngdb}s marry the idea of neural database and graph database where we have a unique latent space for information (including entities, relations, facts). NGDBs perform query planning, execution, and result retrieval all in the latent space using graph representation learning. Such a design choice allows for a unified, more efficient and robust interface for storage and querying. We envision {\ngdb}s to be the next generation of databases tailored for predictive queries over sheer volumes of incomplete graph-structured data.

We also proposed a deep and detailed taxonomy of methods 
tackling certain aspects of the envisioned neural query engine.
Going forward, there is still much room to unlock the full power of NGDB by addressing the open challenges in Neural Query Engines and Neural Storage domains. Pertaining to Neural Query Engines and adhering to the taxonomy in \autoref{sec:taxonomy}, we summarize the challenges in three main areas: Graphs, Modeling, and Queries. 

Along the \textbf{Graph} branch:
\begin{itemize}
    \item \textbf{Modality:} Supporting more graph modalities: from classic triple-only graphs to hyper-relational graphs, hypergraphs, and multimodal sources combining graphs, texts, images, and more.
    \item \textbf{Reasoning Domain:} Supporting logical reasoning and neural query answering over temporal and continuous (\eg, textual and numerical) data -- literals constitute a major portion of graphs as well as relevant queries over literals.
    \item \textbf{Background Semantics:} Supporting complex axioms and formal semantics that encode higher-order relationships between (latent) classes of entities and their hierarchies, \eg, enabling neural reasoning over description logics and OWL fragments.
\end{itemize}

In the \textbf{Modeling} branch:
\begin{itemize}
    \item \textbf{Encoder:} Inductive encoders supporting unseen relation at inference time -- this a key for (1) \textbf{updatability} of neural databases without the need of retraining; (2) enabling the \textbf{pretrain-finetune} strategy generalizing query answering to custom graphs with custom relational schema.
    \item \textbf{Processor:} Expressive processor networks able to effectively and efficiently execute complex query operators akin to SPARQL and Cypher operators. Improving sample efficiency of neural processors is crucial for the \emph{training time vs quality} tradeoff, \ie, reducing training time while maintaining high predictive qualities. 
    \item \textbf{Decoder:} So far, all neural query answering decoders operate exclusively on discrete nodes. Extending the range of answers to continuous outputs is crucial for answering real-world queries.
    \item \textbf{Complexity:} As the main computational bottleneck of processor networks is the dimensionality of embedding space (for purely neural models) and/or the number of nodes (for neuro-symbolic), new efficient algorithms for neural logical operators and retrieval methods are the key to scaling {\ngdb}s  to billions of nodes and trillions of edges.  
\end{itemize}

In \textbf{Queries}:

\begin{itemize}
    \item \textbf{Operators:} Neuralizing more complex query operators matching the expressiveness of declarative graph query languages, e.g., supporting Kleene plus and star, property paths, filters.
    \item \textbf{Patterns:} Answering more complex  patterns beyond tree-like queries. This includes DAGs and cyclic graphs.
    \item \textbf{Projected Variables:} Allowing projecting more than a final leaf node entity, that is, allowing returning intermediate variables, relations, and multiple variables organized in tuples (bindings).
    \item \textbf{Expressiveness:} Answering queries outside simple EPFO and EFO fragments, extending the expressiveness to the union of conjunctive queries with negations (UCQ and $\text{UCQ}_{\text{neg}}$ and aiming for the expressiveness of database languages. 
\end{itemize}

In \textbf{Datasets} and \textbf{Evaluation}:

\begin{itemize}
    \item The need for larger and diverse \textbf{benchmarks} covering more graph modalities, more expressive query semantics, more query operators, and query patterns.
    \item As the existing evaluation protocol appears to be limited (focusing only on inferring \emph{hard} answers) there is a need for a more principled \textbf{evaluation framework} and \textbf{metrics} covering various aspects of the query answering workflow.
\end{itemize}

Pertaining to the Neural Graph Storage and {\ngdb} in general, we identify the following challenges:

\begin{itemize}
    \item The need for a \textbf{scalable retrieval} mechanism to scale neural reasoning to graphs of billions of nodes. Retrieval is tightly connected to the \emph{Query Processor} and its modeling priors (as shown in \autoref{sec:storage}). Existing scalable  ANN libraries can only work with basic L1, L2, and cosine distances that limit the space of possible processors in the neural query engine.
    \item Currently, all complex query datasets listed in \autoref{sec:datasets} provide a hardcoded query execution plan that might not be optimal for neural processors. There is a need for a \textbf{neural query planner} that would transform an input query into an optimal execution sequence taking into account prediction tasks, query complexity, type of the neural processor, and configuration of the Storage layer (that can be federated as well).
    \item Due to encoder inductiveness and updatability without retraining, there is a need to alleviate the issues of \textbf{continual learning}~\citep{cont_learning_thrun, cont_learning_ring}, \textbf{catastrophic forgetting}~\citep{mccloskey1989catastrophic}, and \textbf{size generalization} when running inference on much larger graphs than training ones~\citep{DBLP:conf/icml/YehudaiFMCM21,buffelli2022sizeshiftreg}.
\end{itemize}

Finally, in the era of foundation models, many research works have demonstrated various capabilities hidden inside these large language models often with hundreds of billions of parameters, and most importantly how to unleash these capabilities \citep{chain_of_thought,wang2022self,zhou2022least}. We envision an important future direction that is to design a natural language interface so that we can better harness the reasoning capabilities of these large language models for the \clqa task along the directions we mentioned ahead \citep{drozdov2022compositional}. Besides, NGDB also provides an exciting future opportunity to improve foundation models at various stages, especially inference. Since not all the downstream tasks require a full billion-parameter model call, it's promising to research how NGDB can accelerate or compress foundation models while still keeping the emergent behaviors that are essential for these extremely large models.

\section*{Acknowledgements}

We thank Pavel Klinov (Stardog), Matthias Fey (Kumo.AI), Qian Li (Stanford) and Keshav Santhanam (Stanford) for providing valuable feedback on the draft.


\newpage

\vskip 0.2in
\bibliography{bibliography}
\bibliographystyle{tmlr}

\end{document}